\documentclass[longauth]{aa}  
\usepackage{url}
\usepackage{graphicx}
\usepackage{txfonts}
\usepackage[citecolor=blue]{hyperref} 
\usepackage{diagbox}
\usepackage{multirow}
\usepackage{xcolor}
\newcommand{\emm}[1]{\ensuremath{#1}}
\newcommand{\emr}[1]{\emm{\mathrm{#1}}}
\newcommand{\chem}[1]{\emr{#1}} 
\newcommand{\unit}[1]{\emr{\,#1}}
\newcommand{\e}[1]{\emm{\times 10^{#1}}}
\newcommand{\paren}[1]{\emm{\left(  #1 \right) }} % Parenthesis. 
\newcommand{\bracket}[1]{\emm{\left[  #1 \right] }}
\newcommand{\rbracket}[1]{\emm{\left(  #1 \right) }}
\newcommand{\cbracket}[1] {\emm{\left\{ #1 \right\}}}
\newcommand{\pc}{\unit{pc}} 
\newcommand{\pccm}{\unit{cm^{-3}}} 
\newcommand{\pscm}{\unit{cm^{-2}}} 
 
\newcommand{\ccmps}{\unit{cm^{3}\,s^{-1}}} 
\newcommand{\ps}{\unit{s^{-1}}}
\newcommand{\kms}{\unit{km\,s^{-1}}} 
\newcommand{\K}{\unit{K}}
\newcommand{\mK}{\unit{mK}} 
\newcommand{\Kkms}{\unit{K\,km\,s^{-1}}}
\newcommand{\magn}{\unit{mag}}
\newcommand{\GHz}{\unit{GHz}} 
\newcommand{\el}{\chem{e^{-}}}
\newcommand{\HI}{\chem{H\textsc{i}}}
\newcommand{\HH}{\chem{H_2}}
\newcommand{\HHHp}{\chem{H_3^+}}
\newcommand{\twC}{\chem{^{12}C}}
\newcommand{\thC}{\chem{^{13}C}}
\newcommand{\Cp}{\chem{C^{+}}}

\newcommand{\thCp}{\chem{^{13}C^+}}
\newcommand{\CO}{\chem{CO}}
\newcommand{\thCO}{\chem{^{13}CO}}
\newcommand{\twCO}{\chem{^{12}CO}} 
\newcommand{\twCOp}{\chem{^{12}CO^{+}}} 
\newcommand{\thCOp}{\chem{^{13}CO^{+}}} 
\newcommand{\CeiO}{\chem{C^{18}O}}
\newcommand{\HCOp}{\chem{HCO^{+}}}
\newcommand{\HthCOp}{\chem{H^{13}CO^{+}}}
\newcommand{\Jone}{\chem{(1-0)}}
\newcommand{\Jtwo}{\chem{(2-1)}}
\newcommand{\chisq}{$\chi^2$}
\newcommand{\llog}[1]{\log \emm{{#1}} }
\newcommand{\Av}{\emm{A_V}}
\newcommand{\Tdust}{\emm{T_\emr{dust}}}
\newcommand{\Tkin}{\emm{T_\emr{kin}}}
\newcommand{\Tkinl}[1]{\emm{T_{\emr{kin},\, #1}}} % Tkin of the indexed "l" layer of the cloud
\newcommand{\Pth}{\ensuremath{P_\emr{ther}}}
\newcommand{\Pthl}[1]{\ensuremath{P_{\emr{ther},\, #1}}}
\newcommand{\nHH}{\emm{n_{\HH}}}
\newcommand{\nHHl}[1]{\emm{n_{\HH,\, #1}}}        % nH2 of the indexed "l" layer of the cloud
\newcommand{\CV}{\ensuremath{C_V}}
\newcommand{\CVl}[1]{\ensuremath{C_{V,\, #1}}}
\newcommand{\FWHM}{\ensuremath{\text{FWHM}}}
\newcommand{\FWHMl}[1]{\ensuremath{\text{FWHM}_{#1}}}
\newcommand{\sigmaV}{\ensuremath{\sigma_V}}
\newcommand{\sigmaVl}[1]{\ensuremath{\sigma_{V,\, #1}}}
\newcommand{\Tex}{\ensuremath{T_\text{ex}}}
\newcommand{\Texl}[1]{\ensuremath{T_{\text{ex},\, #1}}} % Tex of the indexed "l" layer of the cloud
\newcommand{\taul}[1]{\ensuremath{\tau_\mathrm{#1}}} % opacity of the indexed "l" layer of the cloud
\newcommand{\TCMB}{\ensuremath{T_\mathrm{CMB}}}
\newcommand{\nin}{\nHHl{\emr{in}}}
\newcommand{\NthCO}{\ensuremath{\emm{N}{({\thCO})}}}
\newcommand{\NthCOl}[1]{\ensuremath{\NthCO_{#1}}} % N13Co of the indexed "l" layer of the cloud
\newcommand{\NtwCO}{\ensuremath{\emm{N({\twCO})}}}
\newcommand{\NCeiO}{\ensuremath{\emm{N({\CeiO})}}}
\newcommand{\NHCOp}{\ensuremath{\emm{N({\HCOp})}}}
\newcommand{\NHthCOp}{\ensuremath{\emm{N({\HthCOp})}}}
\newcommand{\NHH}{\ensuremath{\emm{N({\HH})}}}

\newcommand{\SNR}{S/N}
\newcommand{\CRB}{{CRB}}
\newcommand{\CRBm}{\emr{CRB}}
\newcommand{\bI}{\ensuremath{\boldsymbol{I}}\xspace}
\newcommand{\bs}{\ensuremath{\boldsymbol{s}}\xspace}
\newcommand{\bx}{\ensuremath{\boldsymbol{x}}\xspace}
\newcommand{\btheta}{\ensuremath{\boldsymbol{\theta}}\xspace}
\newcommand{\bthetal}[1]{\ensuremath{\boldsymbol{\theta}_{#1}}\xspace}
\newcommand{\thetal}[1]{\ensuremath{\theta_{#1}}\xspace}
\newcommand{\bvector}[1]{\ensuremath{\boldsymbol{#1}}\xspace} % bold vector
\newcommand{\bvectori}[2]{\ensuremath{\boldsymbol{#1}_{#2}}\xspace} % bold vector of index i (unbolded)
\newcommand{\funci}[2]{\ensuremath{{#1}_{#2}}\xspace} % function of index i (unbolded)

\newcommand{\PDR}{{PDR}}
\newcommand{\CMB}{{CMB}}

\newcommand{\Nin}{\emm{N_\emr{in}}} %
\newcommand{\Nou}{\emm{N_\emr{ou}}} %
\newcommand{\Ntot}{\emm{N_\emr{tot}}} %
\newcommand{\ab}[1]{\emm{\bracket{X}}_{#1}} %
\newcommand{\abmean}{\emm{\left< X \right>}} %

\begin{document} 

\title{Toward a robust physical and chemical characterization of
  heterogeneous lines of sight: The case of the Horsehead nebula}

\titlerunning{Toward a robust physical and chemical characterization of
  heterogeneous lines of sight} \authorrunning{Ségal et al.}

\author{%
  % Paper team
  Léontine Ségal\inst{\ref{IRAM},\ref{IM2NP}} %
  \and Antoine Roueff\inst{\ref{IM2NP}} %
  \and Jérôme Pety\inst{\ref{IRAM},\ref{LERMA/PARIS}} %
  \and Maryvonne Gerin\inst{\ref{LERMA/PARIS}} %
  \and Evelyne Roueff\inst{\ref{LERMA/MEUDON}} %
  \and Javier R. Goicoechea\inst{\ref{CSIC}} %
  % Non-permanent researchers by alphabetical order
  \and Ivana Be\v{s}lic\inst{\ref{LERMA/PARIS}} %
  \and Simon Coud\'e\inst{\ref{WORC},\ref{CfA}}%
  \and Lucas Einig\inst{\ref{IRAM},\ref{GIPSA-Lab}}
  \and Helena Mazurek\inst{\ref{LERMA/PARIS}} %
  \and Jan H. Orkisz\inst{\ref{IRAM}} %
  \and Pierre Palud\inst{\ref{CRISTAL},\ref{LERMA/MEUDON}}
  \and Miriam G. Santa-Maria\inst{\ref{UF}, \ref{CSIC}} %
  \and Antoine Zakardjian\inst{\ref{IRAP}}
  % Reminder of consortium by alphabetical order
  \and S\'ebastien Bardeau\inst{\ref{IRAM}} %
  \and Emeric Bron\inst{\ref{LERMA/MEUDON}} %
  \and Pierre Chainais\inst{\ref{CRISTAL}} %
  \and Karine Demyk\inst{\ref{IRAP}} %
  \and Victor de Souza Magalhaes\inst{\ref{NRAO}} %
  \and Pierre Gratier \inst{\ref{LAB}} %
  \and Viviana V. Guzman\inst{\ref{Catholica}} %
  \and Annie Hughes\inst{\ref{IRAP}} %
  \and David Languignon\inst{\ref{LERMA/MEUDON}} %
  \and François Levrier\inst{\ref{LPENS}} %
  \and Jacques Le Bourlot\inst{\ref{LERMA/MEUDON}} %
  \and Franck Le Petit\inst{\ref{LERMA/MEUDON}} %
  \and Dariusz C. Lis\inst{\ref{JPL}} %
  \and Harvey S. Liszt\inst{\ref{NRAO}} %
  \and Nicolas Peretto\inst{\ref{UC}} %
  \and Albrecht Sievers\inst{\ref{IRAM}} %
  \and Pierre-Antoine Thouvenin\inst{\ref{CRISTAL}}%
}

\institute{%
  IRAM, 300 rue de la Piscine, 38406 Saint Martin d'H\`eres,  France. \label{IRAM} %
  \and Université de Toulon, Aix Marseille Univ, CNRS, IM2NP, Toulon, France,  \email{antoine.roueff@univ-tln.fr}. \label{IM2NP} %
  \and LERMA, Observatoire de Paris, PSL Research University, CNRS, Sorbonne Universit\'es, 75014 Paris, France. \label{LERMA/PARIS} %
  \and Instituto de Física Fundamental (CSIC). Calle Serrano 121, 28006, Madrid, Spain. \label{CSIC} %
  \and National Radio Astronomy Observatory, 520 Edgemont Road, Charlottesville, VA, 22903, USA. \label{NRAO} %
  \and Laboratoire d'Astrophysique de Bordeaux, Univ. Bordeaux, CNRS,  B18N, Allee Geoffroy Saint-Hilaire,33615 Pessac, France. \label{LAB} %
  \and Univ. Grenoble Alpes, Inria, CNRS, Grenoble INP, GIPSA-Lab, Grenoble, 38000, France. \label{GIPSA-Lab} %
  \and Chalmers University of Technology, Department of Space, Earth and Environment, 412 93 Gothenburg, Sweden. \label{Chalmers} %
  \and Univ. Lille, CNRS, Centrale Lille, UMR 9189 - CRIStAL, 59651 Villeneuve d’Ascq, France. \label{CRISTAL} %
  \and LERMA, Observatoire de Paris, PSL Research University, CNRS, Sorbonne Universit\'es, 92190 Meudon, France. \label{LERMA/MEUDON} %
  \and Institut de Recherche en Astrophysique et Planétologie (IRAP), Université Paul Sabatier, Toulouse cedex 4, France. \label{IRAP} %
  \and Instituto de Astrofísica, Pontificia Universidad Católica de Chile, Av. Vicuña Mackenna 4860, 7820436 Macul, Santiago, Chile. \label{Catholica} %
  \and Laboratoire de Physique de l’Ecole normale supérieure, ENS, Université PSL, CNRS, Sorbonne Université, Université de Paris, Sorbonne Paris Cité, Paris, France. \label{LPENS} %
  \and Jet Propulsion Laboratory, California Institute of Technology,  4800 Oak Grove Drive, Pasadena, CA 91109, USA. \label{JPL}
  \and Department of Earth, Environment, and Physics, Worcester State University, Worcester, MA 01602, USA \label{WORC}
  \and Harvard-Smithsonian Center for Astrophysics, 60 Garden Street, Cambridge, MA, 02138, USA. \label{CfA}
  \and School of Physics and Astronomy, Cardiff University, Queen's buildings, Cardiff CF24 3AA, UK. \label{UC} %
  \and Department of Astronomy, University of Florida, P.O. Box 112055, Gainesville, FL
  32611. \label{UF}
} %

\date{Received 19 July 2024 / accepted 23 September 2024}

\abstract %
% Context
{Dense and cold molecular cores and filaments are surrounded by an envelope
  of translucent gas.  Some of the low-\textit{J} emission lines of \CO~and
  \HCOp~isotopologues are more sensitive to the conditions either in the
  translucent environment or in the dense and cold one because their
  intensities result from a complex interplay of radiative transfer and
  chemical properties of these heterogeneous lines of sight (LoSs).} %
% Aims
{We extend our previous single-zone modeling with a more realistic approach
  that introduces multiple layers to take account of possibly varying
  conditions along the LoS. We used the IRAM-30m data from the ORION-B large
  program toward the Horsehead nebula in order to demonstrate our method's
  capability and effectiveness.} %
% Methods
{We propose a cloud model composed of three homogeneous slabs of gas along
  each LoS, representing an outer envelope and a more shielded inner layer.
  We used the non-LTE radiative transfer code RADEX to model the line
  profiles from the kinetic temperature (\Tkin), the volume density (\nHH),
  kinematics, and chemical properties of the different layers. We then used
  a fast and robust maximum likelihood estimator to simultaneously fit the
  observed lines of the \CO~and \HCOp~isotopologues. To limit the variance
  on the estimates, we propose a simple chemical model by constraining the
  column densities.} %
% Results
{A single-layer model cannot reproduce the spectral line asymmetries that
  result from a combination of different radial velocities and absorption
  effects among layers. A minimal heterogeneous model (three layers only)
  is sufficient for the Horsehead application, as it provides good fits of the
  seven fitted lines over a large part of the studied field of view. The
  decomposition of the intensity into three layers allowed us to discuss the
  distribution of the estimated physical or chemical properties along the
  LoS.  About 80\% of the \twCO~integrated intensity comes from the outer
  envelope, while $\sim55\%$ of the integrated intensity of the \Jone{} and
  \Jtwo{} lines of \CeiO~comes from the inner layer.  For the lines of the
  \thCO~and the \HCOp~isotopologues, integrated intensities are more
  equally distributed over the cloud layers.  The estimated column density
  ratio \mbox{\NthCO/\NCeiO} in the envelope increases with decreasing
  visual extinction, and it reaches $25$ in the pillar outskirts.  While the
  inferred \Tkin~of the envelope varies from $25$ to \mbox{$40\,\K$}, that
  of the inner layer drops to \mbox{$\sim 15\,\K$} in the 
  western dense core. The estimated \nHH~in the inner layer is
  \mbox{$\sim 3\times10^4\,\pccm$} toward the filament, and it increases by
  a factor of ten toward dense cores.} %
% Conclusions
{Our proposed method correctly retrieves the physical and chemical
  properties of the Horsehead nebula. It also offers promising prospects for
  less supervised model fits of wider-field datasets.} %


\maketitle{} %

%%%%%%%%%%%%%%%%%%%%%%%%%%%%%%%%%%%%%%%%%%%%%%%%%%%%%%%%%%%%%%%%%%%%%%%%%%%

\newcommand{\FigDataMaps}{%
  \begin{figure*}
    \includegraphics[width=\linewidth]{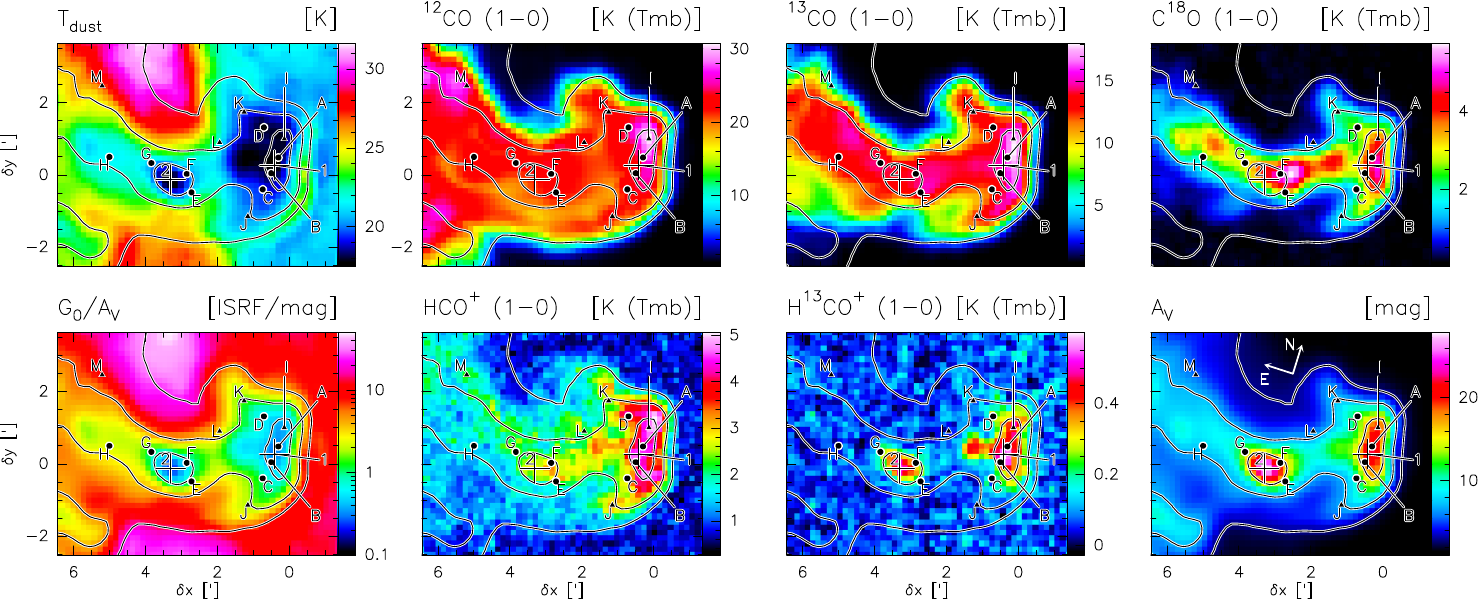}
    \caption{%
      Spatial distribution of the peak main beam temperature
      ($\mathrm{T_{mb}}$) of the \Jone~line of the \CO~and
      \HCOp~isotopologues as well as of the dust temperature
      ($\mathrm{T_{dust}}$), visual extinction ($\Av$), and the ratio of
      the far UV field over the visual extinction ($G_0/\Av$) toward the
      Horsehead nebula. The used projection is the Azimuthal one rotated by
      $14^\circ$ around the center of coordinates
      RA$=05^\emr{h}40^\emr{m}54.27^\emr{s}$, Dec$=-02^\circ{}28'00.00''$
      in the International Celestial Reference System. Contours of
      the dust visual extinction at 3, 7, and 16 magnitudes are overlaid on
      each panel. The crosses and circles show lines of sight that are
      studied in more detail in this paper (see Table~\ref{tab:coord} for
      the corresponding coordinates).}
    \label{fig:data:maps}
  \end{figure*}
}

%%%%%%%%%%%%%%%%%%%%%%%%%%%%%%%%%%%%%%%%%%%%%%%%%%%%%%%%%%%%%%%%%%%%%%%%%%%

\section{Introduction}

\FigDataMaps{}

Star formation takes place in giant molecular clouds (GMCs) with an overall
small efficiency~\citep[][and references
therein]{Shu87,Lada03,Kennicutt12,Schinnerer24}. Current star formation
theory assumes that a bottleneck comes from the formation of dense cores of
density $\ga 10^4\pccm$~\citep[e.g.,][]{Bergin07}, and recent observations
suggest that the galactic environment of a GMC impacts its star formation
efficiency~\citep{Schinnerer24}.

Although GMCs show an internal structure, which can be probed by observing
various molecular lines, a typical representation of a GMC is still
lacking.  This is significant, as it will benchmark the general
understanding of the molecular interstellar medium (ISM) in the Milky Way
and nearby galaxies.  To achieve this representation, realistic large-scale
maps of the cloud physical parameters (column densities, volume density,
kinetic temperature, and kinematics) are needed for a representative sample
of GMCs in the Milky Way. While the current generation of millimeter
observatories, such as the \mbox{IRAM-30m} telescope, NOEMA, or ALMA,
enables surveys of tens of molecular lines over wide fields of
view~\citep[e.g.,][]{Pety17,Kauffmann17,Evans20,Barnes20,
  Tafalla21,Dame23}, there is still a lack of fully relevant methods to
easily extract the physical and chemical information that these lines
encode.

It is thus important to develop fast and rigorous methods to infer 2D maps
of cloud physical parameters from modern-day large-scale, well-resolved
multi-line maps. In order to accurately fit the observed line intensities
and profiles, these methods must take into account that lines of sight
(LoSs) contain gas components with different physical conditions and
chemical compositions, which often interact radiatively.  Previous studies
based on coupled non-local thermodynamic equilibrium (non-LTE) radiative
transfer and chemical models were designed to fit high angular resolution
multi-line observations toward either a single line of sight of interest or
narrow fields of view~\citep[see, e.g.,][toward the Horsehead
PDR]{Goicoechea06,Goicoechea09}. With the ability to conduct large mapping
surveys of molecular gas lines at high spatial resolution, the main
challenge now is to develop robust methods for much larger areas.

The simplest modeling geometry to do this with is a homogeneous LoS
composed of a single component. However, the line emission maps of
\citet{Pety17,Kauffmann17,Barnes20} suggest that the emission comes from
gas with different physical conditions, at least along the LoSs toward
filaments and dense cores. \citet{Roueff24} document inconsistencies in the
inferred gas conditions when simultaneously analyzing different molecular
lines with a non-LTE radiative transfer cloud model of a homogeneous slab
of gas. For instance, simultaneously fitting different sets of lines (such
as \twCO, \thCO, and \HCOp{} on the one hand and \CeiO{} and \HthCOp{} on
the other) in the same field of view yields drastic differences in inferred
volume densities up to one order of magnitude. This suggests that model
misspecifications prevent us from deciphering the information provided by
emission lines. In this study, we develop a three-layer model to consider
the heterogeneous conditions along the LoSs toward filaments or dense cores
while keeping the model as simple as possible. This multi-layer cloud model
is based on routinely used radiative transfer methods. In spite of its
simplicity, it allows us to analyze the contribution of the different
layers to the line emission in addition to delivering an estimation of the
physical quantities with their confidence interval.

The region under study is the Horsehead nebula (Barnard~33), located in the
Orion B cloud. This region is an ideal candidate for evaluating the model's
performance for several reasons. First, this nebula hosts a diversity of
physico-chemical regimes such as photon dissociation regions (\PDR s);
translucent regions (\mbox{$\Tkin>30\,\K$},
\mbox{$\nHH\lesssim10^{4}\,\pccm$}); and dense cores
(\mbox{$\Tkin<15\,\K$}, \mbox{$\nHH\gtrsim10^{5}\,\pccm$}).  Corresponding
density and temperature gradients (e.g., within the
\PDR{}~\cite{Hernandezvera23}) and steep changes of the gas chemical state
detected along the LoS suggest that a standard radiative transfer analysis
within a homogeneous cloud model is a too simple description for this
region. Second, the Horsehead nebula has been the subject of numerous
studies, from the analysis of its
kinematics~\citep[e.g.,][]{Hilyblant05,Gaudel23} to the dissection of the
chemistry and structure of its \PDR~\citep[e.g.,][]{Gerin09,
  Guzman11,Guzman12,Pety12,Gratier13,Guzman13,Guzman14,Guzman15,Fuente17,
  Hernandezvera23,Abergel24}.  Applying our model to this familiar, albeit
complex object, is therefore an opportunity to better diagnose the
relevance of the proposed technique.

The paper is organized as follows. Section~\ref{sec:obs} briefly describes
the observations. The radiative transfer and noise models as well as the
fitting method and the estimation of the confidence intervals are described
in Sect.~\ref{sec:methods}. The physical and chemical assumptions are
justified in Sect.~\ref{sec:assumptions}. Section~\ref{sec:results}
presents the results on the intensity decomposition, the gas velocity
field, the estimated chemical abundances, and the physical conditions
(kinetic temperature and volume density). Section~\ref{sec:discussion}
discusses the estimated abundances. Finally, Sect.~\ref{sec:conclusion}
summarizes our findings.

%%%%%%%%%%%%%%%%%%%%%%%%%%%%%%%%%%%%%%%%%%%%%%%%%%%%%%%%%%%%%%%%%%%%%%%%%%%

\newcommand{\TabLines}{%
  \begin{table}
    \caption{Characteristics of the studied molecular lines.  }
    \label{tab:lines}
    \centering %
    \resizebox{\linewidth}{!}{%
      \begin{tabular}{lcrccr}
        \hline
        \hline
        Species & Transition & Frequency & Vel. Res. & Ang. Res. & $\sigma_b^{(a)}$ \\
                &            & [\GHz]    & [\kms]    & $['']$    & [\mK] \\
        \hline
        \twCO   & \Jone & 115.271202 & 0.5 & 31 &  91 \\ 
        \thCO   & \Jone & 110.201354 & 0.1 & 31 &  74 \\
                & \Jtwo & 220.398684 & 0.1 & 31 & 155 \\
        \CeiO   & \Jone & 109.782173 & 0.1 & 31 &  72 \\
                & \Jtwo & 219.560354 & 0.1 & 31 & 121 \\
        \HCOp   & \Jone &  89.188525 & 0.1 & 31 & 318 \\
        \HthCOp & \Jone &  86.754288 & 0.5 & 31 &  45 \\
        \hline       
      \end{tabular}
    } %
    \tablefoot{%
      \tablefoottext{a}{Median of standard deviations of signal-free
        channels, after resampling and smoothing (see
        Sect.~\ref{sec:noise}). This corresponds to the typical baseline
        noise level at the angular and spectral resolutions listed in the
        table.} %
    }
  \end{table}
}

\newcommand{\TabCoord}{%
  \begin{table}
    \centering %
    \caption{Coordinates of the lines of sight marked in
      Fig.~\ref{fig:data:maps}. DC and LoS refer to dense core and line of
      sight, respectively.}
    \resizebox{\linewidth}{!}{%
      \begin{tabular}{cccc}
        \hline
        \hline
        Name     & R.A.                               & Dec.                 & $(\delta x,\delta y)$\\
        \hline
        DC~1  & $05^\emr{h}40^\emr{m}55.6^\emr{s}$ & $-02^\circ{}27'38''$ & $(+025'',+17'')$ \\
        LoS~A & $05^\emr{h}40^\emr{m}54.8^\emr{s}$ & $-02^\circ{}27'28''$ & $(+018'',+29'')$ \\
        LoS~B & $05^\emr{h}40^\emr{m}56.2^\emr{s}$ & $-02^\circ{}27'50''$ & $(+030'',+03'')$ \\
        LoS~C & $05^\emr{h}40^\emr{m}57.5^\emr{s}$ & $-02^\circ{}28'13''$ & $(+044'',-24'')$ \\
        LoS~D & $05^\emr{h}40^\emr{m}55.7^\emr{s}$ & $-02^\circ{}26'33''$ & $(+042'',+79'')$ \\
        \hline
        DC~2  & $05^\emr{h}41^\emr{m}07.2^\emr{s}$ & $-02^\circ{}27'19''$ & $(+198'',-08'')$ \\
        LoS~E & $05^\emr{h}41^\emr{m}05.3^\emr{s}$ & $-02^\circ{}27'49''$ & $(+163'',-29'')$ \\
        LoS~F & $05^\emr{h}41^\emr{m}05.3^\emr{s}$ & $-02^\circ{}27'17''$ & $(+171'',+02'')$ \\
        LoS~G & $05^\emr{h}41^\emr{m}08.8^\emr{s}$ & $-02^\circ{}26'45''$ & $(+230'',+20'')$ \\
        LoS~H & $05^\emr{h}41^\emr{m}13.2^\emr{s}$ & $-02^\circ{}26'18''$ & $(+300'',+30'')$ \\
        \hline
        LoS~I & $05^\emr{h}40^\emr{m}53.7^\emr{s}$ & $-02^\circ{}27'03''$ & $(+008'',+60'')$ \\
        LoS~J & $05^\emr{h}40^\emr{m}59.8^\emr{s}$ & $-02^\circ{}28'50''$ & $(+069'',-69'')$ \\        
        LoS~K & $05^\emr{h}40^\emr{m}57.4^\emr{s}$ & $-02^\circ{}26'01''$ & $(+075'',105'')$ \\
        LoS~L & $05^\emr{h}41^\emr{m}00.9^\emr{s}$ & $-02^\circ{}26'40''$ & $(+116'',+54'')$ \\
        LoS~M & $05^\emr{h}41^\emr{m}13.2^\emr{s}$ & $-02^\circ{}24'40''$ & $(+324'',126'')$ \\
        \hline
      \end{tabular}
    }
    \label{tab:coord}
  \end{table}
}

%%%%%%%%%%%%%%%%%%%%%%%%%%%%%%%%%%%%%%%%%%%%%%%%%%%%%%%%%%%%%%%%%%%%%%%%%%%

\section{Observations}
\label{sec:obs}

In this study, we use the same data as in~\citet{Roueff24}. As they are
fully described there, we only summarize useful information here. The field
of view contains the Horsehead nebula. Figure~\ref{fig:data:maps} shows the
spatial distribution of various tracers of the interstellar gas and
dust. The data include the low-$J$ transitions of the isotopologues of
carbon monoxide (\twCO, \thCO, and \CeiO) and formylium (\HCOp and
\HthCOp), observed with the IRAM-30m telescope, as part of the ORION-B
large program (PIs: J.~Pety \& M.~Gerin). Table~\ref{tab:lines} lists the
properties of the studied emission lines.

The data also include maps of the dust visual extinction ($\Av$) and of
dust temperature (\Tdust) derived by~\citet{Lombardi14} and~\citet{Pety17}
from the Herschel Gould Belt Survey data~\citep{Andre10}. We used the far
UV illumination parameter $(G_0)$ derived by~\citet{santamaria23} to
compute a map of the ratio $G_0/\Av$. Assuming that there is some relation
between the volume and column densities, $G_0/\Av$ is a proxy of the ratio
of the far UV field by the volume density, which controls to first order
the chemistry of PDRs. In this study, the spatial distribution of \Av,
\Tdust, and $G_0/\Av$ only serve i) as references to locate the different
media that are present in the Horsehead nebula, and ii) to compute the
chemical abundances. In particular, they are not taken into account as
prior knowledge to characterize the gas physical properties along each LoS.

\TabLines{} %

We identify in Fig.~\ref{fig:data:maps} two dense cores where
\mbox{$\Av>16\magn$} and \mbox{$\Tdust<20\K$}. These cores correspond to
the LoSs where the \mbox{\HthCOp\,\Jone} line is clearly detected.  The
maximum of the \mbox{\HthCOp\,\Jone} intensity also corresponds to the
maximum of the visual extinction or gas column density. In each core, the
\mbox{\CeiO\,\Jone} emission peaks are offset compared to the peaks of the
\HH{} column density.  This might be due to \CeiO~depletion from the gas
phase, as CO is known to freeze on grain mantles when the dust grains are
cold enough~\citep[\mbox{$\Tdust \leq 15\K$},
e.g.,][]{Tafalla02,Hilyblant05}.

Outside these dense cores, the line emission from \mbox{\CeiO\,\Jone} is
mostly associated with a filamentary structure where \mbox{$\Av > 7\magn$}.
The \twCO~and \mbox{\thCO\,\Jone} emission is relatively bright down to
\mbox{$\Av \simeq 3\magn$}. Conversely, the emission of these two lines
shows no obvious spatial pattern where the two dense cores are seen at high
visual extinction. The \mbox{\HCOp\,\Jone} emission fills most of the
regions with \mbox{$\Av > 3\magn$}. But it shows a large east-west emission
gradient, probably related to the far UV illumination from the
$\sigma\,$Ori star that excites the IC\,434 H\textsc{ii} region~\citep[see,
e.g.,][]{Ochsendorf2014}. Another striking feature in
Fig.~\ref{fig:data:maps} is the variation of the signal-to-noise ratio
(\SNR) by more than one order of magnitude as a function of the emission
line. The \twCO~and \mbox{\thCO\,\Jone} lines have a maximum S/N of 369 and
242, respectively, while the \HCOp~and \mbox{\HthCOp\,\Jone} maximum
\SNR~are 18.4 and 12.5, respectively. This mostly reflects the relative
line intensities as broadband receivers deliver similar receiver noise over
their bandpass.

\TabCoord{} %

Table~\ref{tab:coord} lists the coordinates of LoSs marked in
Fig.~\ref{fig:data:maps}. These LoSs are studied in greater detail in
Sect.~\ref{sec:results}. The positions DC\,1 and DC\,2 represent the
centers of the eastern and western dense cores that belong to the Horsehead
nebula. The LoSs (A, B, C, D) and (E, F, G, H) allowed us to study
variations inside or in the outskirt of these two dense cores. There is an
embedded young stellar object toward LoS I.  Finally, the LoSs (J, K, L, M)
are representative of the more translucent gas around the Horsehead pillar.

The visible extinction estimated from the dust spectral energy distribution
is a measure of all hydrogen atoms present along each LoS. We needed to
correct it for two effects in order to meaningfully match it with the
results of our line fitting method. First, \citet{Pety17} estimate that
about 1 mag of visual extinction is coming from atomic hydrogen. Second,
the southwestern edge of the Orion B GMC has typically two velocity
components centered on 5 and 10.5\kms{}~\citep{Pety17}, while the method
proposed here assumes that only one velocity component is
present. Appendix~\ref{app:av:correction} details how we corrected the
visual extinction to match the amount of molecular gas associated with the
main velocity component at 10.5\kms{}. From this point on, we only use this
corrected visual extinction, which we call ``molecular'' visual extinction.

%%%%%%%%%%%%%%%%%%%%%%%%%%%%%%%%%%%%%%%%%%%%%%%%%%%%%%%%%%%%%%%%%%%%%%%%%%%

% JP: There is still a notion of top and bottom. Let's keep these words.

\newcommand{\FigModelSketch}{%
  \begin{figure}
    \includegraphics[width=\linewidth]{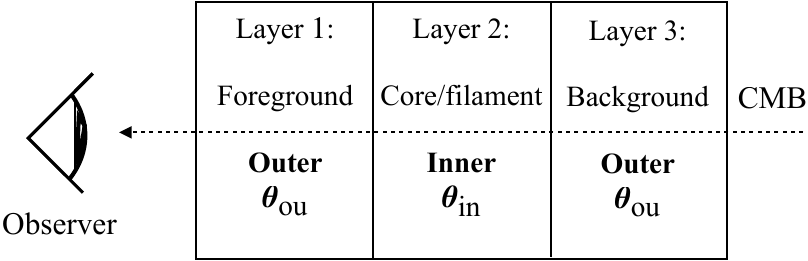}
    \caption{Sketch of the multilayer model of a LoS.  \textbf{Top:} Line
      of sight decomposed into three layers: the foreground, inner, and
      background layers. \textbf{Bottom:} Inner slab (layer 2) surrounded
      by two outer slabs that share the same physical and chemical state.
      Each region has homogeneous physical and chemical conditions (noted
      $\bthetal{\text{in}}$ and $\bthetal{\text{ou}}$, respectively).}
    \label{fig:sandwich}
  \end{figure}
}

\newcommand{\FigDataSpectra}{%
  \begin{figure*}
    \centering %
    \includegraphics[width=0.95\linewidth]{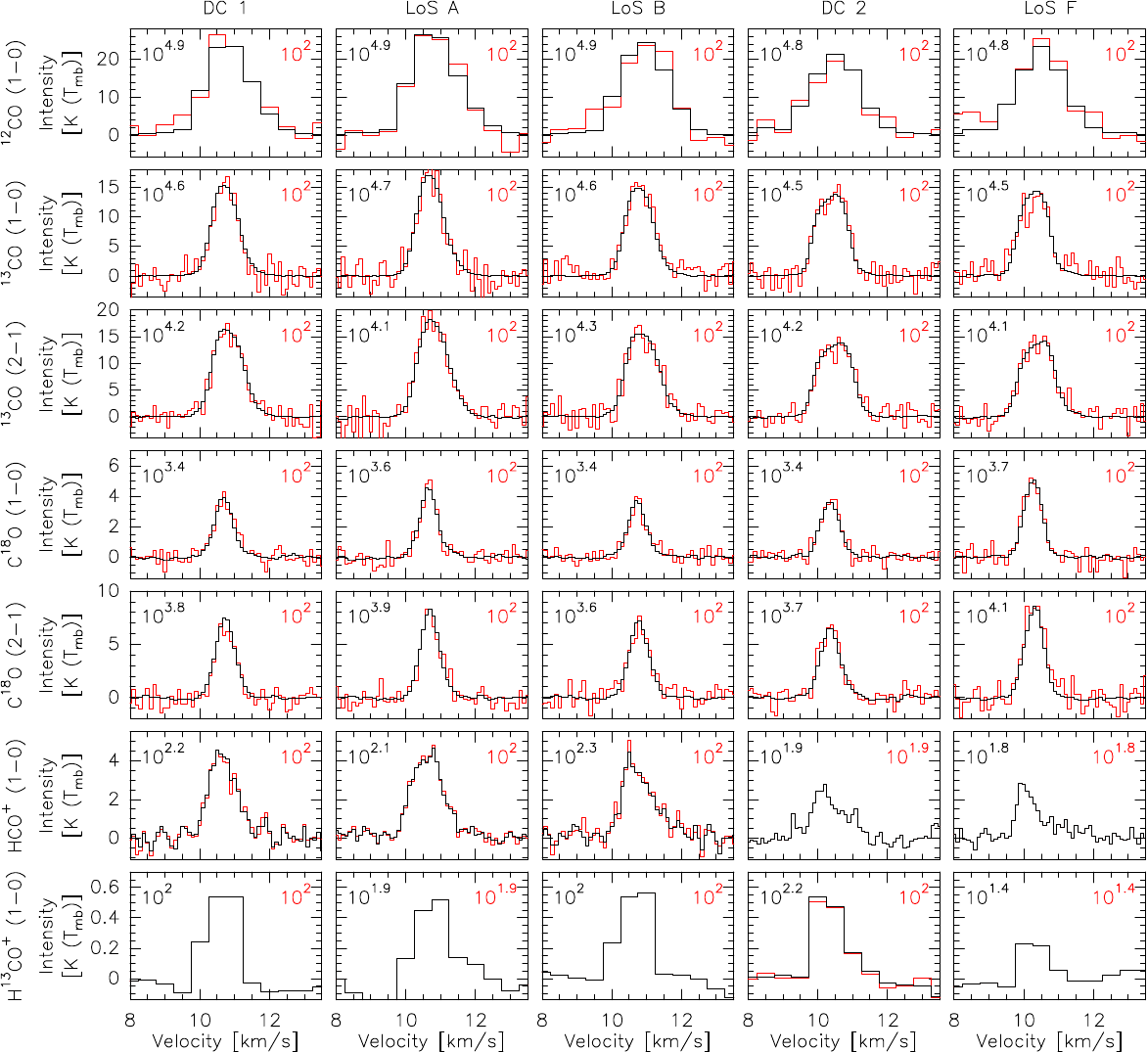}
    \caption{Comparison of five observed spectra (black histograms) and the
      same ones plus the required amount of noise to saturate their peak
      \SNR{} to ten (red histograms). The positions of the spectra are
      marked on Fig.~\ref{fig:data:maps} and listed in
      Table~\ref{tab:coord}. Each row corresponds to a given species and
      transition, while each column corresponds to a given position. In
      each panel, the top left and right numbers give the relative weight
      of a difference of 1\,K in any given channel on the fit \chisq{}
      value before and after the \SNR{} saturation, respectively. The
      \twCO{} \Jone{} and \HthCOp{} \Jone{} lines were observed with a
      resolution about 5 times coarser than the other lines.  When the peak
      \SNR{} is lower than 10, the red spectrum is hidden under the black
      one.}
    \label{fig:data:spectra}
  \end{figure*}
}

\newcommand{\TabCollisionFile}{%
  \begin{table}
    \caption{References of the rate coefficients used in this article to
      compute the collisional excitation.}
    \centering %
    \begin{tabular}{ccll}
      \hline
      \hline
      Species & Collider & Database & Reference \\
      \hline
      $\twCO$   & \HH & LAMDA\tablefootmark{1} &~\citet{yang2010}         \\
      $\thCO$   & \HH & LAMDA                  &~\citet{yang2010}         \\
      $\CeiO$   & \HH & LAMDA                  &~\citet{yang2010}         \\
      $\HCOp$   & \HH & EMAA\tablefootmark{2}  &~\citet{denisalpizar2020} \\
      $\HCOp$   & \el & EMAA                   &~\citet{Fuente:2008}      \\
      $\HthCOp$ & \HH & EMAA                   &~\citet{denisalpizar2020} \\
      $\HthCOp$ & \el & EMAA                   &~\citet{Fuente:2008}      \\
      \hline
    \end{tabular}
    \tablefoot{%
      \tablefoottext{1}{Leiden Atomic and Molecular Database (LAMDA) at
        \url{https://home.strw.leidenuniv.nl/~moldata/}}. %
      \tablefoottext{2}{Excitation of Molecules and Atoms for Astrophysics
        (EMAA) at \url{https://emaa.osug.fr}.}}
    \label{tab:collision:file}
  \end{table}
}

%%%%%%%%%%%%%%%%%%%%%%%%%%%%%%%%%%%%%%%%%%%%%%%%%%%%%%%%%%%%%%%%%%%%%%%%%%%

\section{Methods}
\label{sec:methods}

The emission from the \CO~and \HCOp~isotopologues seems to be sensitive to
the different gas regimes that are present along the LoS throughout the
Horsehead nebula. Cold and dense gas in
cores~\citep[\mbox{$\Tkin\sim15\,\K$},
\mbox{$\nHH\geq10^5\,\pccm,$}][]{WardThompson06} is surrounded by gas with
intermediate temperature and density in filaments
(\mbox{$\Tkin \in \bracket{20, 40}\,\K$},
\mbox{$\nHH \sim 10^{4}\,\pccm$}), and an envelope of warm and translucent
gas ~\citep[\mbox{$\Tkin \in \bracket{40,100}\,\K$},
\mbox{$\nHH \sim 10^{3}\,\pccm$}, see, e.g.,][]{Goicoechea06}.
\citet{Roueff24} show the benefits of simultaneous fitting \CO~and
\HCOp~isotopologue lines to probe the Horsehead nebula environments, but
highlight the limitations and biases induced by a homogeneous model for
describing these very molecular lines. This paper studies how these
limitations are overcome by fitting the same lines with a heterogeneous
model. In this section we introduce the model for an heterogeneous line of
sight and the methods to fit it efficiently.

\subsection{Modeling a heterogeneous line of sight}
\label{sec:model}

\FigModelSketch{} %

We aim to model the emission from a dense core or a filament surrounded by
translucent gas. To do this, we model each LoS as three successive layers
starting from the observer as illustrated in the top part of
Fig.~\ref{fig:sandwich}: The foreground layer (layer 1), the inner layer
(layer 2), the background layer (layer 3). The contribution from the Cosmic
Microwave Background (CMB) is considered separately as it is not emitted by
the ISM cloud. We nickname this specific multilayer structure ``sandwich
model'' because the foreground and background layers are part of the same
outer layer and share the same physical and chemical conditions (see the
bottom part of Fig.~\ref{fig:sandwich}). In contrast, the inner layer has
physical and chemical conditions that are distinct from those of the outer
layer.

Following~\citet{Myers96}, we assume 1) that the line excitation is local
and the level populations in a slab do not depend on the physical
conditions of other layers, but 2) that the emission of one layer is
absorbed by the layers located between it and the observer. In other words,
we neglect the non-local radiative coupling between layers, and the
scattering of line photons of the core by the envelope, that is, the
absorption and reemission of line photons coming from the core. This is a
good approximation for CO and for optically thin lines, but it will be less
exact for very subthermal $(\Tex \ll \Tkin)$ optically thick lines.  These
assumptions are consistent with the escape probability radiative transfer
method that is used to model the emission from each layer~\citep[see,
e.g.,][]{VanDerTak07}.  This approximation allows us to easily decompose
the contribution of the different layers to the observed spectrum.

The brightness temperature $s$ at observed frequency $\nu$ can be written
as
\begin{equation}
  s(\nu)= s_\emr{fore}(\nu) + s_\emr{inner}(\nu) + s_\emr{back}(\nu) + s_\emr{CMB}(\nu),
  \label{eq:s:total}
\end{equation}
where $s_L(\nu)$ is the brightness temperature of the layer
\mbox{$L\in\{\emr{fore}, \emr{inner}, \emr{back}\}$} attenuated by the gas
in front of it. For any layer $L$, it can be written as
\begin{equation}
  \label{eq:s:layer}
  s_L(\nu) %
  = J(\Texl{L},\nu) \, %
  \bracket{ 1 - \exp \rbracket{-\Psi_{L}(\nu)}} \, %
  \exp \rbracket{- \sum_{i = 1}^{L-1} \Psi_{i}(\nu)}, %
\end{equation}
where $\Psi_i(\nu)$ is the opacity profile of layer $i$ and the specific
intensity of layer $L$, expressed in temperature units, is
\begin{equation}
  \label{eq:J}
  J(\Texl{L}, \nu) = \frac{h \nu}{k} \frac{1}{\exp \rbracket{\frac{h \nu}{k \Texl{L}}} -1}.
\end{equation}
The constants $h$ and $k$ in Eq.~\ref{eq:J} are the Planck and Boltzmann
constants, respectively.  The specific intensity of layer $L$,
\mbox{$J(\Texl{L}, \nu)$}, is affected by the layer itself through the term
\mbox{$\bracket{ 1 - \exp \rbracket{-\Psi_L(\nu)}}$}, and by the gas of the
layers separating the layer $L$ from the observer through the term
\mbox{$\exp \rbracket{- \sum_{i = 1}^{L-1} \Psi_i(\nu)}$}.  The excitation
temperature $\Texl{L}$ between the upper and lower energy levels is related
to the ratio of the populations of these energy levels through
\begin{equation}
  \label{eq:Tex}
  \frac{n_\emr{up}}{n_\emr{low}} = \frac{g_\emr{up}}{g_\emr{low}} \exp \rbracket{- \frac{h \nu}{k \Texl{L}}},
\end{equation}
where $g_\emr{up}$ and $g_\emr{low}$ are the degeneracies of the associated
levels~\citep[][Chap.~7, Eq.~7.8]{Draine2011}.

Each molecular transition provides a spectrum around the rest frequency
$\nu_{l}$ of the emission line $l$ in the source frame. When dealing with
several molecular lines simultaneously, it is more convenient to consider
opacity profiles as a function of velocity $V$ rather than frequency $\nu$.
These are related through the Doppler effect expressed with the radio
convention as
\begin{equation}
  \label{eq:Doppler}
  \frac{\nu}{\nu_l} = 1 - \frac{V}{c},
\end{equation}
where $c$ is the velocity of light.  Assuming that the distribution of the
gas velocities, which results from turbulent and thermal motions, is a
Gaussian, centered on the centroid velocity \CVl{L}, and of velocity
dispersion $\sigmaVl{L}$, we can express the opacity profile for layer $L$
and line $l$ as a function of the velocity as
\begin{equation}
  \label{eq:opacity:profile}
  \Psi_{l,L}(V) = \tau_{l,L} \exp \bracket{ - \frac{\rbracket{V - \CVl{L}}^2}{2 \sigmaVl{L}^2} },
\end{equation}
where $\tau_{l,L}$ is the optical depth at the center of line $l$ for the
layer $L$.  We assume that the centroid velocity \CVl{L}~and velocity
dispersion \sigmaVl{L}~are identical for all molecular lines in this layer.
They thus only depend on the layer $L$. The full width at half maximum
(FWHM) is related to the velocity dispersion \sigmaV~by
\begin{equation}
  \label{eq:fwhm}
  \text{FWHM} = \sqrt{8 \ln{2}} \, \sigmaV.
\end{equation}
Finally, as the lines observed in the ISM have narrow widths, we can assume
that \mbox{$J(\Texl{L},\nu)\approx J(\Texl{L}, \nu_l)$}. For the sake of
readability, we hereafter note $J(\Texl{L},\nu)$ as $J(\Texl{L})$ and
\mbox{$\Psi_L(V) = \Psi_L$}.

In summary, the spectrum observed at the output of the simplified sandwich
model can be written as
\begin{equation}
  s(V)= s_\emr{fore}(V) + s_\emr{inner}(V) + s_\emr{back}(V) + s_\emr{CMB}(V),
  \label{eq:sandwich:total}
\end{equation}
with
\begin{eqnarray}
  s_\emr{fore}  & = & J(\Texl{1}) \bracket{ 1 - \exp \rbracket{-\Psi_1}},                                   \label{eq:fore} \\ 
  s_\emr{inner} & = & J(\Texl{2}) \bracket{ 1 - \exp \rbracket{-\Psi_2}} \exp \rbracket{- \Psi_1},          \label{eq:inter}\\
  s_\emr{back}  & = & J(\Texl{1}) \bracket{ 1 - \exp \rbracket{-\Psi_1}} \exp \rbracket{- \Psi_1 - \Psi_2}, \label{eq:back} \\ 
  s_\emr{CMB}   & = & J(\TCMB)    \bracket{\exp \rbracket{- 2 \Psi_1 - \Psi_2}} - J(\TCMB),                 \label{eq:CMB}
\end{eqnarray} 
where the $J(\Tex)$ and $\Psi$ terms depend on the line $l$, and the $\Psi$
terms in addition depends on the velocity $V$, but $J(\TCMB)$ only slowly
depends on the frequency. In these equations, we used the symmetry of the
slabs to set \mbox{$\Psi_1=\Psi_3$} (see
Fig.~\ref{fig:sandwich}). Moreover, the contribution from the CMB has a
different expression as compared to the other layers. The CMB is a
continuum emission from a source located at an ``infinite'' distance from
the observer. This CMB emission is thus absorbed by the inner and outer
layers in front of it (i.e.,
\mbox{$J(\TCMB) \bracket{\exp \rbracket{- 2 \Psi_1 - \Psi_2}}$}).  However,
the calibrated spectrum obtained with the IRAM-30m radiotelescope results
from the subtraction of the observation toward a reference LoS devoid of
signal coming from the source (named OFF-position) from the observation
toward the ON-source LoS. Assuming that the CMB emission is homogeneous and
isotropic on the sky, the signal from the OFF-source LoS only contains the
CMB emission. Furthermore, we assume that the \CMB{} along the OFF-source
LoS is not affected by any absorption.  The ON-OFF process thus effectively
subtracts $J(\TCMB)$ from the observed line signal.  We fold this into our
model by subtracting $J(\TCMB)$ from the \CMB~contribution to the modeled
line emission (i.e., $s_\emr{CMB}$ in Eq.~\ref{eq:CMB}). Finally, we use as
\CMB~temperature \mbox{$\TCMB \simeq 2.73\K$}~\citep{Mather94}.
Appendix~\ref{app:m-layers} generalizes this formalism to an arbitrary
number of layers.

\subsection{Using the RADEX code to compute the emission and opacity of
  each layer}
\label{sec:radex}

\TabCollisionFile{} %

We used the non-LTE RADEX code~\citep{VanDerTak07} to compute the
excitation temperature (\Tex) and opacity $(\tau)$ for each layer and each
line. RADEX allows one to compute large grids of models in reasonnable
times.  To do this, RADEX uses a local approach, namely the escape
probability radiative transfer method, to derive the populations of the
different energy levels of the chosen molecular species. The references of
the collisional excitation rate coefficients are listed in
Table~\ref{tab:collision:file}. We assume that the collisional partners are
molecular hydrogen for all species, plus electrons for the high-dipole
moment species \HCOp~and \HthCOp~\citep{Goldsmith17}.  The detailed
prescriptions for the ortho-to-para ratio of the molecular hydrogen and for
the electron density can be found in Sect.~2.1 of~\citet{Roueff24}.

We have chosen to use the escape probability for a static uniform sphere.
A spherical geometry appears better suited than a plane parallel geometry,
because it allows photons to escape in all directions, and molecular clouds
have a complex self-similar structure.  In the chosen multilayer model, the
velocity field is independently fitted in each layer and the layers are not
radiatively coupled (i.e., the radiation field of a layer does not
contribute to the level population of the gas in another layer).
Therefore, we have chosen to consider each layer as a uniform medium.
Using the expanding sphere geometry is less appropriate because this
physical description requires no discontinuity for the centroid velocity
between the layers. The expanding sphere model also implicitly assumes that
the outer layer has a larger velocity dispersion than the inner layer
because the expansion velocity field reaches higher values further away
from the center.  For a single-layer model, the influence of the choice of
the escape probability law has been discussed by~\cite{VanDerTak07}. At low
column densities, the uniform sphere model and the expanding spherical
model lead to similar emergent intensities within about 10\% while the
plane parallel slab predicts higher line opacities and intensities. The
difference is more significant at high column densities.  This means that
the choice of the escape probability law impacts the estimation of the
physical conditions in a systematic way, but the difference between the
estimations remain moderate (less than 50\% for the density) for lines with
low to moderate opacities (less than $\sim10$). Anyway, the proposed method
can be used with another choice of the escape probability, depending on the
science context.

\subsection{Estimated input parameters}
\label{sec:fit:unknowns}

\FigDataSpectra{}

For each species and each layer, the physical conditions are characterized
by five unknown parameters: the kinetic temperature ($\Tkin$), the volume
density of molecular hydrogen ($\nHH$), the column density ($N$) of the
considered species, the velocity dispersion ($\sigma_V$) expressed as a
full width at half maximum (FWHM), and the centroid velocity ($C_V$). Since
the foreground and background layers are part of the same outer layer, they
are characterized by the same five parameters. For each species, the
sandwich model is therefore characterized by ten unknown parameters,
\begin{eqnarray}
  \label{eq:theta:sandwich}
  \btheta &= \{&\llog{\Tkinl{\text{ou}}}, \llog{\nHHl{\text{ou}}},
                 \llog{N_{\text{ou}}}, \text{FWHM}_{\text{ou}}, \CVl{\text{ou}}, \notag \\
          &    &\llog{\Tkinl{\text{in}}}, \llog{\nHHl{\text{in}}},
                 \llog{N_\text{in}}, \text{FWHM}_{\text{in}}, \CVl{\text{in}} \}.
\end{eqnarray}
We assumed that the five species are co-located inside each layer. The
species thus share the same $\Tkin$, $\nHH$, $C_V$, and FWHM per layer.

Constraining the column density of each species in each layer is complex
because we only have one or two transitions per species. On the one hand,
estimating all these column densities simultaneously is inefficient because
there is not enough information to get accurate estimations. On the other
hand, \citet{Roueff24} show that fixing the ratio of column densities can
bias the estimation of both the column densities and the other physical
parameters such as the volume density when the injected constraints are
invalid. To work around this issue, we discuss in
Sect.~\ref{sec:assumptions} the choices of chemical assumptions made to
decrease the number of fitted parameters.

\subsection{Noise model}
\label{sec:noise}

A noise model is required 1) to quantify the precision of the estimates,
and 2) to derive the maximum likelihood estimator (MLE), which is used to
fit the data. We design two slightly different noise models to meet the
specific needs of each goal. From the first perspective, we take all noise
sources into account in order to avoid underestimating the achievable
precision (i.e., estimating too small error bars). In particular, our model
takes into account not only the thermal additive noise, but also a
calibration multiplicative noise that has a significant effect at high
S/N~\citep[see][]{einig23}. However, when deriving the MLE, we omit the
calibration multiplicative noise because \citet{Roueff24} observe that
taking it into account leads to correct results on Monte Carlo simulations,
but poorer and inconsistent fits on actual data. They suggest that, in the
case of real data, this additional flexibility of the fit is used to deal
not only with calibration errors but also with model misspecifications, the
latter effect being more significant than the former one.  We now detail
the full noise model used to quantify the precision on estimates. An
observed spectrum $\bvectori{x}{l}$ for line $l$ is a realization of the
signal (noted $\bvectori{s}{l}$) from Eq.~\ref{eq:sandwich:total} regularly
sampled in velocity affected by an additive thermal noise $\bvectori{b}{l}$
and a multiplicative calibration noise $c$
\begin{equation}
  \label{eq:x}
  \bvectori{x}{l} = c \, \bvectori{s}{l}  \,+\, \bvectori{b}{l}.
\end{equation}
The additive noise $\bvectori{b}{l}$ is drawn from a centered Gaussian
distribution with standard deviation $\sigma_{b,l}$, whose typical values
are listed in Table~\ref{tab:lines}. The calibration noise is drawn from a
Gaussian distribution of mean $1.0$ and standard deviation $\sigma_{c,l}$
with
\begin{equation}
  \label{eq:calibartion:noise}
  \sigma_{c,l} = 
  \begin{cases}
    5 \% & \text{for 3\,mm lines,} \\
    10 \% & \text{for 1\,mm lines.}
  \end{cases}
\end{equation}
The value quoted at 3\,mm was estimated on the ORION-B dataset in
Appendix~D of~\citet{einig23}.

\subsection{Saturation of the \SNR{} during the fit}
\label{sec:fit:nll}

Our proposed estimate $\widehat{\btheta}$ of the vector of unknowns
$\btheta$ is obtained by minimizing the negative log-likelihood (NLL):
\begin{equation}
  \widehat{\btheta} = \arg\min_{\theta} \bracket{-\llog{\mathcal{L}(\btheta)}},
\end{equation}
where
\begin{equation}
  -\llog{\mathcal{L}(\btheta)}=
  \frac{1}{2}\sum_{l}
  \frac{\|\bvectori{x}{l}-\bvectori{s}{l}\|^2}{\sigma^2_{b,l}} +
  \emr{constant}.
  \label{eq_NLL}
\end{equation}
In this equation, $\bvectori{x}{l}$ and $\bvectori{s}{l}$ are the observed
and modeled spectra for line $l$, and $\sigma^2_{b,l}$ is the associated
noise level. The NLL, $-\llog{\mathcal{L}}$, is a function of the physical
input parameters $\btheta$, and we search for the value $\widehat{\btheta}$
that minimizes it. Following~\citet{Roueff24}, we only take into account
the additive thermal noise in this expression. Indeed, implementing a MLE
that takes into account the calibration noise leads to better performances
on Monte Carlo simulations, especially when the \SNR~is high. However, we
observe that it results on actual data in artificially large calibration
factors to compensate for the oversimplified model assumptions, and thus in
unrealistic estimations of the physical and chemical parameters (see
Sect.~\ref{sec:noise}).

We also saturate the peak \SNR~of all lines to a value of ten by adjusting
the $\sigma_{b,l}$ value in the likelihood function when needed. This
saturation ensures that the \HCOp~isotopologues have similar weights in the
NLL as the \CO~isotopologues and thus significantly contribute to the
estimation of the parameters. This saturation is needed because the use of
broadband receivers means that the integration time is the same for all
lines, resulting in different \SNR~for lines of different intensities.  We
are thus limited by the line that has the minimum \SNR, that is the
\mbox{\HthCOp{} \Jone{}} line in our case.  Figure~\ref{fig:data:spectra}
compares some observed spectra with their saturated noise versions. This
figure illustrates the potential effect of the S/N saturation on the line
profiles. We actually just adjust the $\sigma_{b,l}$ value during the fit
without changing the data themselves. In each panel, the top left and right
numbers give the relative weight of a difference of 1\,K in any given
channel, that is,
\begin{equation}
  \paren{\frac{1\K}{\sigma_{b,l}}}^2,
\end{equation}
on the fit \chisq{} value before and after the \SNR{} saturation,
respectively. Before the saturation (left numbers), the CO isotopologues
completely dominate the variations of the \chisq{}. In contrast, after
saturation (right numbers), the \HCOp{} isotopologues can significantly
influence the variations of the \chisq.

\subsection{Constrained optimization}
\label{sec:constrained:optimization}

Compared to~\citet{Roueff24}, the challenge in this study is to estimate
about twice as many parameters by going from a homogeneous model to a
heterogeneous one. To deal with this challenge, our optimizer starts with a
grid search followed by a gradient descent algorithm. The grid of RADEX
models is computed once and for all, while RADEX is run on the fly to
evaluate the line models during the gradient descent.  In addition, we
chose to optimize the likelihood under the constraint that the column
density ratios can vary within some intervals based on a priori
astrophysical knowledge. In our implementation, the column density ratios
are estimated within these constraints on abundance ratios during the grid
search with a finite resolution of 0.05 in the logarithm of the abundance
ratio (that is a factor of 1.12 in the abundance ratio).  The column
density ratios are then fixed during the gradient descent to the optimum
value found during the grid search. This choice allows us to regularize our
gradient, and thus to avoid convergence issues that occur when trying to
apply a gradient descent to refine the estimations of the column density
ratios.  Indeed, the more unknowns, the higher the probability of obtaining
singular Fisher matrices. Appendix~\ref{appendix:likelihood:optimization}
details this optimization process.

\subsection{Removing inaccurate estimations with the Cramér-Rao bound}
\label{sec:crb}

Once the fit has converged (i.e., after the grid search and the gradient
descent), we filter out estimations with too large $[-\sigma,+\sigma]$
confidence intervals. It happens that the square root of the Cramér-Rao
bound (CRB) provides a lower bound on the standard deviation of estimated
parameters for any unbiased estimators~\citep{Garthwaite95}.
Mathematically speaking,
\begin{equation}
  \sigma\left(\widehat\theta_i\right) \geq {\left[\CRBm(\btheta)\right]_{ii}}^{1/2} %
  \quad \mbox{with} \quad %
  \CRBm(\btheta)=\bI_F^{-1}(\btheta),
  \label{eq:CRB}
\end{equation}
where $\bI_F$ is the Fisher information matrix defined by
\begin{equation}
  \forall (i,j) \quad
  [\bI_F(\btheta)]_{ij}=
  \mathbb{E}\left[
    \frac{\partial \llog{\mathcal{L}(\btheta)}}{\partial \theta_i}
    \frac{\partial \llog{\mathcal{L}(\btheta)}}{\partial \theta_j}
  \right].
  \label{eq:Fisher}
\end{equation}
In the previous equation, $\mathbb{E}$ is the statistical expectation among
the different realizations of $\bx$, defined in Eq.~\ref{eq:x}.  As in
\citep{Roueff24}, and since the true value of $\btheta$ is unknown, we
compute $\CRBm(\widehat{\btheta})$ instead of $\CRBm(\btheta)$ to provide
reference precisions of our estimations. Therefore, if $\widehat\theta_i$
is the estimate of $\theta_i \in \btheta$ (see~Eq.\ref{eq:theta:sandwich}),
we only retain $\widehat\theta_i$ for LoSs that comply with
\begin{equation}
  \begin{array}{l}
    \CRBm\left(\widehat\CV\right) \leq 0.1\kms,
    \\
    \CRBm^{1/2}\left(\widehat{\FWHM}\right) \leq 0.1\kms,
    \\
    \CRBm^{1/2}\left(\widehat{\llog\Tkin}\right) \leq 0.1,
    \\
    \CRBm^{1/2}\left(\widehat{\llog N}\right) \leq 0.3,
    \\
    \CRBm^{1/2}\left(\widehat{\llog\nHH}\right) \leq 2.0,
    \\
    \CRBm^{1/2}\left(\widehat{\llog\Pth}\right) \leq 2.0,
  \end{array}
\end{equation}
where \Pth{} is the thermal pressure.  When computing these CRB reference
precisions, we take into account both the thermal and calibration noises to
avoid underestimating the credibility interval (see Sect.~\ref{sec:noise}).
In the previous section, we have noted that some abundance ratios are only
estimated during the grid search within astrophysically chosen fixed
bounds, in order to stabilize the optimization problem. As the estimated
abundance ratios are constrained within 12\%, we simply neglect the
increase of uncertainty due to this procedure.  We therefore only have
access to the accuracy on the estimation of one column density per layer
with a Fisher information matrix of size $10\times 10$, and we have no
access to the accuracy on the estimation of the ratio of the column
densities.

The gradients required to calculate the Fisher matrix and the CRB for any
multilayer model, including the sandwich model, are given in
Appendix~\ref{appendix:Fisher:matrix}.

%%%%%%%%%%%%%%%%%%%%%%%%%%%%%%%%%%%%%%%%%%%%%%%%%%%%%%%%%%%%%%%%%%%%%%%%%%%

\newcommand{\FigtwCOspectra}{%
  \begin{figure}
    \includegraphics[width=\linewidth]{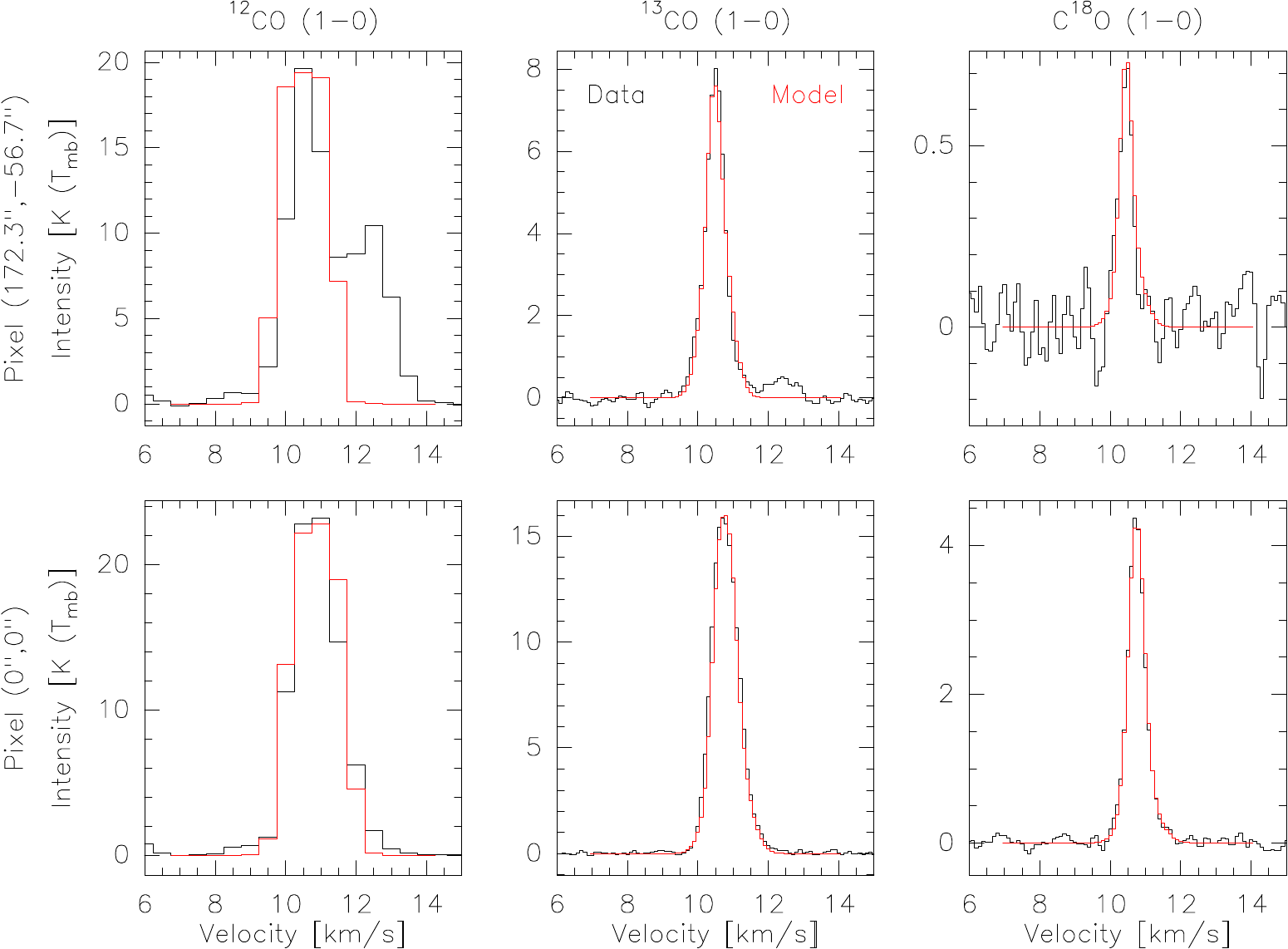}
    \caption{Comparison of the observed (black) and modeled (red)
      \Jone~line of the \twCO, \thCO, and \CeiO~species for two different
      LoSs.}
    \label{fig:12co10:spectra}
  \end{figure}
}

\newcommand{\FigHomogeneousVsSandwich}{%
  \begin{figure}
    \includegraphics[width=\linewidth]{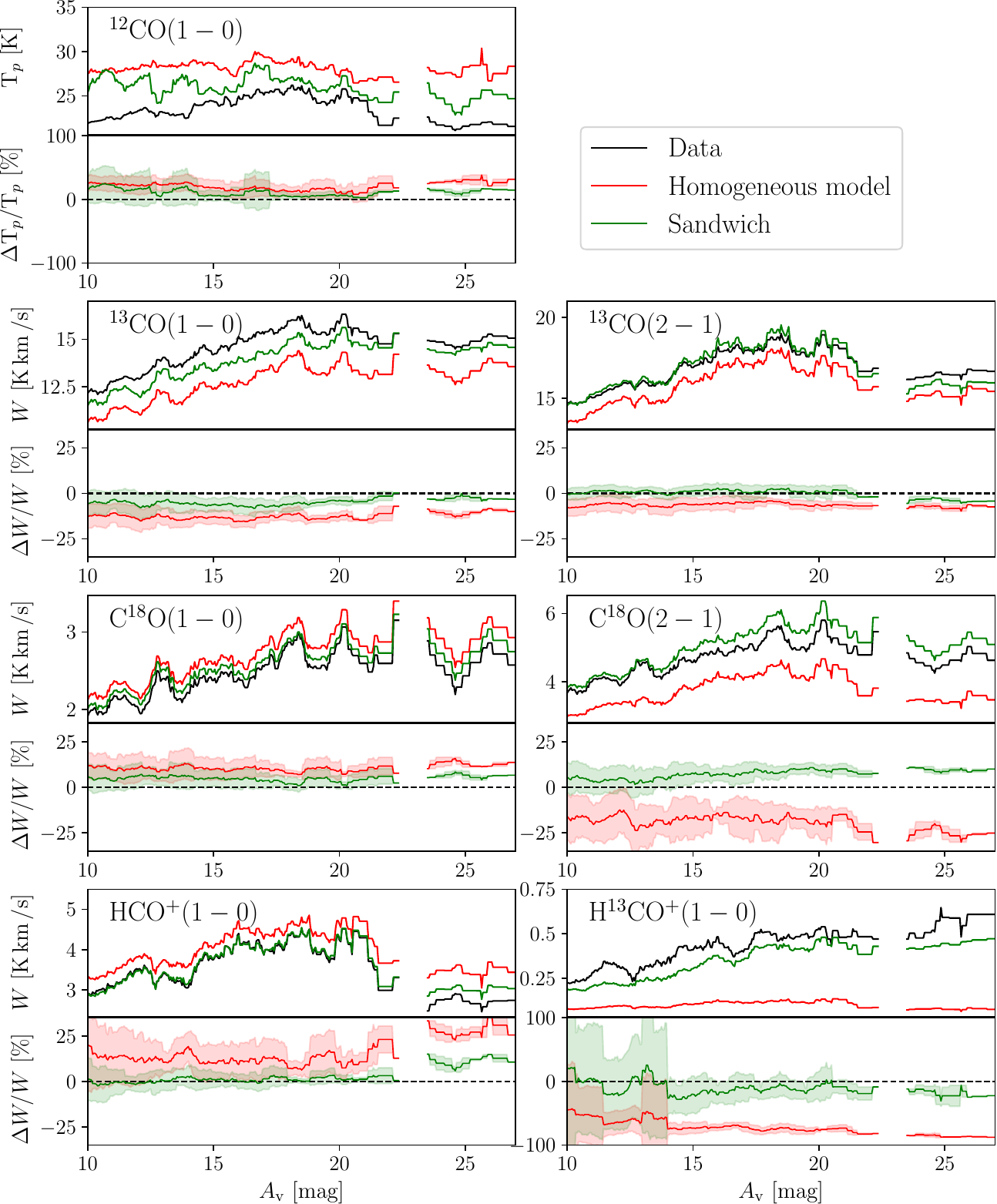}
    \caption{Comparison of the reconstruction of all integrated line
      intensities for the best fit obtained with either a homogeneous (in
      red) or a simple sandwich model (in green). The integrated line
      intensity or peak temperature are plotted as a function of the visual
      extinction ($\Av$).  More precisely, the integrated line intensities
      (zero-order moment) or peak temperature are averaged over all the
      pixels whose visual extinction belongs to the
      $[\Av-0.5,\Av+0.5]\magn$ interval, where \Av~is plotted on the x-axis
      of the panels. The lack of values for $\Av \in \bracket{23, 24}\magn$
      results from the fact that there is no pixel in this range of
      \Av~over the studied field of view. In each panel, the top one shows
      the line integrated intensity or peak intensity, while the bottom one
      shows the reconstruction relative error in percentage. The colored
      areas on the relative error panels show the standard deviation over
      all the pixels that belong to the 1\magn{}-\Av~sliding window. The
      horizontal dashed line indicates a perfect reconstruction.  While the
      simple sandwich model already better fit the data than the
      homogeneous model, the reconstruction will still be improved by
      refining the assumptions.}
    \label{fig:homogeneous:vs:sandwich}
  \end{figure}
}

\newcommand{\FigLineAsymmetry}{%
  \begin{figure}
    \includegraphics[width=\linewidth]{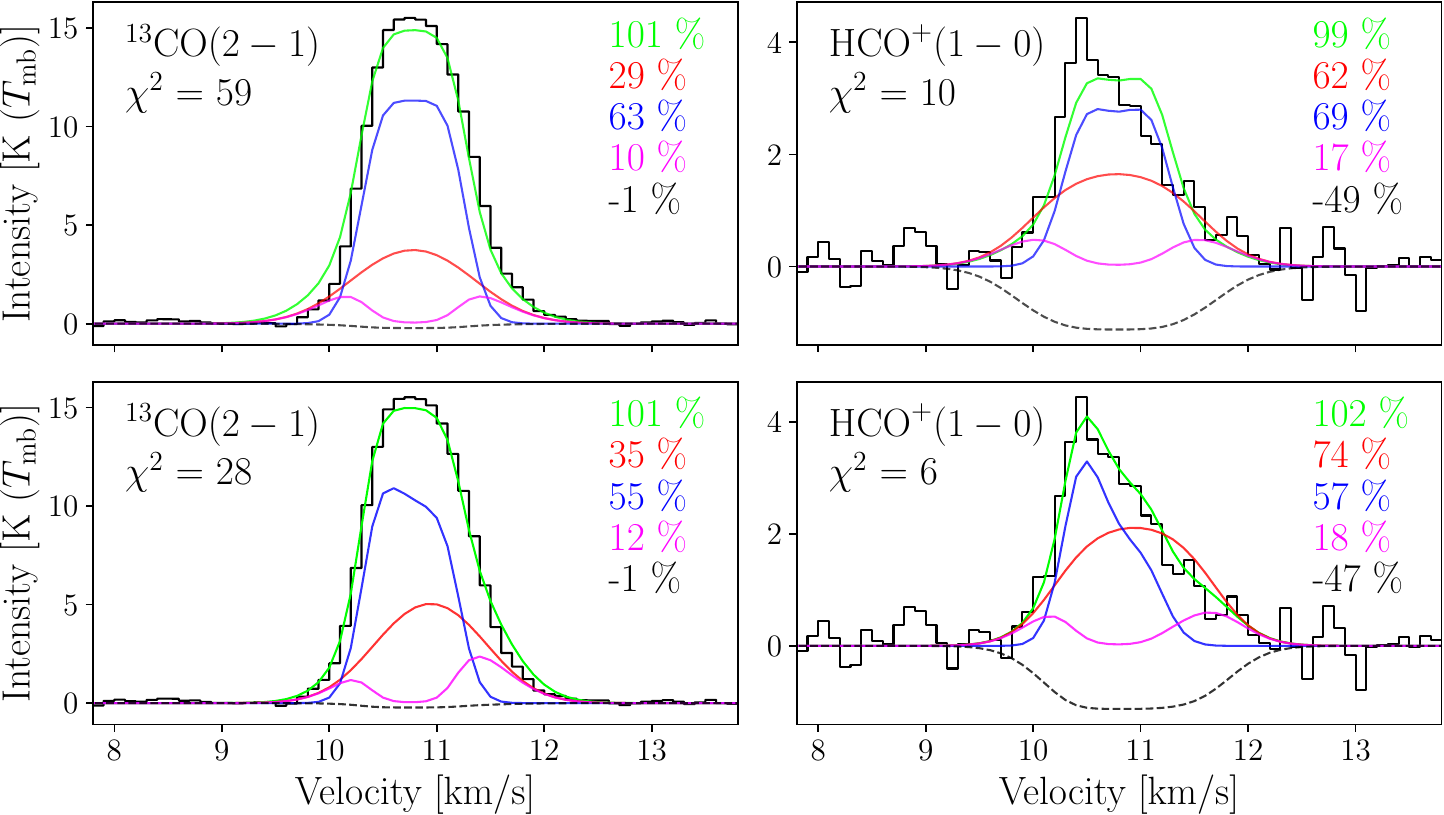}
    \caption{Comparison of the data and the fit of the sandwich model when
      either the centroid velocities of all layers are assumed to be
      identical \textbf(top row) or they can have different values in the
      inner and outer layers (bottom row). The left and right columns show
      this comparison for the \mbox{\thCO\,\Jtwo} and \mbox{\HCOp\,\Jone}
      lines, respectively. These spectra correspond to the LoS~B (see
      Table~\ref{tab:coord}). Data and fitted spectra are shown in black
      and green lines, respectively. The contributions of the different
      layers are shown as colored lines: red for the foreground layer (see
      Eq.~\ref{eq:fore}), blue for the inner layer (see
      Eq.~\ref{eq:inter}), pink for the background layer (see
      Eq.~\ref{eq:back}), and dashed black for the CMB (see
      Eq.~\ref{eq:CMB}).}
    \label{fig:line:asymmetry}
  \end{figure}
}

\newcommand{\FigCOisoFit}{%
  \begin{figure}
    \includegraphics[width=\linewidth]{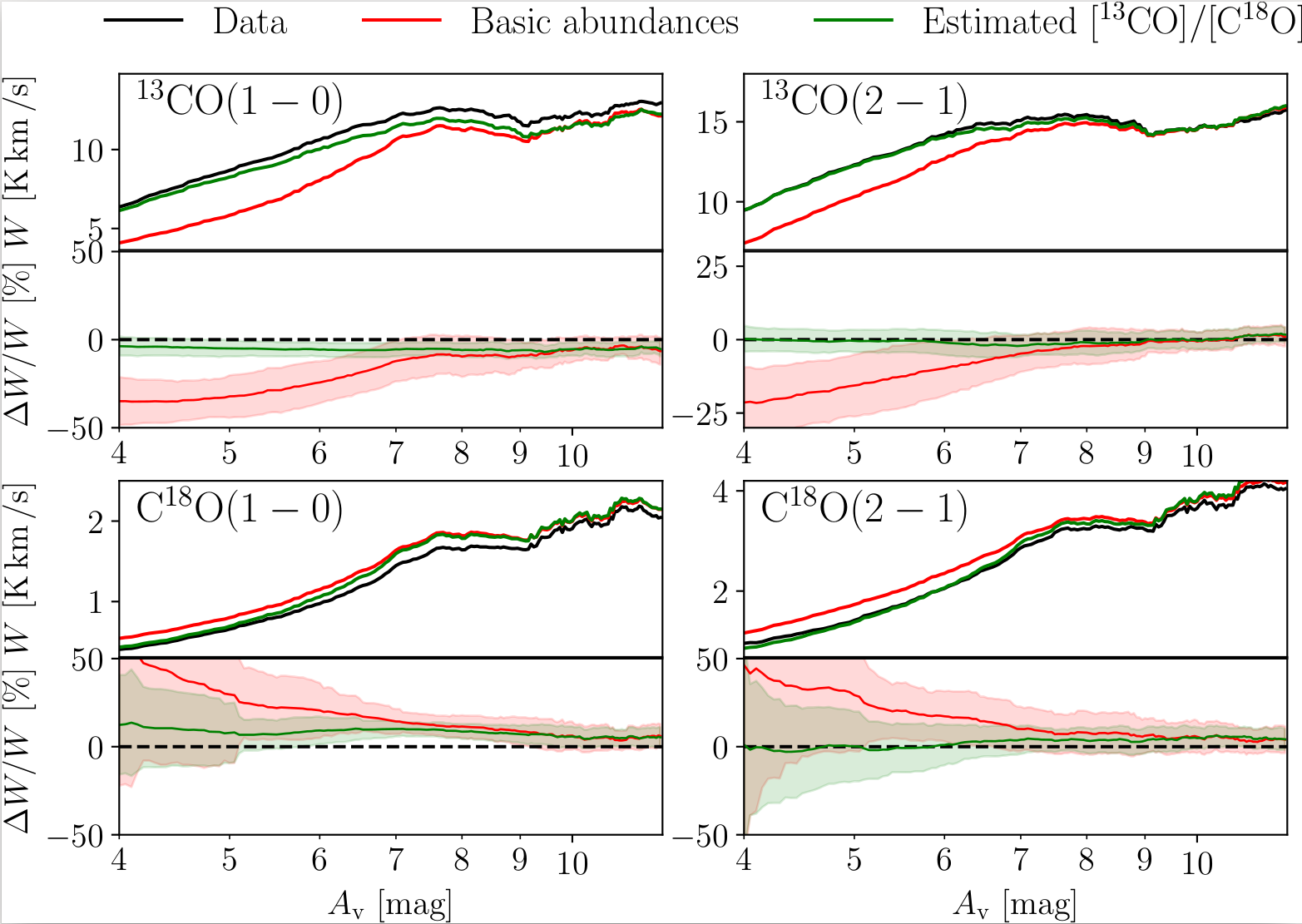}
    \caption{%
      Comparison of the reconstruction of the \thCO{} and \CeiO{} line
      integrated intensities for the sandwich model fit obtained with
      either a fixed relative abundance of $[\thCO]/[\CeiO]$ (in red) or an
      estimated relative abundance of this ratio (in green).  The figure
      layout is similar to Fig.~\ref{fig:homogeneous:vs:sandwich}.  } %
    \label{fig:coiso:fit:1}
  \end{figure}
}

\newcommand{\FigNLLvsnHHone}{%
  \begin{figure*}
    \sidecaption %
    \includegraphics[width=12cm]{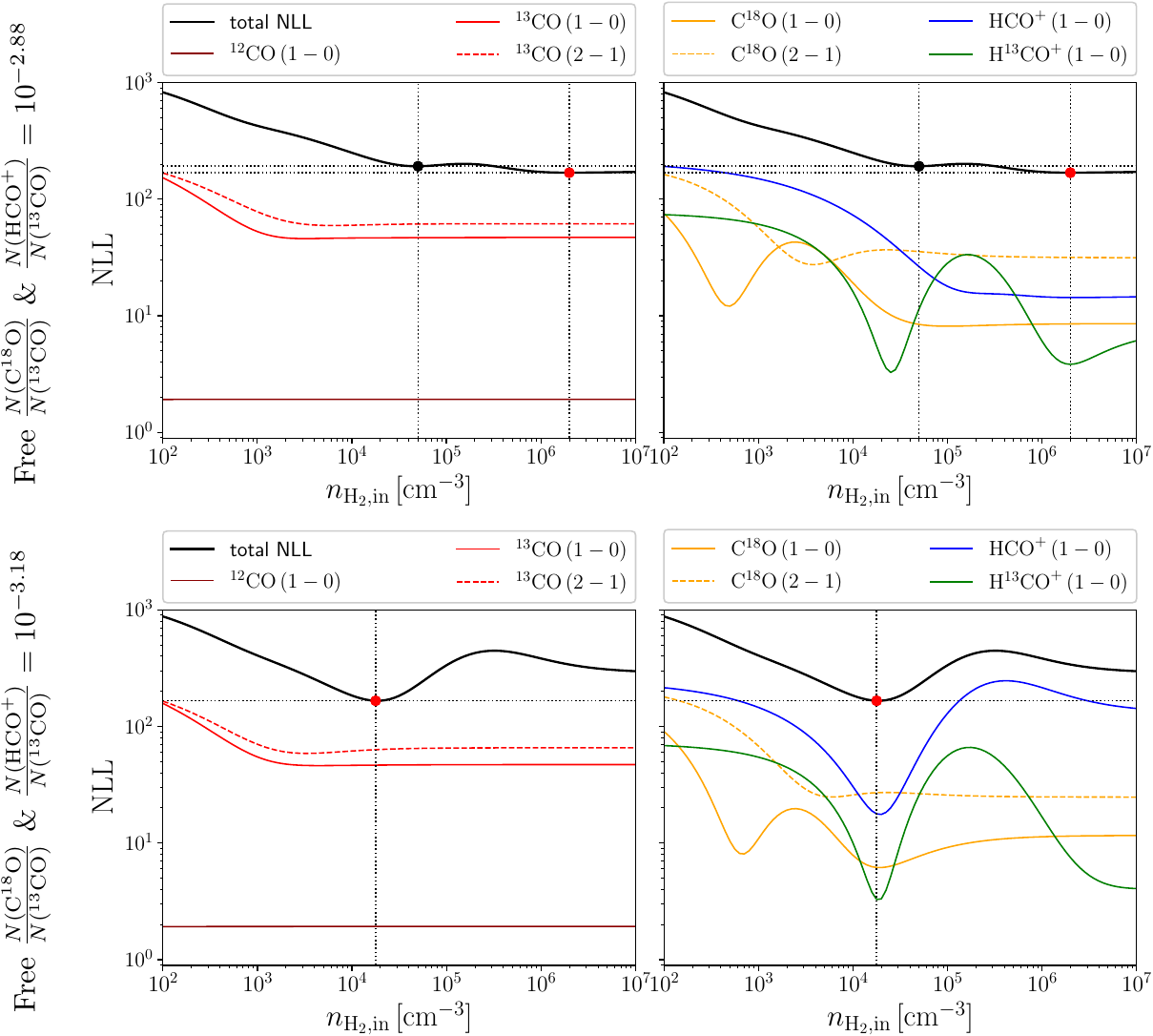}
    \caption{Comparison of the variations of the NLL as a function of the
      volume density of the inner layer for the LoS~A. We used a sandwich
      model where $N(\thCO)$ and $N(\CeiO)$ were fitted, while the column
      densities of \twCO{}, \mbox{\HCOp}, and \HthCOp{} were computed
      assuming fixed abundances relative to \thCO. The top and bottom rows
      show the NLL variations for two values of $N(\HCOp)/N(\thCO)$ varying
      by a factor of 2.0.  The left panel shows the NLL variations for the
      \twCO{} and \thCO{} lines, while the right panel show that of the
      \CeiO{}, \mbox{\HCOp}, and \HthCOp{} lines. In both panels, the black
      plain lines shows the variations of the NLL when fitting all the
      lines. The top row shows two local minima at similar NLL values while
      the bottom row only shows one local minimum. The vertical and
      horizontal black dotted lines intersect at the lowest NLL value,
      indicated by the red circle, and at the second local minimum that is
      indicated by the black circle, when it exists.}
    \label{fig:NLLvsnH2:13co-c18o}
  \end{figure*}
}

\newcommand{\FigCOvsHCOp}{%
  \begin{figure}
    \includegraphics[width=\linewidth]{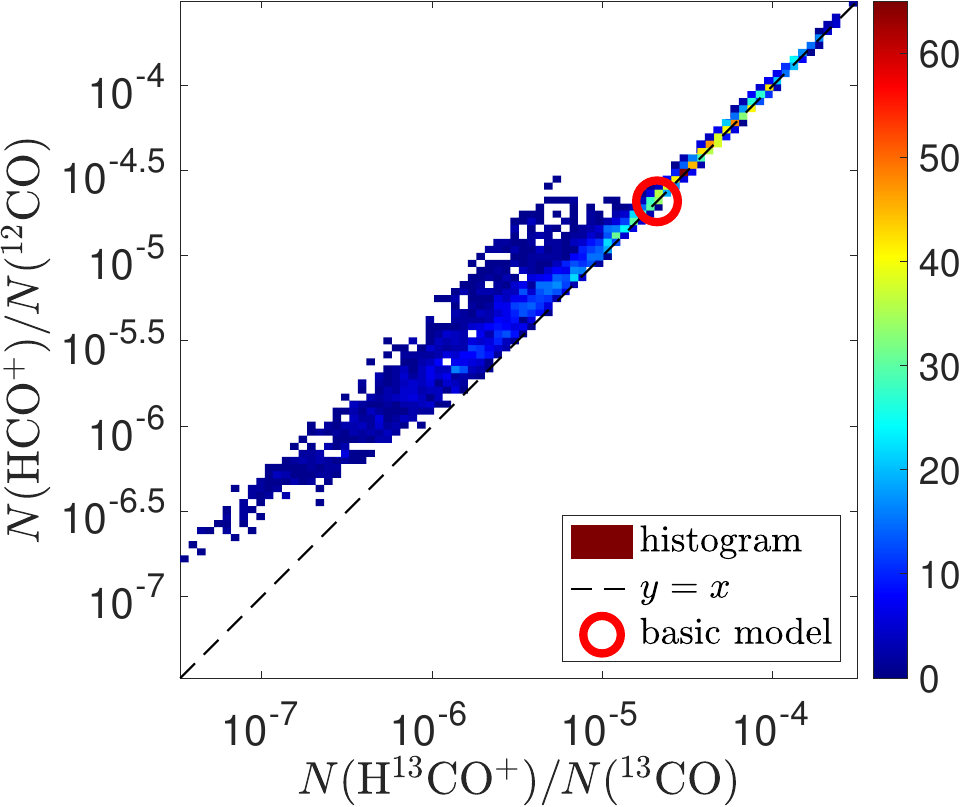}
    \caption{Joint histogram of $N(\HCOp)/N(\twCO)$ versus
      $N(\HthCOp)/N(\thCO)$ in log space for the 3224 chemical models
      sampling various physical conditions in translucent gas.  The red
      circle shows the ratio derived from the basic abundances listed in
      Eq.~\ref{eq:abundance}. The dashed blue line has a slope of $1.0$.}
    \label{fig:COvsHCOp}
  \end{figure}
}

%%%%%%%%%%%%%%%%%%%%%%%%%%%%%%%%%%%%%%%%%%%%%%%%%%%%%%%%%%%%%%%%%%%%%%%%%%%

\newcommand{\TabVelocityStudy}{%
  \begin{table}
    \centering{} %
    \caption{Fitted parameters when assuming that the inner and outer
      layers have the same or different centroid velocities and velocity
      dispersions.}
    \resizebox{\linewidth}{!}{%
      \begin{tabular}{cc|cc|cc}
        \hline\hline               
        Parameter & Unit & \multicolumn{2}{c|}{$\CVl{\text{in}} = \CVl{\text{ou}}$} & \multicolumn{2}{c}{Free \CV}\\
                  &      & \text{Inner} & \text{Outer} & \text{Inner} & \text{Outer}\\
        \hline                      
        \CV{}                   & \kms{} & $10.78 \pm 0.01$  & $10.78 \pm 0.01$ & $10.76 \pm 0.01$ & $10.94 \pm 0.03$ \\
        \FWHM{}                 & \kms{} &  $0.59 \pm 0.03$  &  $1.22 \pm 0.11$ &  $0.54 \pm 0.02$ &  $1.10 \pm 0.08$ \\
        \Tkin{}                 & \K{}   &    $22 \pm 3$     &    $30 \pm 4$    &    $21 \pm 2$    &    $30 \pm 3$    \\
        $\llog{(\nHH/\pccm)}$   & $-$    &  $5.36 \pm 0.41$  &  $3.78 \pm 0.14$ &  $5.41 \pm 0.41$ &  $3.93 \pm 0.08$ \\
        $\llog{(N_\thCO/\pscm)}$ & $-$    & $16.36 \pm 0.03$  & $15.45 \pm 0.06$ & $16.33 \pm 0.03$ & $15.54 \pm 0.05$ \\
        \hline
      \end{tabular}
    }
    \label{tab:velocity:study}
  \end{table}
}

\newcommand{\TabAbundanceRatios}{%
  \begin{table}
    \label{tab:abundance:ratios}
    \centering %
    \caption{Column density ratios (or abundance ratios) set between
      tracers.  Fraction of LoSs for which each column density ratio is
      assigned is indicated in the third column.}
    \resizebox{\linewidth}{!}{%
      \begin{tabular}{ccc}
        \hline
        \hline
        Species & Column density ratios & Lines-of-sight [\%]  \\
        \hline                    
        \NtwCO / \NthCO & $10^{1.8} \sim 63.1$\tablefootmark{$\star$} & $100$ \\     
        \hline                     
                & $10^{0.9} \sim 7.94$\tablefootmark{$\star$} & $22.3$ \\ 
                & $10^{0.95} \sim 8.91$ & $11.5$ \\
                & $10$ & $9.37$ \\
        \NthCO / \NCeiO & $10^{1.05} \sim 11.22$ & $7.72$ \\
                & $10^{1.1} \sim 12.58$ & $8.10$ \\
                & $10^{1.15} \sim 14.13$ & $10.95$ \\
                & $10^{1.2} \sim 15.85$ & $30.06$ \\
        \hline                          
        \NthCO / \NHCOp & $10^{2.88} \sim 758.6$\tablefootmark{$\star$} & $100$\\
        \hline                          
        \NHCOp / \NHthCOp & $10^{1.8} \sim 63.1$\tablefootmark{$\star$} & $100$\\     
        \hline
        \hline
      \end{tabular}
    } \tablefoot{\tablefoottext{$\star$}{Isotopic abundances.}}
  \end{table}
}

\newcolumntype{C}[1]{>{\centering\arraybackslash}p{#1}} % centered version of 'p' column type

\newcommand{\TabChemicalBasicModel}{%
  \begin{table*}
    \caption{%
      Assumptions on column density ratios or relative abundances for the
      basic and improved sandwich models.}
    \centering %
    \begin{tabular}{cc|c|cc}
      \hline
      \hline
      \multicolumn{2}{c|}{Column density ratio}
      & Basic model
      & \multicolumn{2}{c}{Improved model} \\
      \hline 
      & 
      & Inner \& Outer layers
      & Inner layer
      & Outer layer
      \\
      \hline 
      \multirow{4}*{Assumed}
      & $\twCO/\thCO$
      & $10^{1.8} \sim 63$
      & $10^{1.8}$
      &  $[10^{1.3}, 10^{1.9}] \sim [20, 80]$
      \\
      & $\thCO/\CeiO$
      & $10^{0.9} \sim 7.9$
      & $10^{0.9}$
      &  $[10^{0.75}, 10^{1.4}] \sim [5.6, 25.1]$
      \\
      & $\thCO/\HCOp$
      & $10^{2.88} \sim 759$
      & $[10^{2.88}, 10^{3.78}]$
      & $[10^{2.88}, 10^{3.78}]$
      \\
      & $\thCO/\HthCOp$
      & $10^{4.68}$
      & $\twCO/\HCOp$
      & $\twCO/\HCOp$
      \\
      \hline
      \multirow{3}*{Derived}
      & $\twCO/\CeiO$ 
      & $10^{2.7} \sim 500$
      & $10^{2.7}$
      & $[10^{2.05}, 10^{3.3}]$
      \\
      & $\CeiO/\HthCOp$
      & $10^{3.78}$
      & $[10^{3.78}, 10^{4.68}]$
      & $[10^{2.78}, 10^{4.93}]$
      \\
      & \mbox{$\twCO/\HCOp$} 
      & $10^{4.68}$
      & $[10^{4.68}, 10^{5.58}]$
      & $[10^{4.18}, 10^{5.68}]$
      \\
      \iffalse
      & \mbox{$\twCO/\HH$} 
      & $1.45\e{-4}$
      & 
      & 
      \\
      & \mbox{$\thCO/\HH$} 
      & $2.3\e{-6}$
      & 
      & 
      \\
      & \mbox{$\CeiO/\HH$} 
      & $2.9\e{-7}$
      & 
      & 
      \\
      & \mbox{$\HCOp/\HH$} 
      & $3\e{-9}$
      & 
      & 
      \\
      & \mbox{$\HthCOp/\HH$} 
      & $2.3\e{-10.68}$
      & 
      & 
        \fi
        \hline
    \end{tabular}
    \tablefoot{We write $X/Y$ the column density ratio $N(X)/N(Y)$, for
      each couple of species \cbracket{\mbox{$X$, $Y$}}.  For the improved
      model, intervals correspond of the explored ranges when estimating
      the corresponding ratios.  Moreover, $N(\thCO)/N(\HthCOp)$ is deduced
      from \mbox{$N(\thCO)/N(\HthCOp) = N(\HCOp)/N(\twCO)$} (see
      Sect.~\ref{sec:assumption:HCOp-vs-CO}).}
    \label{tab:chemical:basic:model}
  \end{table*}
}

%%%%%%%%%%%%%%%%%%%%%%%%%%%%%%%%%%%%%%%%%%%%%%%%%%%%%%%%%%%%%%%%%%%%%%%%%%%

\section{Physical and chemical assumptions}
\label{sec:assumptions}

Having a large number of fitted parameters leads to large uncertainties. To
reduce these uncertainties, we need to add constraints on estimated
parameters with physical and chemical assumptions. \citet{Roueff24} discuss
in depth the possible occurrence of biases linked to these additional
assumptions, and highlight that a single unreasonable constraint can
dramatically bias the results.  This section details and justifies the
proposed hypotheses.  First, we explain how we use the information carried
by the \twCO~\Jone~line of high opacity.  Then, we compare the sandwich
model with a simple homogeneous model consisting of a single layer, to
illustrate the gain in signal reconstruction quality with the latter model.
We continue by discussing the differences in the kinematic properties of
the different layers, namely their velocity centroids and velocity
dispersions.  Finally, we introduce the hypotheses on the chemical
abundances in the inner and outer layers.

\subsection{Specific treatment for the \mbox{\twCO\,\Jone} line}
\label{sec:assumptions:12co10}

\FigtwCOspectra{} %

Figure~\ref{fig:12co10:spectra} compares the observed and modeled spectra
of the \mbox{\Jone} lines of the three main \CO~isotopologues for two lines
of sight. The main difference between these LoSs is the existence of a
second velocity component between $12$ and $14\kms$ that contributes more
than $10\%$ of the total integrated intensity of the \mbox{\twCO\,\Jone}
line and much less than this for the \thCO~and \mbox{\CeiO\,\Jone}
lines. While the model does a correct reconstruction of the main velocity
component for the \mbox{\twCO\,\Jone} line when we only fit the peak
intensity of this line, the second velocity component perturbs the fit when
we try to fit the full \mbox{\twCO\,\Jone} profile. Since at this stage our
algorithm has not been designed to fit multiple velocity components, we
thus fit only the \mbox{\twCO\,\Jone} peak temperature as proposed by
\citet{Roueff24}. Fitting multiple velocity components will be the subject
of a forthcoming paper.

\subsection{On the importance of a heterogeneous model}
\label{sec:assumptions:sandwich}

\FigHomogeneousVsSandwich{} %

The simplest possible geometry is to assume a homogeneous cloud composed of
a single layer emitting in front of the CMB. However, the maps of the line
emission (see Fig.~\ref{fig:data:maps}) suggest that the emission comes
from gas with different physical conditions, at least along the LoSs toward
filaments and dense cores.  We therefore introduce a three-layer model to
take into account this inhomogeneity while keeping the model as simple as
possible.

In this section we compare the integrated intensity computed on the data
with that derived from the molecular line fitting with two different
models. The first one is a homogeneous model described with
\begin{equation}
  s = s_\emr{fore} + s_\emr{CMB}.
  \label{eq:homogeneous:model}
\end{equation}
The second one is a three-layer model described with
\begin{equation}
  s = s_\emr{fore} + s_\emr{inner} + s_\emr{back} + s_\emr{CMB}.
  \label{eq:sandwich:model}
\end{equation}
The components used in these equations are defined in Eq.~\ref{eq:fore}
to~\ref{eq:CMB}.

To simplify the comparison, we kept in each experiment the same fixed
relative abundances, defined by using the following isotopic ratios for
Orion \mbox{$^{12}$C/$^{13}$C$ = 63,$ $^{16}$O/ $^{18}$O$ = 510,$} and
\mbox{{N(\HCOp)}/N(H$_2$)=$3\times 10^{-9}$} and
\mbox{$\NthCO/\NHH=2.3\e{-6}$}~\citep[and references
therein]{Gerin19,Roueff24}.  Those give
\begin{equation}
  \begin{array}{ll}
    N(\twCO)/N(\thCO) = 10^{1.8}  \sim 63,
    \\
    N(\thCO)/N(\CeiO) = 10^{0.9}  \sim 8,
    \\
    N(\thCO)/N(\HCOp) = 10^{2.88} \sim 800,
    \\
    N(\CeiO) / N(\HthCOp) = 10^{3.78} \sim 6000.
  \end{array}
  \label{eq:abundance}
\end{equation}
We refer to these column density ratios collectively as ``basic'' in the
remainder of this work. In contrast, the centroid velocities, velocity
dispersions, column densities, kinetic temperatures, and volume densities
were estimated independently for the inner and outer layers.

Figure~\ref{fig:homogeneous:vs:sandwich} compares the reconstruction of all
the line profiles for the best fits obtained with either the homogeneous or
the sandwich model. The total integrated line is plotted as a function of
the visual extinction between 10 and 27\magn, where the probability of
having different gas components along a single LoS is high.  Intensities
are averaged over all pixels that belong to a 1\magn~sliding window of the
visual extinction. The colored areas around the relative error curves show
the associated standard deviations for all the pixels belonging to the
sliding window. The vertical width of this $\pm\sigma$-interval mostly
decreases from low to high visual extinction because the number of pixels
quickly decreases with \Av.

The first obvious feature in Fig.~\ref{fig:homogeneous:vs:sandwich} is that
the sandwich model better reconstructs the line total integrated
intensities than the homogeneous model. This is expected since the number
of fitted parameters doubles from 5 to 10 when going from a homogeneous to
a sandwich model. More interestingly, the sandwich model better
reconstructs the lines that mostly originate from one type of medium with
specific physical conditions: either translucent or dense and cold gas. The
most striking example is the \HthCOp~line whose relative error decreases
from \mbox{$\sim -80\%$} to less than \mbox{$-25\%$} for
\mbox{$\Av>15\magn$}. The second example is the \mbox{\CeiO\,\Jtwo} line
whose typical relative error decreases from \mbox{$\sim -20\%$ to less than
  $\sim 10\%$} over the plotted \Av{} range.
% The changes proposed by the language editor modify the science
% meaning. We thus came back to the original formulation.
The second clear feature is that the sandwich model allows to decrease the
intensities of the lines that are known to be optically thick (namely
\mbox{\twCO\,\Jone} and \mbox{\HCOp\,\Jone}) and thus mostly sensitive to
the external layer, while it increases the intensity of the optically
thinnest line (e.g., \mbox{\HthCOp\,\Jone}). Finally, the sandwich model
becomes unbiased (i.e., its relative error is compatible with zero within
the $\pm\sigma$-interval) for large ranges of \Av{} and for many lines. The
most obvious example is the \mbox{\thCO\,\Jtwo} line.  Whatever the
\Av~range, the absolute value of the relative reconstruction error almost
always remains below \mbox{$10\%$}, a value consistent with the fact that
we saturate the peak \SNR~to $10$.  Hereafter, we only discuss variants of
the sandwich model.

\subsection{On the importance of having independent centroid velocities to
  reproduce the spectrum asymmetries}
\label{assumption:centroid:velocity}

\FigLineAsymmetry{} %
\TabVelocityStudy{}%

A straightforward way to decrease the number of fitted parameters would be
to use the same velocity distribution (centroid velocity and velocity
dispersion) for the inner and outer layers. However, it is well known that
the lines mostly tracing the dense and shielded gas (e.g., the \CeiO~lines)
are narrower than lines (e.g., the \twCO~lines) probing the surrounding
lower density, warmer, and more turbulent gas. This result is for example
highlighted by~\citet{Orkisz19,Roueff21,Gaudel23}.  We here explore the
importance of estimating independently the centroid velocities for the
inner and outer layers. To do this, we compare the reconstruction of the
spectra for the LoS~B that shows asymmetric line profiles for, e.g., the
\mbox{\HCOp\,\Jone} line (see Table~\ref{tab:coord} and
Fig.~\ref{fig:data:spectra}).

Figure~\ref{fig:line:asymmetry} shows the fit of the \mbox{\thCO\,\Jtwo}
and \mbox{\HCOp\,\Jone} lines for a sandwich model with the same centroid
velocity for all the layers (top row) or with independent centroid
velocities for the inner and outer layers (bottom row). The conditions for
the fit of the other parameters are unchanged between these two
experiments. In particular, the centroid velocity for each layer is
identical for the seven fitted lines, and the column density ratios are
fixed to their basic values listed in Eq.~\ref{eq:abundance}.

Fitting independently the centroid velocities of the inner and outer layers
better reproduces the wings of the \mbox{\thCO\,\Jtwo} line, explaining the
decrease by more than a factor of two of the \chisq~value for this line.
But, more importantly, it is the only way to reproduce the large asymmetry
of the \mbox{\HCOp\,\Jone}~line.  Indeed, when using the same centroid
velocity for all layers, the contribution of each layer shows a symmetrical
profile.  In contrast, allowing the simultaneous fit of the outer and inner
centroid velocities delivers a symmetrical profile contribution of the
foreground layer (red spectrum), and asymmetrical ones for all other layers
(inner, background, and CMB). These asymmetrical line profiles are not just
the result of adding symmetrical components shifted in velocity. Instead,
marked asymmetries mainly result from the absorption of the intrinsically
symmetrical emission term ($J$ in Eq.~\ref{eq:J}) of one layer by the
velocity-shifted absorption of another layer in front of it. Taking into
account different centroid velocities for the inner and outer layers
therefore allowed us to model this coupling between emission and
absorption.

Table~\ref{tab:velocity:study} compares the fitted parameters for the inner
and outer layers for these two experiments. The differences are of the
order of \mbox{$10\%$}, except for the volume densities of the inner and
outer layers, which increase by $0.05$ to $0.1$ dex
(\mbox{$12\text{ to }25\%$}).  When comparing the homogeneous model to the
sandwich model in Sect.~\ref{sec:assumptions:sandwich}, the centroid
velocities of the inner and outer layers are fitted independently. This is
probably a reason why the \mbox{\HCOp\,\Jone} integrated line profile is
much better reconstructed with the sandwich model than with the homogeneous
model.

Hereafter, we only discuss models where the centroid velocities for the
inner and outer layers are independently estimated.

\subsection{On the importance of fitting the \mbox{\NCeiO/\NthCO} column
  density ratio}
\label{sec:assumption:C180/13CO}

\citet{Roueff24} show that it is in principle better to avoid constraining
the column densities of the studied species because this lowers potential
biases. However, fitting here one column density per species and per layer
would imply to estimate ten additional parameters (five species for the
inner, and five for the outer layers). Such a fit would lead to large
uncertainties because of the limited number of measured transitions for the
different species.  We therefore search for the most suitable chemical
hypotheses for the Horsehead nebula to limit the number of fitted column
densities.

We start from the experiment described in
Sect.~\ref{sec:assumptions:sandwich}, namely, fitting a sandwich model with
column density ratios following the basic abundance ratios given in
Eq.~\ref{eq:abundance}.  While the relative error on the integrated
intensity is within $\pm 10\%$ for the \thCO~and \CeiO~lines for
\mbox{$\Av > 10\magn$}, the fit almost systematically underestimates the
integrated intensity of the \thCO~lines, and overestimates that of the
\CeiO~lines, i.e., the relative errors are anticorrelated for these two
species (see the green curves of
Fig.~\ref{fig:homogeneous:vs:sandwich}). This behavior that is more salient
for the \Jone~lines, questions the value chosen for the basic column
density ratio. We therefore carried out a series of experiments in which we
refitted the spectra over the whole field of view for several values of the
\mbox{\NthCO/\NCeiO} ratio. For each pixel, we thus obtained 14 solutions
for which the only difference is the value of the \mbox{\NthCO/\NCeiO}
ratio that goes from $5.62$ to $25.12$ in multiplicative steps of
$1.12$.\footnote{The basic value of this ratio is $7.94$ (see
  Eq.~\ref{eq:abundance}).}  In all of these experiments, we used the same
value of the \mbox{\NthCO/\NCeiO} ratio for both the inner and outer
layers. As the solutions are obtained with the same data and same number of
fitted parameters, comparing the NLL values (Eq.~\ref{eq_NLL}) allowed us
to select the best value of the \mbox{\NthCO/\NCeiO} ratios for each
pixel. As this manual approach is akin to a fit of the column density
ratio, we incorporated it in the grid search step of the proposed MLE
estimator for the remainder of this work.

Figure~\ref{fig:coiso:fit:1} compares the reconstruction of the \thCO~and
\CeiO~line integrated intensities obtained with either a fixed basic
\mbox{\NthCO/\NCeiO} ratio for all pixels, or an estimated
\mbox{\NthCO/\NCeiO} ratio per pixel. The reconstructed integrated
intensities are shown as a function of the visual extinction in the
$[4,13]$ magnitude range, where we expect deviations of the
\mbox{\NthCO/\NCeiO} ratio from its basic value because of the selective
photodissociation of \CeiO~combined with enrichment of \thCO~by isotopic
fractionation~\citep{Langer84,Kong15}. We do not show the results for lower
\Av~values because the \CeiO~lines have either very low \SNR{} or are not
detected at all, implying that the data are no longer sensitive to the
\mbox{\NthCO/\NCeiO} ratio.  We find that fitting the \mbox{\NthCO/\NCeiO}
ratio considerably decreases the negative (resp. positive) bias when
reconstructing the \thCO~(resp. \CeiO) integrated line intensity. In
particular, the \Jtwo~lines now have unbiased fitted integrated
intensities.

Hereafter, we continue to fit the \mbox{\NthCO/\NCeiO} column density
ratio.

\FigCOisoFit{} %

\subsection{On the importance of allowing the \mbox{\NHCOp/\NthCO} column
  density ratio to vary in the outer translucent layer}
\label{sec:assumption:HthCOp:HCOp}

\FigNLLvsnHHone{} %

In this subsection we first show that fixing the value of the
\mbox{\NHCOp/\NthCO} column density ratio to its basic value listed in
Eq.~\ref{eq:abundance} leads to an ambiguity in the estimation of the
volume density of the inner layer because the NLL shows two local minima of
similar depth around $10^4$ and \mbox{$10^6\pccm$}. The estimator thus
oscillates between these two values depending on the noise level. We
therefore considered fitting the column densities of the
\HCOp~isotopologues in addition to the column densities of the
\CO~isotopologues (see Sect.~\ref{sec:assumption:C180/13CO}).  However,
fitting the column densities of both \HCOp~and \HthCOp~for all lines of
sight will result in a high variance because of the higher number of
unknowns and relatively low \SNR~of the \HCOp~isotopologue lines. As a
compromise, we thus propose 1) to take \mbox{\NHCOp/\NtwCO =
  \NHthCOp/\NthCO} for the outer layer, as this is plausible for
translucent gas following the analysis described in
Sect.~\ref{sec:assumption:HCOp-vs-CO}, and 2) to keep the basic abundance
ratios in the inner layer made of colder, denser gas.

\subsubsection{On the impact of the \mbox{\NHCOp/\NthCO} value on the
  number of local minima of the NLL}
\label{sec:one:or:two:local:minima}

The model described in Sect.~\ref{sec:assumption:C180/13CO} fits the column
densities of \thCO~and \CeiO, but it fixes the column densities of \twCO,
\HCOp, and \HthCOp~by assuming that they have constant abundances relative
to \thCO. Fitting all the lines of sight toward the Horsehead nebula with
this model shows that the fit oscillates between two solutions for the
inner layers: one with a density around \mbox{$10^4\pccm$}, and another one
with the density around \mbox{$10^6\pccm$}. This situation often happens
when fitting low-$J$ lines of \CO~and \HCOp~isotopologues ~\citep[see, for
instance, Fig.~10][]{Roueff24}.  As previous studies of the Horsehead
nebula suggest that the volume density should be about
\mbox{$\sim 2\e{5}\pccm$}~\citep{Pety05,Habart05,Goicoechea06,Pety07}, it
is tempting to limit the range of volume densities to this upper value, but
it is better to solve this issue based on chemical assumptions.

To understand why the fit fluctuates between two values of the volume
density separated by two orders of magnitude, we have selected one line of
sight for which the volume density of the inner layer is fitted to
\mbox{$10^6\pccm$}, namely LoS~A. The top row of
Fig.~\ref{fig:NLLvsnH2:13co-c18o} shows the variations of the NLL as a
function of the volume density of the inner layer, while all other
estimated parameters remain fixed at the values found during the fit. The
minimum of the NLL for this LoS happens at \mbox{$2\e{6}\pccm$}, but a
second local minimum occurs at \mbox{$5\e{4}\pccm$} with only an increase
of the NLL of $14\%$.  The NLL of all optically thick (\twCO, \thCO, and
\HCOp) lines show a decrease followed by a plateau when the volume density
increases. A plateau implies that the line intensity is no longer sensitive
to the volume density of the inner layer. The NLL of the two optically thin
lines (\CeiO~and \HthCOp) show one or two more or less deep wells at
various relatively low \nin~values followed by a plateau at higher
\nin~values.  Only the \Jone~lines of \HCOp~and \HthCOp~have not yet
reached the NLL plateau at \mbox{$\nin \sim 10^{4}\pccm$}. The variations
of the NLL of these two lines above \mbox{$10^{4}\pccm$} thus discriminate
between the fitted \nin~values.  This is consistent with the higher
critical density of the \HCOp~isotopologue lines
(\mbox{$\sim10^{5\pm 0.3}\pccm$}) than those of the \CO~isotopologue lines
(\mbox{$\sim 10^{3.3\pm 0.2}\pccm$}, see Tables~6 and~7
of~\citealt{Roueff24}).

Next, we decreased the abundance of \HCOp{} with respect to \thCO{} by a
factor of two.  The bottom row of Fig.~\ref{fig:NLLvsnH2:13co-c18o} shows
the variations of the NLL as a function of the volume density in this
case. The NLL variations are significant only for the \mbox{\HCOp\,\Jone}
line, which now shows a deep well aligned with the much more optically thin
\mbox{\HthCOp\,\Jone} line at \mbox{$\nin \sim 1.8\e{4}\pccm$}.  This
translates into a marked well of the total NLL at the same value of \nin.
For \mbox{$\nin \ga 10^4\pccm$}, the fitted value of \nin~is governed by
the variations of the NLL of the \Jone~lines of the two main
\HCOp~isotopologues.  Varying the abundance of \HCOp~relative to
\thCO~yields or lifts a degeneracy of the solution for \nin. It is thus
important to fit the column density of \HCOp~and \HthCOp~in order to avoid
biasing the value of the fitted volume densities. However, fitting the
column density of \HCOp~and \HthCOp~in both layers increases considerably
the number of unknowns to be estimated. We thus turn to chemical
considerations to assess whether we can make a more relevant hypothesis
than assuming constant abundances of \HCOp~and \HthCOp~relative to
\twCO~and \thCO~respectively.

\subsubsection{The \mbox{\NHCOp/\NtwCO=\NHthCOp/\NthCO} hypothesis for the
  outer layer}
\label{sec:assumption:HCOp-vs-CO}

\FigCOvsHCOp{} %
\TabChemicalBasicModel{} %

The outer layer consists primarily of translucent gas with moderate
temperatures in the presence of ultraviolet radiation. The forward
fractionation reaction involved in the
\begin{equation}
  \thC^+ + \twCO \rightleftharpoons \thCO + \twC^+ + \Delta E \, (\sim 35 \K)
  \label{eq:CO:fractionation}
\end{equation}
equilibrium is moderately efficient in lukewarm gas~\citep{langer:77}.  The
left-to-right reaction is privileged when the gas kinetic temperature is
$\la 35\K$ because the right-to-left reaction is endothermic.  Moreover,
selective photodissociation tends to destroy more efficiently \thCO~and
\CeiO~than \twCO.  The actual chemical balance between these two opposite
trends requires a detailed radiative transfer treatment which can be
performed for example by the Meudon PDR code.

We now discuss the main formation and destruction pathways for \HCOp{} and
\HthCOp. Both molecular ions are essentially formed through two reactions
that involve \twCOp{} or \thCOp{}. The first reaction,
\begin{eqnarray*}
  \twCOp + \HH & \rightarrow & \HCOp + \rm{H}, \\
  \thCOp + \HH & \rightarrow & \HthCOp + \rm{H},
\end{eqnarray*}
has the same reaction rate coefficient for both isotopologues,
$k=7.5 \times 10^{-10}\ccmps$~\citep{Scott97}. The second reaction also
have the same reaction rate coefficient,
\mbox{$k=1.7 \times 10^{-9}\ccmps$}~\citep[computed with the Langevin
approximation][]{Gioumousis58}:
\begin{eqnarray*}
  \twCO + \HHHp & \rightarrow &\HCOp   + \HH,     \\
  \thCO + \HHHp & \rightarrow &\HthCOp + \HH.
\end{eqnarray*}
In addition, both molecular ions are principally destroyed through
dissociative recombination with a reaction rate coefficient of
$1.4 \times 10^{-6}\ccmps$ at 30\K~\citep{Hamberg2014}:
\begin{eqnarray*}
  \HCOp   + \el & \rightarrow &\twCO  + \rm{H},   \\
  \HthCOp + \el & \rightarrow &\thCO  + \rm{H}.
\end{eqnarray*}
With an electronic fraction of about $10^{-4} - 10^{-6}$ in translucent
environments, as found in~\citet{bron21}, the destruction rate of \HCOp{}
and \HthCOp{} is in the $(10^{-10} - 10^{-12}) \times n_{\HH} \ps$ range.

The \HCOp{} and \HthCOp{} destruction rates via the isotopic exchange
reactions~\citep{Mladenovic2014},
\begin{equation}
  \thCO + \HCOp \rightleftharpoons \HthCOp + \twCO + \Delta E (= 17.8\K),
\end{equation}
have a significantly smaller value of a few $10^{-14} \times n_{\HH} \ps$,
as the forward reaction rate coefficient is $6.5\times10^{-10}\ccmps$ at
30\K{}, and the relative abundance of \twCO{} is a few $10^{-5}$. In the
conditions where the main formation channel of \HCOp{} and \HthCOp{} is via
the reactions of \twCO{} and \thCO{} with \HHHp, we obtain the
$\HCOp/\HthCOp=\twCO/\thCO$ relation at steady state.

We checked how the 0D models used in~\cite{bron21}, which sample a range of
densities and illuminations appropriate to translucent conditions, probe
the \mbox{\NtwCO/\NHCOp} ratio as a function of the \mbox{\NthCO/\NHthCOp}
ratio.  We introduced selective photodissociation effects in
\CO~isotopologues by introducing different approximate analytical
expressions of the photodissociation rates obtained from Fig.~9
of~\citet{visser09}, in the chemical network, including the various
isotopic exchange reactions of carbon and oxygen.  Given these assumptions,
we have computed 4000 models that uniformly sample the parameter space
appropriate to translucent environments as in~\citep{bron21}, and filtered
the simulations that lead to \mbox{$\NthCO <
  10^{14}\pscm$}. Figure~\ref{fig:COvsHCOp} shows the joint histograms of
\mbox{\NHCOp/\NtwCO} vs \mbox{\NHthCOp/\NthCO} in log-space for the 3224
remaining models. The models sample more than three orders of magnitude for
these abundance ratios but the histogram is relatively well approximated by
a line of slope one going through the basic abundance ratio listed in
Eq.~\ref{eq:abundance}.

For the sake of clarity, Table~\ref{tab:chemical:basic:model} compares the
chemical assumptions and their implications for column density ratios for
the two main chemical models presented so far.

\subsubsection{Intermediate summary}
\label{section:chemical:model}

In the remainder of this study, we only consider the following hypotheses
that we refer to as the ``improved'' model. We fit the peak intensity of
the \mbox{\twCO\,\Jone} line and the line profiles of all the other
lines. We consider a sandwich model (foreground, inner, background, and
CMB) in which the foreground and background layers (collectively called the
outer layer) share the same physical and chemical properties. In contrast,
the physical properties (velocity centroid and velocity dispersion, kinetic
temperature, and volume density) are estimated independently for the inner
and outer layers. In each layer, all the lines share the same physical and
chemical properties, that is, their emissions are considered co-located
inside each layer.

From the chemical viewpoint, the inner and outer layers are modeled
differently. In the inner dense layer, we estimate the \thCO~and
\HCOp~column densities, and we derive the column densities of the other
species by assuming that they scale with the elemental abundances of the
corresponding carbon and oxygen isotopes, that is,
\begin{eqnarray}
  N(\twCO)   &=& 63\,N(\thCO)\\ 
  N(\CeiO)   &=& N(\thCO) / 7.9,\\
  N(\HthCOp) &=& N(\HCOp) / 63.
\end{eqnarray}
In the outer translucent layer, we fit the column density of \twCO, \thCO,
\CeiO, and \HCOp, and we deduce the column density of \HthCOp~with the
constraint
\begin{equation}
  \label{eq:twco:thco:constraint}
  \frac{N(\HthCOp)}{N(\thCO)} = \frac{N(\HCOp)}{N(\twCO)}.
\end{equation}
We thus have 14 parameters to estimate: four physical parameters (\nHH,
\Tkin, \CV, \FWHM) per layer plus four and two column densities for the
outer and inner layers, respectively.

%%%%%%%%%%%%%%%%%%%%%%%%%%%%%%%%%%%%%%%%%%%%%%%%%%%%%%%%%%%%%%%%%%%%%%%%%%%

\newcommand{\FigIntensityDecompositionChemical}{%
  \begin{figure*}
    \includegraphics[width=\linewidth]{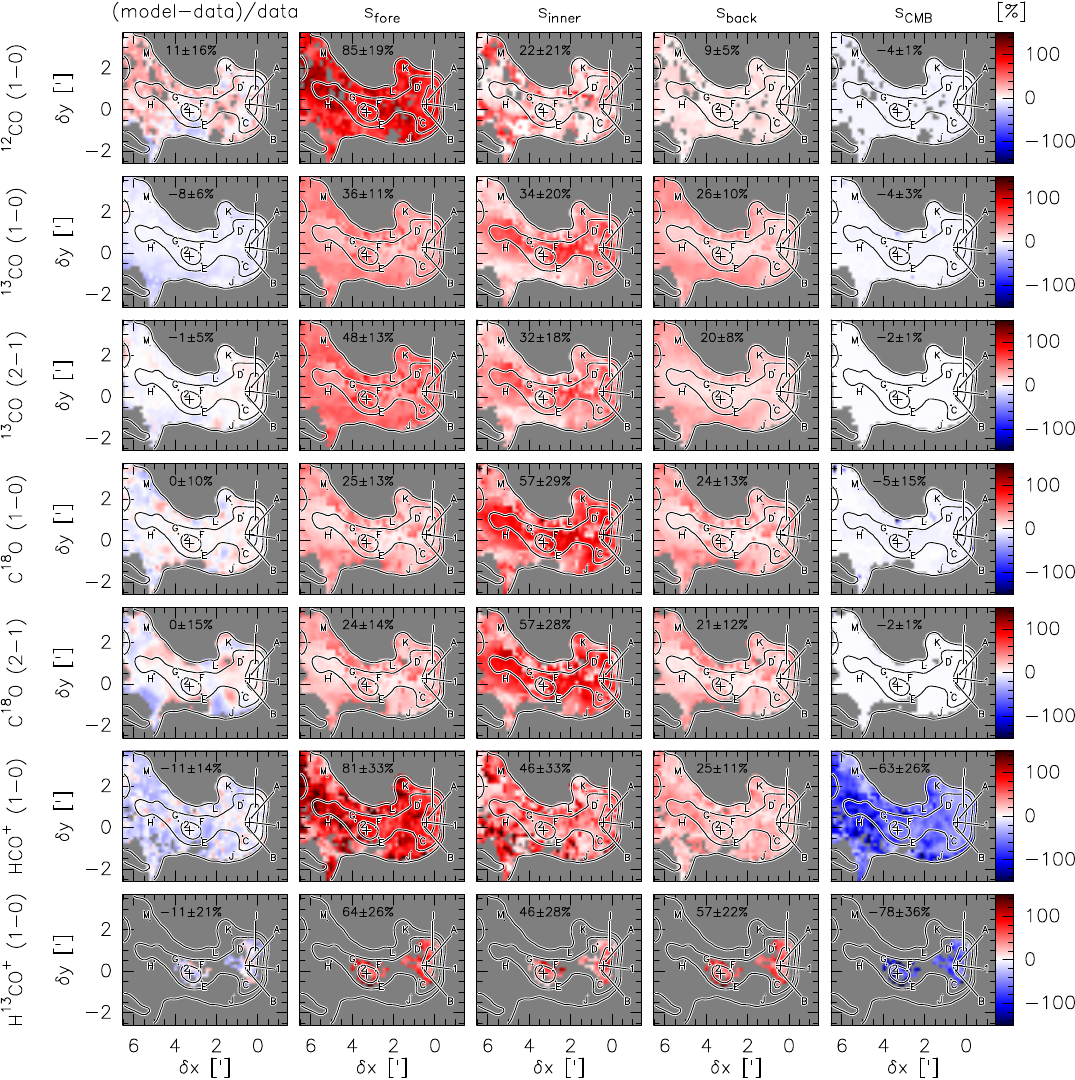}
    \caption{Maps of the contributions of the different layers to the line
      intensity integrated between 8 and 13.5\kms.  \textbf{First column:}
      Relative error of the model with respect of the data in percentage.
      \textbf{Second to fifth columns:} Contribution of the foreground,
      inner, background, and CMB layers to the line integrated intensity in
      percentage of that measured on the data. Each row corresponds to a
      given line: from top to bottom, \mbox{\twCO\,\Jone},
      \mbox{\thCO\,\Jone} and \Jtwo, \mbox{\CeiO\,\Jone} and \Jtwo,
      \mbox{\HCOp\,\Jone}, and \mbox{\HthCOp\,\Jone}.  The percentage on
      the top center of each panel is the mean$\pm$rms value computed over
      the associated maps. Positive and negative relative errors or
      contributions are shown in red and blue, respectively.}
    \label{fig:intensity:decomposition:chemical}
  \end{figure*}
}

\newcommand{\FigSpectraDecompositionOneChemical}{%
  \begin{figure*}
    \centering %
    \includegraphics[width=\linewidth]{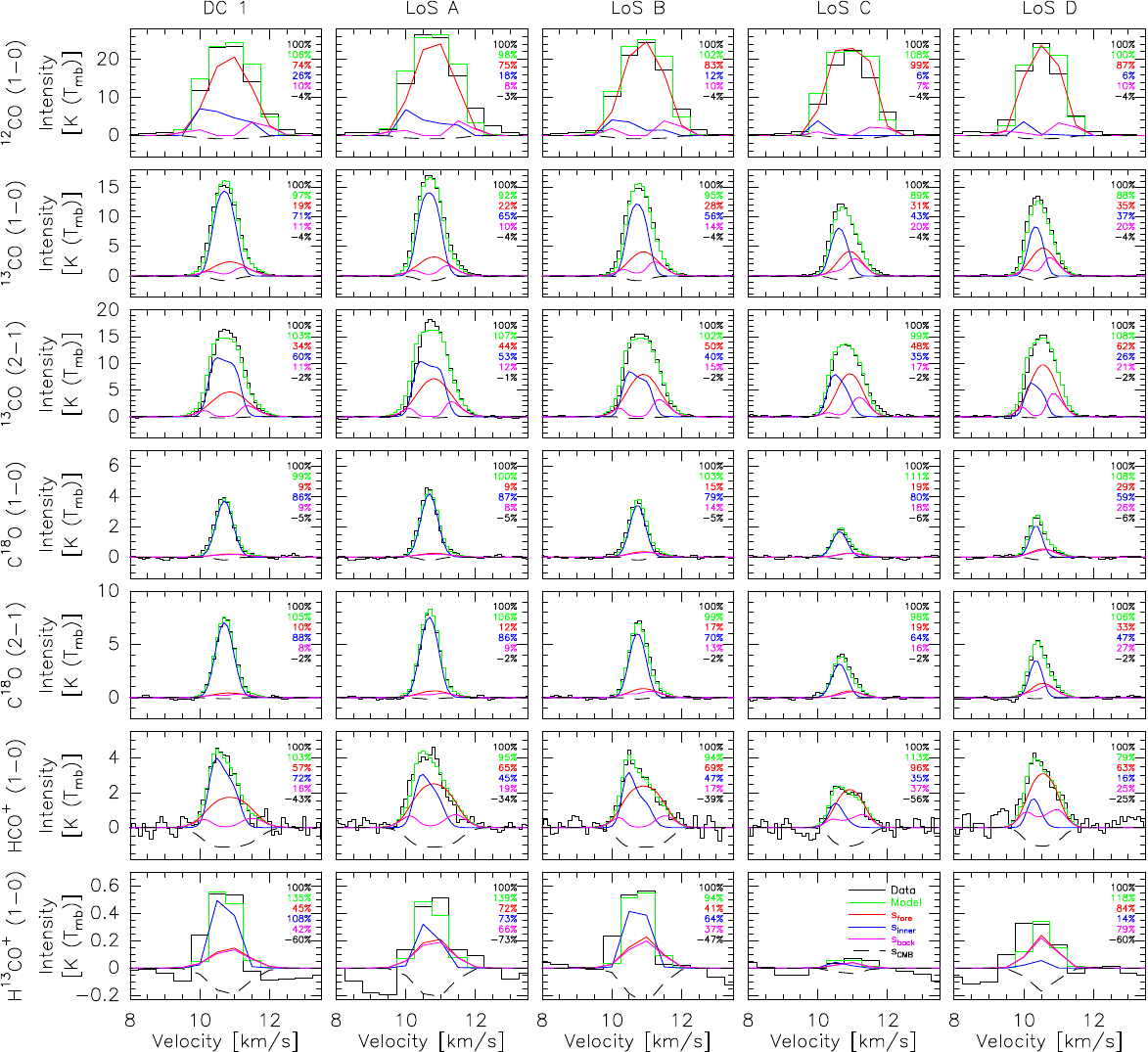}
    \caption{Modeled decomposition of some spectra around the first dense
      core.  Each column corresponds to the spectra for one of the lines of
      sight listed in Table~\ref{tab:coord}. Each row corresponds to a
      given line, from top to bottom: \mbox{\twCO\,\Jone},
      \mbox{\thCO\,\Jone} and \Jtwo, \mbox{\CeiO\,\Jone} and \Jtwo,
      \mbox{\HCOp\,\Jone}, and \mbox{\HthCOp\,\Jone}. We show the
      decomposition of the full spectrum of \mbox{\twCO\,\Jone} even though
      we only used its peak intensity during the fit.  In each panel, the
      data and model spectra are shown as the black and green histograms,
      respectively. The contributions of the foreground, inner, background
      layers are shown as the plain lines in red, blue, pink, respectively,
      and the CMB layer is displayed as the dashed black line.}
    \label{fig:spectra:decomposition:1:chemical}
  \end{figure*}
}

\newcommand{\FigChemistryChemical}{%
  \begin{figure*}
    \centering %
    \includegraphics[width=\linewidth]{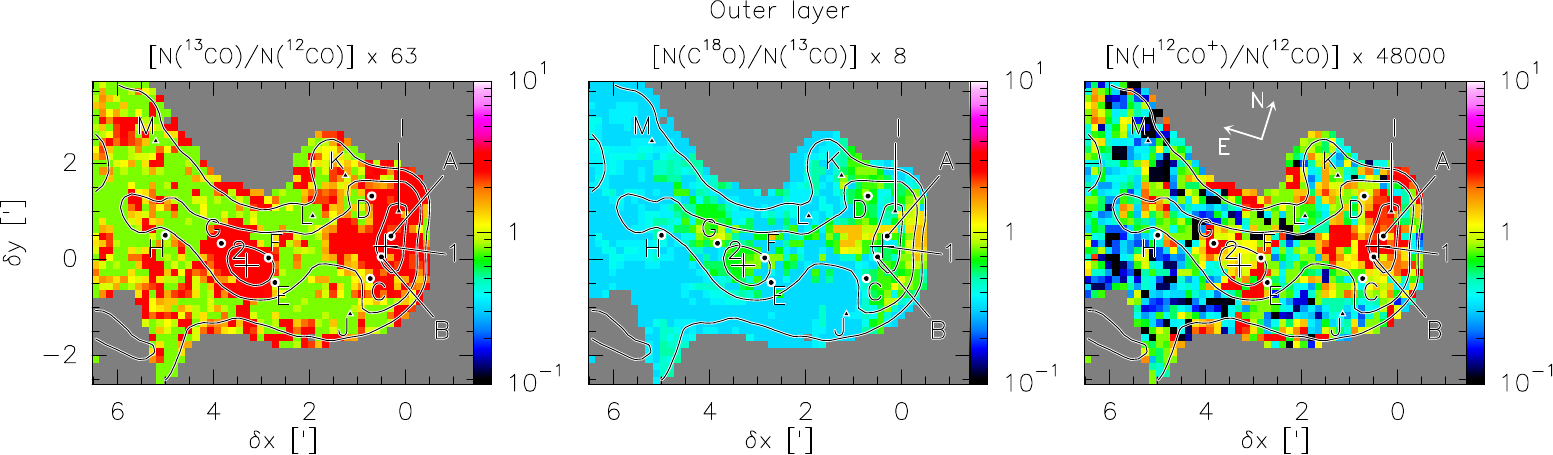}
    \caption{Maps of the ratios of the column densities for the outer
      layer. These ratios have been normalized by the basic values
      established using the elemental isotopic ratios of carbon and oxygen
      (see Eq.~\ref{eq:abundance}).  By construction
      $N(\HCOp)/N(\twCO)=N(\HthCOp)/N(\thCO)$ (see
      Sect.~\ref{sec:assumption:HthCOp:HCOp}).  Contours are drawn at
      visual extinctions of 3, 7, and 16\magn.}
    \label{fig:chemistry:chemical}
  \end{figure*}
}

\newcommand{\FigAbundancesMapsChemical}{%
  \begin{figure*}
    \centering %
    \includegraphics[width=\linewidth]{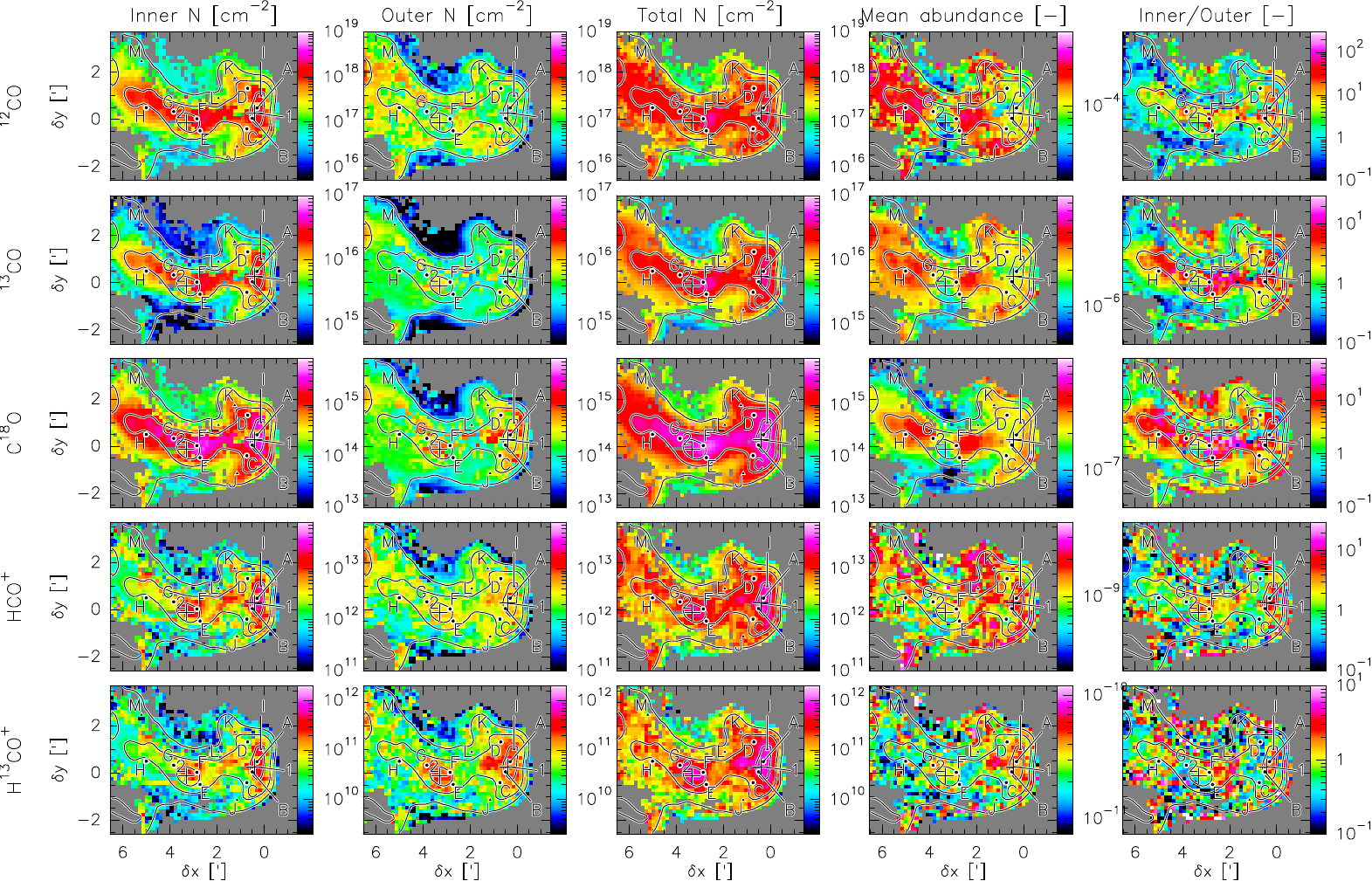}
    \caption{Maps of the chemical properties for \twCO{}, \thCO{}, \CeiO{},
      and \HCOp{}, from top to bottom. The three first columns show the
      column densities for the inner, outer (foreground or background), and
      total (inner plus twice outer) column densities, while the last
      column shows the mean abundance, namely, the total column density
      divided by the \HH{} column density deduced from the visual
      extinction. The last column is the ratio of the inner by the outer
      column density.  Contours are drawn at visual extinctions of 3, 7,
      and 16\magn.}
    \label{fig:abundance:maps:chemical}
  \end{figure*}
}

\newcommand{\FigAbundancePDFsOneChemical}{%
  \begin{figure*}
    \centering %
    \includegraphics[width=\linewidth]{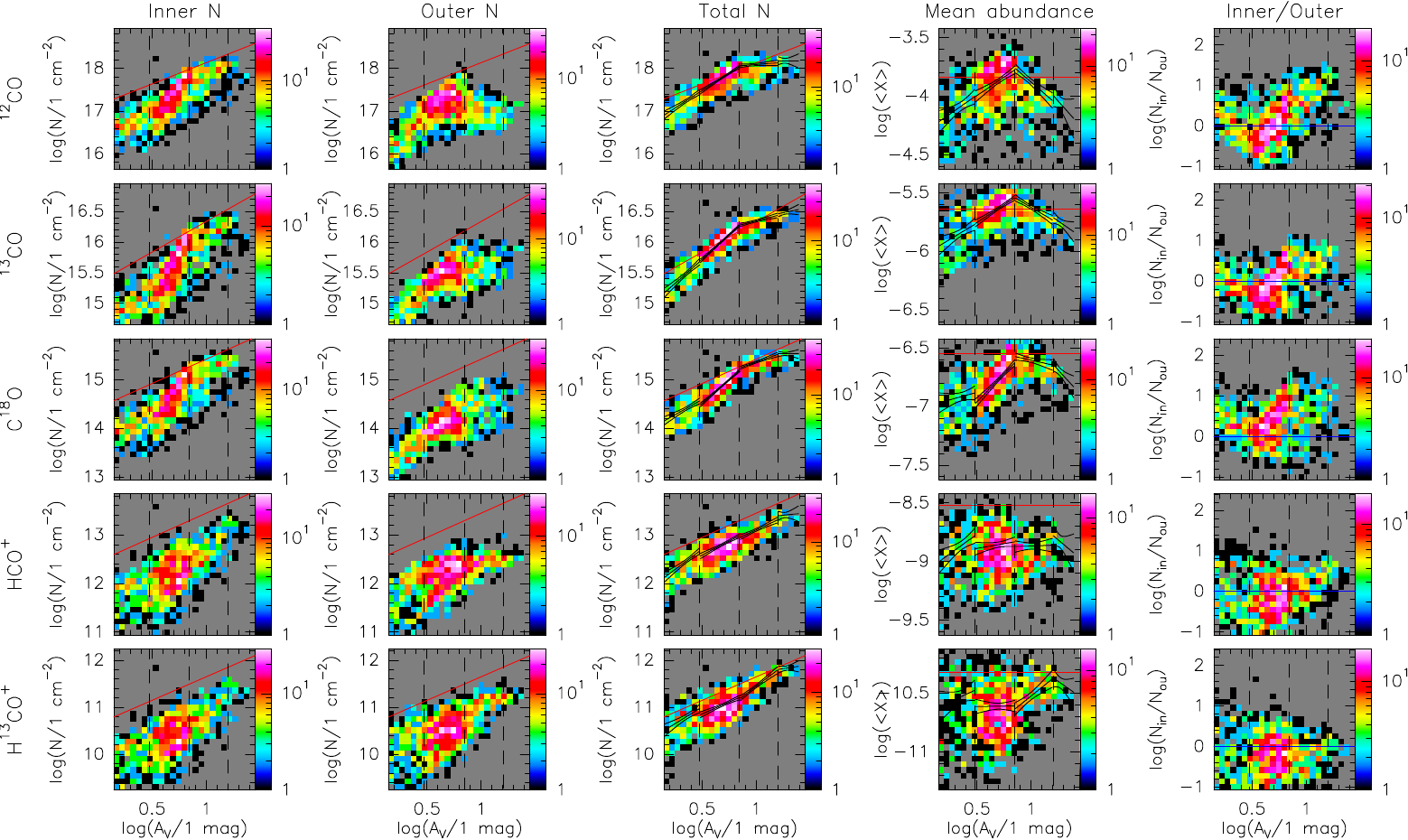}
    \caption{Joint histograms of the molecular column densities,
      abundances, or their ratios as a function of the visual extinction
      for \twCO{}, \thCO{}, \CeiO{}, \HCOp{}, and \HthCOp{}, from top to
      bottom. The three first columns show the column densities for the
      inner, outer (foreground or background), and total (inner plus twice
      outer) column densities, while the last column shows the mean
      abundance, namely, the total column density divided by the \HH{}
      column density deduced from the visual extinction. The dashed
      vertical lines mark the visual extinction at 3, 7, and 16\magn. The
      red plain lines show the species basic abundance.  The black lines
      show the piecewise power-law fits of the column densities within the
      \Av{} ranges specified above (see Tab.~\ref{tab:wmse}) and the
      $\pm 3\sigma$ intervals of the fits. These fits can be directly
      converted to an abundance vs visual extinction relation.}
    \label{fig:abundance-vs-av:pdfs:chemical}
  \end{figure*}
}

\newcommand{\FigAbundancePDFsTwoChemical}{%
  \begin{figure*}
    \centering %
    \includegraphics[width=\linewidth]{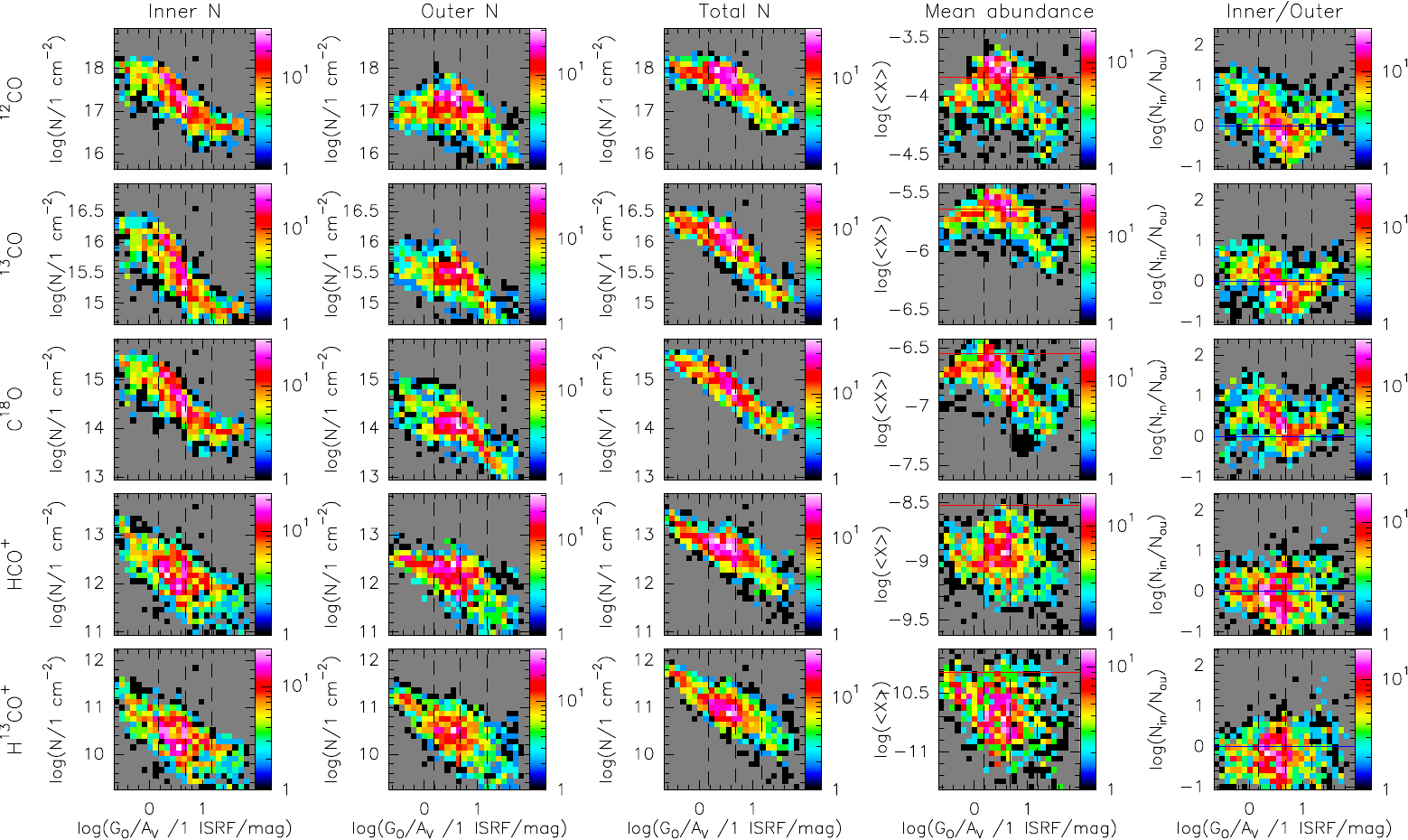}
    \caption{Joint histograms of the column densities, their abundances, or
      their ratios as a function of the visual extinction for the studied
      species. The dashed vertical lines mark the $G_0/\Av$ values at 1.5,
      4.75, and $15\,\emr{ISRF}/\magn$.  The remainder of the figure layout
      is identical to Fig.~\ref{fig:abundance-vs-av:pdfs:chemical}.}
    \label{fig:abundance-vs-go-over-av:pdfs:chemical}
  \end{figure*}
}

\newcommand{\FigKinematicsChemical}{%
  \begin{figure*}
    \includegraphics[width=\linewidth]{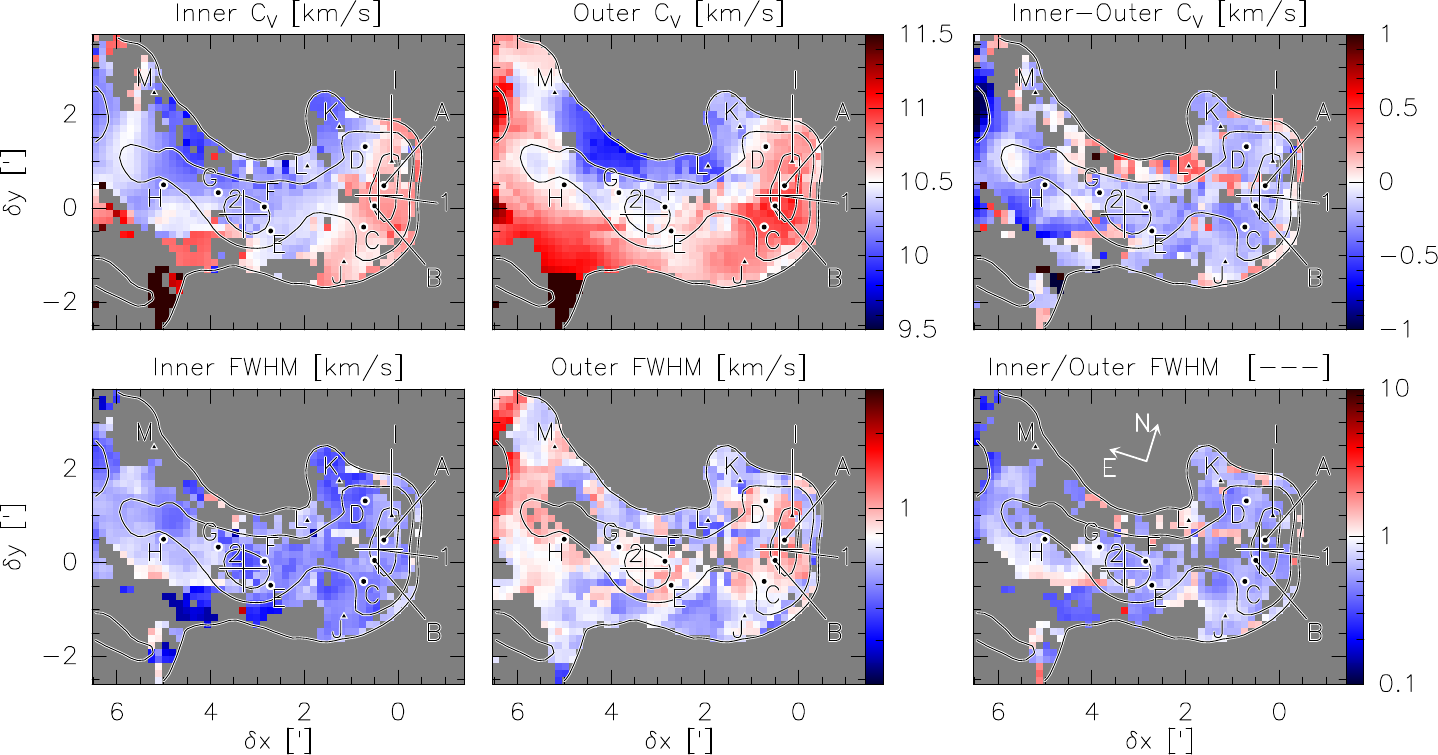}
    \caption{Maps of the kinematic properties. \textbf{Top:} Centroid
      velocity (\CV).  \textbf{Bottom:} Velocity dispersion (\FWHM). The
      two first columns show the kinematic property for the inner and outer
      (foreground or background) layers, while the last column shows either
      the subtraction or the ratio of the of the inner physical property
      divided by the outer one.}
    \label{fig:kinematics:chemical}
  \end{figure*}
}

\newcommand{\FigStateChemical}{%
  \begin{figure*}
    \includegraphics[width=\linewidth]{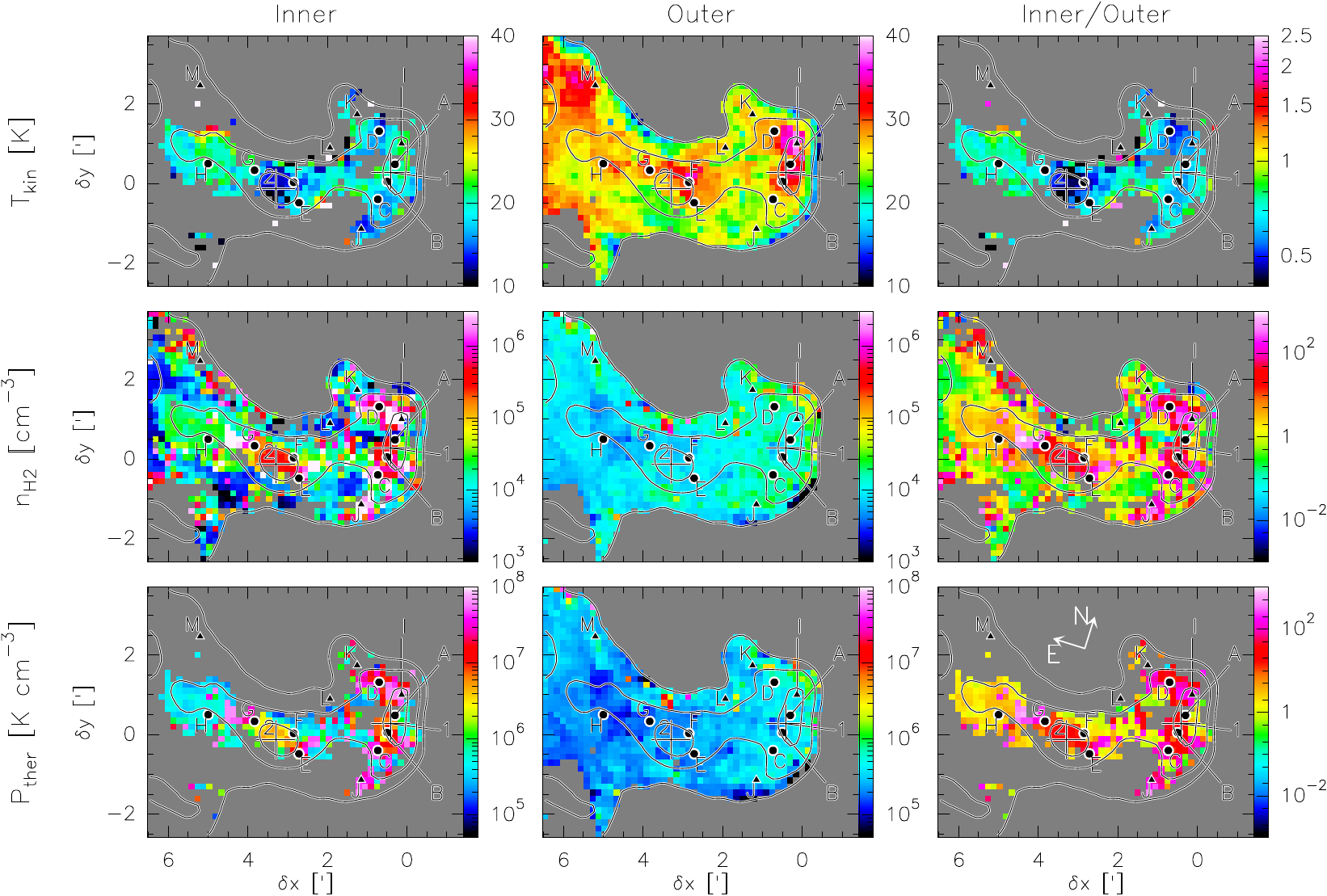}
    \caption{Maps of the gas physical conditions. From top to bottom:
      Kinetic temperature, volume density, and thermal pressure. The first
      two columns display the values for the inner and outer (foreground or
      background) layers, while the last column shows their ratio.}
    \label{fig:state:chemical}
  \end{figure*}
}

%%%%%%%%%%%%%%%%%%%%%%%%%%%%%%%%%%%%%%%%%%%%%%%%%%%%%%%%%%%%%%%%%%%%%%%%%%%

\newcommand{\TabPowerLawFits}{%
  \begin{table*}
    \caption{Parameters of the piecewise power-law fits
      $(\log(\Ntot)=\alpha\,\bracket{\log(\Av)-\log(\Av)_0}+\beta)$ for
      each studied species.}
    \begin{tabular}{c|cc|cc|cc|cc}
      \hline
      \hline
      & \multicolumn{2}{|c|}{$0.15\leq \log(\Av) \leq 0.48$}
      & \multicolumn{2}{|c|}{$0.48\leq \log(\Av) \leq 0.85$}
      & \multicolumn{2}{|c|}{$0.85\leq \log(\Av) \leq 1.2$}
      & \multicolumn{2}{|c }{$1.2 \leq \log(\Av) \leq 1.4$}
      \\
      \hline
      $\log(\Av)_0$
      & \multicolumn{2}{|c|}{0.315}
      &\multicolumn{2}{|c|}{0.665}
      &\multicolumn{2}{|c|}{1.025}
      &\multicolumn{2}{|c }{1.3}
      \\
      \hline
      $\mathrm{^{12}CO}$ & $\left(\begin{array}{c}1.65 \\ 17.18\end{array}\right)$ & $\left(\begin{array}{c}0.01904 \\ 0.00019 \\ -0.00084\end{array}\right)$ & $\left(\begin{array}{c}1.63 \\ 17.75\end{array}\right)$ & $\left(\begin{array}{c}0.00430 \\ 0.00005 \\ -0.00019\end{array}\right)$ & $\left(\begin{array}{c}0.20 \\ 18.10\end{array}\right)$ & $\left(\begin{array}{c}0.00749 \\ 0.00010 \\ 0.00044\end{array}\right)$ & $\left(\begin{array}{c}-0.03 \\ 18.18\end{array}\right)$ & $\left(\begin{array}{c}0.16839 \\ 0.00054 \\ 0.00552\end{array}\right)$ \\
      $\mathrm{^{13}CO}$ & $\left(\begin{array}{c}1.56 \\ 15.42\end{array}\right)$ & $\left(\begin{array}{c}0.01287 \\ 0.00013 \\ -0.00054\end{array}\right)$ & $\left(\begin{array}{c}1.60 \\ 15.98\end{array}\right)$ & $\left(\begin{array}{c}0.00155 \\ 0.00002 \\ -0.00007\end{array}\right)$ & $\left(\begin{array}{c}0.45 \\ 16.36\end{array}\right)$ & $\left(\begin{array}{c}0.00293 \\ 0.00003 \\ 0.00011\end{array}\right)$ & $\left(\begin{array}{c}0.35 \\ 16.50\end{array}\right)$ & $\left(\begin{array}{c}0.04363 \\ 0.00014 \\ 0.00133\end{array}\right)$ \\
      $\mathrm{C^{18}O}$ & $\left(\begin{array}{c}1.26 \\ 14.36\end{array}\right)$ & $\left(\begin{array}{c}0.01556 \\ 0.00017 \\ -0.00063\end{array}\right)$ & $\left(\begin{array}{c}1.96 \\ 14.84\end{array}\right)$ & $\left(\begin{array}{c}0.00258 \\ 0.00003 \\ -0.00013\end{array}\right)$ & $\left(\begin{array}{c}0.79 \\ 15.38\end{array}\right)$ & $\left(\begin{array}{c}0.00401 \\ 0.00005 \\ 0.00016\end{array}\right)$ & $\left(\begin{array}{c}-0.01 \\ 15.53\end{array}\right)$ & $\left(\begin{array}{c}0.05222 \\ 0.00017 \\ 0.00164\end{array}\right)$ \\
      $\mathrm{HCO^{+}}$ & $\left(\begin{array}{c}1.67 \\ 12.40\end{array}\right)$ & $\left(\begin{array}{c}0.02108 \\ 0.00020 \\ -0.00096\end{array}\right)$ & $\left(\begin{array}{c}1.16 \\ 12.78\end{array}\right)$ & $\left(\begin{array}{c}0.00584 \\ 0.00006 \\ -0.00019\end{array}\right)$ & $\left(\begin{array}{c}1.03 \\ 13.13\end{array}\right)$ & $\left(\begin{array}{c}0.00787 \\ 0.00009 \\ 0.00015\end{array}\right)$ & $\left(\begin{array}{c}0.22 \\ 13.42\end{array}\right)$ & $\left(\begin{array}{c}0.15840 \\ 0.00049 \\ 0.00516\end{array}\right)$ \\
      $\mathrm{H^{13}CO^{+}}$ & $\left(\begin{array}{c}1.43 \\ 10.74\end{array}\right)$ & $\left(\begin{array}{c}0.03755 \\ 0.00038 \\ -0.00172\end{array}\right)$ & $\left(\begin{array}{c}1.01 \\ 11.03\end{array}\right)$ & $\left(\begin{array}{c}0.00791 \\ 0.00008 \\ -0.00026\end{array}\right)$ & $\left(\begin{array}{c}1.75 \\ 11.49\end{array}\right)$ & $\left(\begin{array}{c}0.01084 \\ 0.00013 \\ 0.00039\end{array}\right)$ & $\left(\begin{array}{c}0.35 \\ 11.84\end{array}\right)$ & $\left(\begin{array}{c}0.09248 \\ 0.00029 \\ 0.00235\end{array}\right)$
      \\
      \hline
    \end{tabular}
    \tablefoot{In each sub-table, we provide on the left $\alpha$ and
      $\beta$ coefficents and on the right $\mathrm{var}(\alpha)$,
      $\mathrm{var} (\beta)$ and $\mathrm{cov} (\alpha,\beta)$. The error
      bars of the line fit are given by
      $\sqrt{\mathrm{var}(\alpha)\,(\log(\Av)-\log(\Av)_0)^2 +\mathrm{var}
        (\beta) +2 \,\mathrm{cov} (\alpha,
        \beta)\,(\log(\Av)-\log(\Av)_0)}$.}
    \label{tab:wmse}
  \end{table*}
}

%%%%%%%%%%%%%%%%%%%%%%%%%%%%%%%%%%%%%%%%%%%%%%%%%%%%%%%%%%%%%%%%%%%%%%%%%%%

\section{Results}
\label{sec:results}

\FigIntensityDecompositionChemical{}%

Our goal in this section is two-fold. We first want to check whether the
used hypotheses allow us to estimate chemical and physical parameters that
are consistent with previous findings. Second, most of the previous results
about the Horsehead pillar were derived with a single-layer model, while
the sandwich model allows us to characterize the chemical and physical
structure of the gas along the LoS. We first focus on the structure along
the LoS (Sect.~\ref{sec:results:intensity} and \ref{sec:results:spectra}):
We study whether the integrated intensity of the different lines is
dominated by one of the three layers or whether it is distributed between
the different layers. We search for the origin of the asymmetry of the line
profiles. We compare the line excitation temperature with the CMB.  We then
look at the abundances of the CO and \HCOp{} isotopologues along the LoS as
a function of \HH{} column density and far UV radiation field
(Sect.~\ref{sec:results:chemistry}): We try to quantitatively relate the
variations of the abundances with known chemical processes such as
photodissociation, isotopic fractionation, or freezing onto dust grains.
Finally, we investigate the relationship of the derived physical parameters
with the star formation activity (Sect.~\ref{sec:results:kinematics}
and~\ref{sec:results:physical:state}): We search for a kinematic signature
of infall or expansion of the gas onto the filaments and dense cores. We
discuss whether the molecular gas properties (kinetic temperature, volume
density, and thermal pressure) are affected by the nearby presence of a PDR
or of a young stellar object.

\subsection{Contribution of the different layers to the integrated
  intensity}
\label{sec:results:intensity}

For each studied line, Fig.~\ref{fig:intensity:decomposition:chemical}
shows 1) the relative error of the integrated intensity estimated by the
model with respect to the observed value, integrated between $8.0$ and
$13.5\kms$ (first column), and 2) the contributions of the different layers
(foreground, inner, background, and CMB, from the second to the fifth
column) to the line integrated intensity.  While our estimator only fits
the \mbox{\twCO\,\Jone} peak intensity, we still provide the relative error
and contributions of each layer to the integrated intensity of this
line. We have filtered out sight-lines where the relative error on the
integrated line intensity is larger than $40\%$ or the visual extinction is
lower than $3\magn$ for all lines, except for the faint
\mbox{\HthCOp\,\Jone} line where we used a threshold of $10\magn$. The
percentage listed in the top right corner of each panel is the mean$\pm$rms
value computed over the selected sight-lines for the associated map. As
expected, a negative contribution is ascribed to the CMB layer because of
the ON-OFF observation procedure that implies the subtraction of a black
body emission at $2.73\K$ as indicated in Eq.~\ref{eq:sandwich:total}.
Moreover, the sum of contributions of the layers to the integrated
intensities can be larger than $100\%$ when the estimated signal exceeds
the observed integrated line intensity.

The integrated intensity of all the lines is reconstructed within $20\%$
overall (\mbox{$= \sqrt{\emr{mean}^2+\emr{rms}^2}$}).  The rms of the
relative error on the integrated intensity decreases when the \SNR{}
increases, except for the \mbox{\twCO\,\Jone} line. The \mbox{\thCO\,\Jtwo}
and \mbox{\CeiO\,\Jone} and \Jtwo~lines are reconstructed with much less
bias ($\pm 1\%$) than the \twCO, \thCO, \HCOp, and \mbox{\HthCOp\,\Jone}
lines. The bias for the \mbox{\twCO\,\Jone} line can be explained by the
fact that we only use the peak temperature in the fit.

The contributions of the different layers to the integrated intensity
depend much on the considered species. The \mbox{\twCO\,\Jone} integrated
intensity mainly comes from the foreground layer ($85\%$) with a second
contribution from the inner layer ($22\%$), while the integrated intensity
of the two \CeiO~lines mainly comes from the inner layer ($57\%$) with an
approximately equal contributions from the foreground and background layers
(\mbox{$\sim24\%$} each). The \thCO~lines have an intermediate
behavior. The \mbox{\thCO\,\Jone} integrated intensity comes from the
foreground, inner, and background layers at approximately equal shares. The
foreground layer contributes more than the background layer for the
\mbox{\thCO\,\Jtwo} line.  This is consistent with the higher optical depth
for the \Jtwo~line as compared with the \Jone~one.

The \Jone~line of the \HCOp~isotopologues shows a very different behavior
that is mostly related to the fact that its excitation temperature is close
to the CMB temperature. While the subtraction of the CMB has a small effect
on the CO isotopologues, it is significant for the \HCOp~and \HthCOp~lines
and can lead to a significant negative contribution to the total integrated
line intensity, beyond $-50\%$. The foreground, inner, and background
layers must therefore provide contributions significantly larger than the
detected integrated line intensity to compensate, implying that the
percentage of the combined layer contributions is larger than 100\%.  The
inner layer contribution to the \Jone~line is about $50\%$ for both
\HCOp~and \HthCOp. However, the foreground contribution largely dominates
the background contribution for the \HCOp~optically thick line, while both
layers have approximately equal contributions for the optically thin
\HthCOp~line.

Looking at the spatial distribution of the contributions for the \thCO~and
\CeiO~lines, the inner layer contributes much more than the outer layers
when the visual extinction is larger than $7\magn$, and vice-versa. We
relate this to the change of geometry of the Horsehead pillar that happens
approximately at this visual extinction. Above $7\magn$, the geometry is
mostly face-on and the inner layer dominates for optically thin
lines. Below $7\magn$, the geometry becomes edge-on, and the contribution
of the inner layer decreases and sometimes vanishes compared to that of the
outer layers.

\subsection{Contribution of the different layers to the spectra}
\label{sec:results:spectra}

\FigSpectraDecompositionOneChemical{}%

To get a better understanding of how the different layers contribute to the
observed spectra, Fig.~\ref{fig:spectra:decomposition:1:chemical},
\ref{fig:spectra:decomposition:2:chemical},
and~\ref{fig:spectra:decomposition:3:chemical} show the modeled
decomposition of spectra selected around the dense cores 1, and 2, and then
in the outskirts of the Horsehead nebula, respectively.  In all cases, the
background emission is attenuated by the inner layer in its foreground.
Hence, the contribution of the background layer to the emergent spectra
mostly matters in the line wings. Similarly, the emission from the inner
layer is absorbed by the foreground layer, and this radiative interaction
allows one to fit the asymmetries of the profiles of the optically thick
lines of \twCO, \thCO, and \HCOp.

Figure~\ref{fig:spectra:decomposition:1:chemical} first shows the
decomposition at the peak of the \HthCOp~emission (DC 1) for the dense core
located at the top of the Horsehead. Positions A and B are two nearby LoSs
that have already been used to justify some of the assumptions made in our
modeling (see Sect.~\ref{assumption:centroid:velocity}
and~\ref{sec:assumption:HthCOp:HCOp}). Positions C and D are located at the
edge of the dense core away from the $\sigma\,$Ori star that photo-excites
the Horsehead PDR, where the \mbox{\HCOp\,\Jone} peak emission abruptly
decreases from above $3.5\K$ to below $2.5\K$ on
Fig.~\ref{fig:data:maps}. The inner layer emission contributes the most to
the integrated line intensity near the dense core projected center (e.g.,
positions 1 and A). This is true even for the \mbox{\twCO\,\Jone} line
where the inner emission contributes about $20\%$ of the total integrated
line intensity. When going further away from the dense core center
(positions C and D), the inner layer continues to provide the dominant
contribution for the optically thin \CeiO~lines, while the contributions
from the foreground and background layers have either similar integrated
line intensities as that from the inner layer, or even dominate for the
\twCO~or \HCOp~lines. Position B seems to be an intermediate case.

Figure~\ref{fig:spectra:decomposition:2:chemical} shows the decomposition
at the peak of the \HthCOp~emission (DC 2) for the dense core located in
the neck of the Horsehead. Positions E and G are located where the
\mbox{\HthCOp\,\Jone} peak emission abruptly decreases from above $0.25$ to
below $0.10\K$. Position F is extensively studied
in~\citet{Roueff24}. Position H is located at a LoS where the
\mbox{\CeiO\,\Jone} emission drops below $3\K$. In contrast with the
analysis of the previous dense core, the inner layer contributes less to
the overall integrated line intensity of the optically thick lines at
position 2 near the dense core projected center than at position F located
further away. Positions E and G at the edge of the dense core shows a
strong asymmetry of the most optically thick lines for which the foreground
and background contributions may even dominate the integrated line
intensity. This effect is less prominent for the position H located at the
transition between the \CeiO~filament and its environment.

\FigChemistryChemical{}%

Figure~\ref{fig:spectra:decomposition:3:chemical} shows the decomposition
for LoSs located at the outskirt of Horsehead pillar. There is an embedded
young stellar object at position I.  Positions J and K are located in the
Horsehead mane and muzzle where the \mbox{\CeiO\,\Jone} peak intensity is
still $\sim 2\K$, but the \mbox{\HthCOp\,\Jone} is not clearly detected
anymore. Finally, positions L and M are located where the
\mbox{\thCO\,\Jone} peak intensity drops below $10\K$. Except for position
I that has a similar behavior as the positions near the dense core 1,
positions J to M have no clearly detected \HthCOp, and faint \HCOp~and
\CeiO~lines. The optically thick lines of \HCOp~and \twCO~at the K and L
positions have inner and outer contributions whose radiative interaction
(absorption) almost cancel each other. This is accompanied by an inversion
of the centroid velocities of the two components (see
Sect.~\ref{sec:results:kinematics}). Finally, position M shows a spectral
decomposition where the fit seems to have inverted the properties of the
inner and outer layers as compared with the rest of the map.  The
integrated line intensity is completely dominated by the outer layers, and
the associated linewidth is much wider than for the previous positions. We
interpret this as the transition between the face-on and edge-on geometry
at the outskirts of the Horsehead pillar. In the edge-on geometry the
chemistry is better represented by assumptions enforced in the outer
layer. Position M is a clear example of the strong coupling between the
chosen chemical hypotheses as a function of the geometry and the inferred
kinematics properties.

\subsection{Chemical abundances}
\label{sec:results:chemistry}

\subsubsection{Relative abundances}

Figure~\ref{fig:chemistry:chemical} shows the spatial distribution of
ratios of different column densities derived for the outer layer.  These
ratios are normalized by the basic value derived from elemental abundances
of carbon and oxygen isotopes (see Eq.~\ref{eq:abundance}) and previous
studies of the targeted molecules as explained in
Sect.~\ref{sec:assumptions:sandwich}. By construction, these basic values
are enforced in the inner layer, and \mbox{\NHCOp/\NtwCO=\NHthCOp/\NthCO}
in both layers (see Sect.~\ref{sec:assumption:HthCOp:HCOp}).

The mean value of the \mbox{\NthCO/\NtwCO} ratio in the outer layer is
$\sim 1.6$ times the ``basic'' value used in the inner layer. Moreover,
\thCp{} is detected by~\citet{Pabst17}, and the temperature remains
moderate in the outer layer, as shown in
Sect.~\ref{sec:results:physical:state}. All this allows the forward
reaction of the fractionation reaction (Eq.~\ref{eq:CO:fractionation}) to
predominate in the outer layer~\citep[see also][]{Liszt07,Rollig13}.  The
\mbox{\NthCO/\NtwCO} ratio is also enhanced toward the DC\,2 LoS, which is
in line with the \HthCOp{} detection in that environment.

Two effects help explain the spatial variations of the \mbox{\NCeiO/\NthCO}
ratio in the outer layer. First, the abundance of \thCO~is enhanced due to
the \Cp~fractionation described above.  Second, \CeiO{} is more easily
photodissociated than \thCO{} (a process named selective
photodissociation). Indeed, the shift between the \CeiO{} and \twCO{}
pre-dissociated UV absorption transitions is significantly larger than the
one between the \thCO{} and \twCO{} transitions. Shielding effect by
\twCO{} are thus reduced for \CeiO{} with respect to \thCO{}.  Both effects
are important over most of the field of view, in particular for visual
extinctions lower than $7\magn$.

The spatial variations of the \HCOp~abundance relative to \CO~are larger
than those of the two other abundance ratios. The \mbox{\NHCOp/\NtwCO} and
\mbox{\NHthCOp/\NthCO} ratios are one to four times larger than the basic
value toward dense core LoSs and two to ten times lower than this value for
LoSs where \mbox{$3 \le \Av \le 7\magn$}. The \HCOp~abundance relative to
\CO~reaches values within a factor of two from the basic value in regions
where \Av~is around $10\magn$.

\subsubsection{Column densities and mean abundances relative to \HH{}}

\FigAbundancesMapsChemical{}%
\FigAbundancePDFsOneChemical{}%
\TabPowerLawFits{} %

For each species, Fig.~\ref{fig:abundance:maps:chemical} shows the inner,
outer (foreground plus background), and total (inner plus twice outer)
column densities, as well as its mean abundance along the LoS defined as
\begin{equation}
  \abmean = \frac{N_\emr{tot}(X)}{N_\emr{tot}(\HH)},
  \label{eq:Ntot:1}
\end{equation}
where
\begin{equation}
  \label{eq:Ntot}
  N_\emr{tot}(X) = \sum_L N_{L}(X), 
  \quad \mbox{and} \quad
  N_\emr{tot}(\HH) = \sum_L N_{L}(\HH).
\end{equation}
Following~\citep{Pety17}, we estimated the total column density of \HH~from
the ``molecular'' visual extinction (defined in Sect.~\ref{sec:obs}
and~\ref{app:av:correction}) as
\begin{equation}
  N_\emr{tot}(\HH) = 0.9\e{21}\pscm \,\frac{\Av^\emr{mol}}{1\magn}.
  \label{eq:NHH:from:av}
\end{equation}
We have also added the ratio of column densities in the inner and outer
layers, \mbox{\Nin/\Nou} ratio for each species in this figure to
illustrate whether one layer dominates or whether the mean abundance
results from just an average of both layers.  Indeed, if we define the
abundance of a species $X$ with respect to \HH~in the layer $L$ as
\begin{equation}
  \ab{L} = \frac{N_{L}(X)}{N_{L}(\HH)},
\end{equation}
it is straightforward to show that
\begin{equation}
  \frac{\abmean}{\ab{\text{in}}} %
  = \frac{1 + 2\, \frac{\Nou(X)}{\Nin(X)}}
  {1 + 2\, \frac{\ab{in}}{\ab{ou}}\,\frac{\Nou(X)}{\Nin(X)}}.
  \label{eq:ab:over:ab-in}
\end{equation}
In this equation, the mean abundance of species $X$ converges to
$\ab{\text{in}}$ when \mbox{$\Nou(X) \ll \Nin(X)$}, and to $\ab{\text{ou}}$
when $\Nou(X) \gg \Nin(X)$. The convergence rate depends on the value of
the \mbox{$\ab{\text{in}}/\ab{\text{ou}}$} ratio.

The maps of the total column density for the three main \CO~isotopologues
clearly show the filamentary structure inside the Horsehead pillar. From
this viewpoint, they look more like the column density of the inner layer
than that of the outer layer. This is consistent with the fact that the
ratio of the inner and outer column densities is much larger than one for
extinctions above $7\magn$, and around or lower than one when
\mbox{$3 \le \Av \le 7\magn$}. Even though the \mbox{\twCO\,\Jone} line is
highly optically thick, the simultaneous fit of different chemical species
allowed us to recover reasonable column densities of \twCO~for the inner
layer.

In contrast, the maps of the total column density for the two main
\HCOp~isotopologues show a more balanced combination of column densities
from the inner and outer layers, in agreement with the less contrasted maps
of the ratio of the inner and outer column densities.  For \HCOp, the
column density of the inner layer delineates the filamentary structure with
a peak of the column density toward the dense cores. The map of the outer
column density shows less structure. The total column density map of
\HthCOp~clearly shows higher column densities in the regions of higher
extinctions. Overall the relative abundance seems to be somewhat higher
along the LoSs toward dense cores and fairly constant elsewhere given the
limited \SNR{} of the line.

The most striking feature of the mean abundance maps is the depletion of
the CO isotopologues toward the two dense cores. This is obvious for the
dense core in the Horsehead neck. For the dense core at the top of the
Horsehead, it is clearer for the mean abundance of \twCO~than for those of
\thCO~and \CeiO. To quantitatively confirm these trends,
Fig.~\ref{fig:abundance-vs-av:pdfs:chemical} shows the joint histograms of
the column densities and mean abundances as a function of the ``molecular''
visual extinction. We have fitted independently in each range of \Av~a
power-law to the scatter plots of \Ntot~versus \Av. We have used the CRB
values to take into account the uncertainty on \Ntot{} and we have assumed
that the visual extinction are perfectly
known. Appendix~\ref{appendix:crb:derived:quantities} explains the
computation of the uncertainty on \Ntot{}, and Appendix~\ref{appendix:wmse}
details the fit algorithm.  For the sake of reproducibility, the values of
the fitted coefficients are listed in Table~\ref{tab:wmse} and shown as the
black lines in the third column of
Fig.~\ref{fig:abundance-vs-av:pdfs:chemical}. The abundance of \CeiO{} with
respect to \HH{} reaches $2.0\e{-7}$ at the DC\,1 position and $1.3\e{-7}$
at DC\,2. As the used ``basic'' abundance of \CeiO{} is $2.84\e{-7}$, this
implies depletion factors of 1.4 for DC\,1 and 2.2 for DC\,2. Such values
are in agreement with abundances and depletion factors obtained in other
studies of molecular cores.  For instance, \citet{Tafalla02}~found a strong
depletion of \CeiO{} in the inner regions of five Taurus starless cores
(less than $100''$ or $0.07\pc$ from their center) using a reference
\CeiO{} abundance of $1.7\e{-7}$.  In a more recent study of four dense
cores located in infrared dark clouds, \citep{feng20}~also measure
depletion factors of a few for \CeiO{}.

As the far UV radiation field plays an important role at least in the
eastern part of the Horsehead pillar,
Fig.~\ref{fig:abundance-vs-go-over-av:pdfs:chemical} shows the joint
histograms of the column densities and mean abundances as a function of the
$G_0/\Av$ ratio. We use this ratio as a proxy of the ratio of the far UV
field over the volume density that controls to first order the chemistry of
PDRs. Using this proxy assumes that there is some relation between the
volume and column densities.

The mean abundances of \twCO~and \thCO~increases up to a visual extinction
of about $3\magn$. At this visual extinction, they reach the values
corresponding to the elemental abundance of the associated carbon
isotope. For \mbox{$\Av \ga 7\magn$} depletion sets in, and the column
density does not increase much anymore, which leads to a decreased
abundance.  The abundance of \CeiO~increases up to
\mbox{$\Av \sim 7\magn$}, and then decreases.  Looking at the $G_0/\Av$
ratio, the column densities increase when $G_0/\Av$ decreases from 15 to
1.5 \unit{ISRF\,\magn^{-1}}, and then they remain mostly constant.  When
$G_0/\Av \le 4.75\unit{ISRF\magn^{-1}}$, the inner column density clearly
dominates the column density budget. Above this value the total column
density is dominated by the outer layer for \twCO~and \thCO, but it remains
dominated by the inner layer for \CeiO.

The column densities of the two \HCOp~isotopologues studied increase more
steadily with increasing visual extinction or decreasing $G_0/\Av$
values. The associated mean abundances are quite scattered. Their typical
values seem to be lower than the basic values (see Eq.~\ref{eq:abundance})
by a factor between two and three (or between 0.3 and 0.5 \unit{dex}). For
both isotopologues, the total column density includes contributions from
both layers.

Looking at the inner and outer column densities of the \CO~isotopologues as
a function of $G_0/\Av$, we can distinguish another regime. For
\mbox{$1.5 \le G_0/\Av \le 4.75\unit{ISRF\magn^{-1}}$}, the column
densities of \CO~isotopologues clearly decrease with increasing $G_0/\Av$
in the inner layer, while they seem to saturate in the outer layer.  The
opposite behavior is seen for
\mbox{$4.75 \le G_0/\Av\unit{ISRF\magn^{-1}}$}, namely the column densities
of the \CO~isotopologues do not change much with increasing $G_0/\Av$ in
the inner layer, while the column densities of the outer layer decrease.
This behavior may be caused by the increased photodissociation of the
\CO~isotopologues in a higher far UV radiation field.

\subsection{Kinematics}
\label{sec:results:kinematics}

\FigKinematicsChemical{}%

In Fig.~\ref{fig:kinematics:chemical}, we compare the spatial distribution
of the centroid velocity and velocity dispersion between the inner and
outer layers. The centroid velocity of the inner and outer layers displays
to first order an increasing gradient from north to south of the
Horsehead. \citet{Hilyblant05} already described this behavior and
interpreted this as a global rotation of the Horsehead pillar around its
east-west axis.

We concentrate on LoSs where \mbox{$\Av \ge 7\magn$}. The difference of the
centroid velocities for the inner and outer layers confirms that the
centroid velocity of the outer layer follows that of the inner layer,
shifted by about \mbox{$-0.25\kms$}. To check whether this shift is
significant, we divide the absolute value of
$\CVl{\text{in}}-\CVl{\text{ou}}$ by its standard deviation computed as the
square root of the CRB. We obtain that the centroid velocity difference is
larger than the standard deviation for 90\% of the lines of sight, and it
is four times larger for 50\% of the lines of sight, including those toward
the dense cores. This suggests that the foreground layer moves toward the
inner layer and away from the observer. This could hint to accretion of
translucent gas onto the denser inner filament. However, the proposed model
attributes the same centroid velocity to the foreground and background
layers. A model that allows for the description of the infall or expansion
motions along the LoS would mirror the centroid velocities of the
foreground and background layers with respect to the inner layer one. More
precisely, it would use different centroid velocities of the foreground and
background layers with a velocity offsets of opposite signs with respect to
the centroid velocity of the inner layer. This model variant will be tested
in a forthcoming paper. In the meanwhile, we note that the background layer
contributes little to the total integrated line intensity. Moreover, for
the same LoSs \mbox{($\Av \ge 7\magn$)}, the velocity dispersion is about
twice as large in the outer layer than in the inner one ($\sim 0.5$ and
$\sim0.25\kms$, respectively). The outer layer thus appears more turbulent
than the inner one, as might be expected when the gas becomes denser from
translucent to dense gas through filamentary gas.

\FigStateChemical{}%

For LoSs where \mbox{$\Av < 7\magn$}, an inversion of the centroid velocity
shift and of the velocity dispersion ratio sometimes happens between the
inner and outer layers (see, for instance,
\mbox{$\delta x \in [1.5',3.0']$} and \mbox{$\delta y \in
  [0.5',1.3']$}). This could be due to the change of geometry from face-on
to edge-on for which our chemical assumptions for the inner and outer
layers incorrectly lead to an inversion of their respective positions along
the LoS and associated line profile properties.  We plan to address this
issue in a future paper.

\subsection{Physical state}
\label{sec:results:physical:state}

Figure~\ref{fig:state:chemical} shows the spatial variations of the gas
physical conditions (kinetic temperature, volume density, and thermal
pressure) for the inner and outer layers. It also shows the maps of the
ratio of these three parameters in the inner and outer layers.

As expected, the kinetic temperature of the inner layer is almost always
lower than that of the outer layer when \mbox{$\Av \ge 7\magn$}, and the
inner layer has a higher volume density than the outer layer for the same
LoSs.  However, while the temperatures of the inner and outer layers are
different by a factor of a few, the volume density of the inner layer
increases by more than one order of magnitude toward the two dense cores.
This explains why the map of the ratio of the thermal pressures
\mbox{$(\Pth = n\,\Tkin)$} in the inner to outer layers resembles more that
of the ratios of the volume densities than that of the ratio of kinetic
temperatures.

The inner layer of the eastern dense core is colder than that of the
western dense core.  This is consistent with the fact that \CO~depletion
onto dust grains is much more prominent toward the eastern dense core
LoS. The western dense core lies close to an embedded Young Stellar Object
(YSO) (Position I). The presence of this YSO could provide an additional
source of heating of the inner layer around this position.

The outer layer shows a decreasing gradient of both the volume density and
thermal pressure from west to east.  In particular, toward the eastern
dense core, the thermal pressure of the inner layer is 1.9 times lower than
that of the same layer toward the western dense core. Such a gradient could
be related to the compression associated with the irradiation of the cloud
by the $\sigma\,$Ori star~\citep[see][]{WardThompson06}, creating a PDR on
the east side.
% DC1 Pth(in) $\sim 6.1.10^6$, DC2 Pth(in) $\sim 3.8.10^6$

The volume density and thermal pressure maps show several outliers. For
instance, regions around the LoSs E and G have a high volume density and
thus thermal pressure. These two LoSs belong to transition zones between
the dense cores and their surroundings.

%%%%%%%%%%%%%%%%%%%%%%%%%%%%%%%%%%%%%%%%%%%%%%%%%%%%%%%%%%%%%%%%%%%%%%%%%%%

\newcommand{\FigTafallaComparison}{%
  \begin{figure*}
    \centering %
    \includegraphics[width=\linewidth]{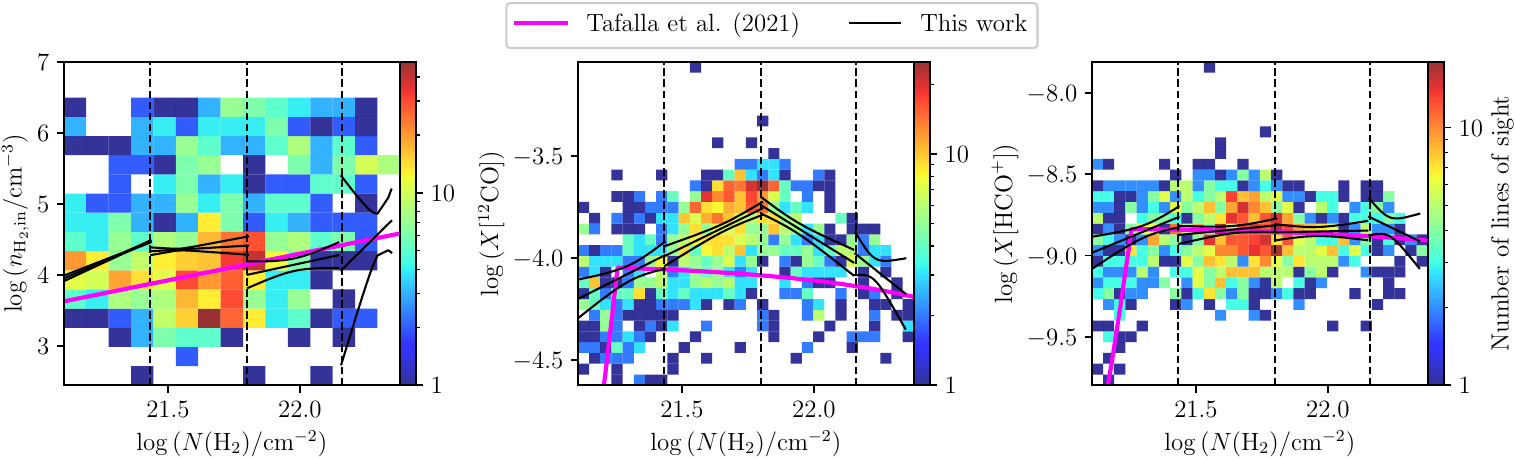}
    \caption{Comparison between the sandwich and TUH models.
      \textbf{Left:} Volume density of the inner layer $\nHHl{\text{in}}$
      estimated with the sandwich model as a function of the \HH{} column
      density. \textbf{Middle and Right:} Comparison of the mean abundance
      estimated with the sandwich model as a function of the \HH{} column
      density and the model proposed by~\citet{Tafalla21}. The second and
      third panels show this comparison for \twCO{} and \HCOp{},
      respectively.  In all panels, joint histograms show our data. The
      black lines show the power law fits and their uncertainty
      intervals. The pink lines show the \mbox{Tafalla et al.'s} model. The
      dashed vertical lines mark the \NHH~corresponding to a visual
      extinction at 3, 7, and 16\magn.}
    \label{fig:tafalla:comparison}
  \end{figure*}
}%

%%%%%%%%%%%%%%%%%%%%%%%%%%%%%%%%%%%%%%%%%%%%%%%%%%%%%%%%%%%%%%%%%%%%%%%%%%%

\FigTafallaComparison{}%

\section{Discussion: Comparison of the estimated mean abundances with the
  model proposed by~\citet{Tafalla21} }
\label{sec:discussion}

To model the line emission toward the Perseus, California, and Orion A
molecular clouds, \citet{Tafalla21,Tafalla23} introduce and apply a model
of the abundance of molecular species as a function of the \HH{} column
density (a summary is given in Appendix~\ref{app:tafalla}).  We refer to
this model as the TUH one from this point on. The left panel of
Fig.~\ref{fig:tafalla:comparison} overlays TUH model onto the joint
distributions of the estimated volume density of the inner layer as a
function of the \HH{} column density. The two models are consistent for
$\llog({\NHH/\pscm}) > 21.8$, but they show differences of about a factor
of two for lower column densities.

The middle and right panels of Fig.~\ref{fig:tafalla:comparison} overlay
the TUH model onto the joint distributions of the estimated mean abundance
of \twCO{} and \HCOp{} versus \HH{} column density toward the Horsehead
nebula. These joint distributions are identical to these displayed in
Fig.~\ref{fig:abundance-vs-av:pdfs:chemical}, except that the molecular
visual extinctions were scaled to the \HH{} column density using
Eq.~\ref{eq:NHH:from:av}. For an easier comparison, we also show the fits
listed in Table~\ref{tab:wmse}.

The \twCO{} mean abundance profile that we derive toward the Horsehead
nebula is different from TUH one. Our power-law fit reaches its maximum for
\mbox{$\llog({\NHH/\pscm}) = 21.8$}, while TUH model starts to decline at
$\llog({\NHH/\pscm}) \sim 21.3$. Moreover, the maximum abundance of \twCO{}
is two to three times higher than that predicted by the TUH model.

In contrast, the agreement between our estimated \HCOp{} mean abundance
profile and TUH model is good for
\mbox{$21.4 < \llog({\NHH/\pscm}) < 22.2$}. Both profiles show a slow
decrease as \NHH{} increases.  Above \mbox{$\llog({\NHH/\pscm}) = 22.2$},
our estimation yields a steeper fall of the \HCOp{} abundance than TUH
model, but TUH model still lies within our fit confidence interval.

For both species, TUH model predicts a sharp drop of the abundance below
$\llog({\NHH/\pscm}) \lesssim 21.3$ or $\Av = 2\magn$, resulting from the
molecule photodissociation. The Horsehead nebula contains few LoSs in this
\Av{} regime at the very outskirts of the pillar. It is thus difficult to
draw a robust conclusion for this regime.

%%%%%%%%%%%%%%%%%%%%%%%%%%%%%%%%%%%%%%%%%%%%%%%%%%%%%%%%%%%%%%%%%%%%%%%%%%%

\section{Conclusions}
\label{sec:conclusion}

In this paper, we have proposed a multilayer model that allows us 1) to fit
the low$-J$ lines of different species toward a heterogeneous LoS and 2) to
recover information about the chemical abundances, the kinematics, and the
physical state (kinetic temperature and volume density) of the gas at
different depths along the LoS. To do this, we introduced a fast and
efficient maximum likelihood estimator that simultaneously fits all the
lines available for each pixel independently.

We also took special care in selecting adequate physical and chemical
assumptions in order to avoid biasing the results. Following the
recommendations of \citet{Roueff24}, we not only searched for small and
unbiased fit residuals but also coherent physical and chemical results.
More precisely, in addition to fixing the LoS geometry to three layers
(similar to a core surrounded by an envelope), we approximated a centroid
velocity gradient along the LoS by allowing the layers to have different
velocities in order to be able to reproduce the observed line
asymmetries. We also proposed a varying chemistry along the LoS in order to
take into account the different processes (e.g., fractionation or molecular
freeze-out) at play when the gas is more or less exposed to far UV or more
or less dense and cold.  While we assumed that the relative abundances
$\twCO/\thCO$, $\CeiO/\thCO$, and $\HthCOp/\HCOp$ scale with the elemental
abundances of the corresponding carbon and oxygen isotopes in the inner
dense layer, we fit the column densities of \twCO{}, \thCO, \CeiO, and
\HCOp{} in the outer translucent layer, and we deduced the column density
of \HthCOp{} with the constraint $N(\HthCOp)/N(\thCO) = N(\HCOp)/N(\twCO)$.

We used this model to fit the \Jone{} and \Jtwo{} lines of the CO and
\HCOp{} isotopologues toward the translucent, filamentary, and dense core
LoSs in the Horsehead pillar. Our main astrophysical findings are as
follows:
\begin{itemize}
\item A precise modeling of the chemistry and physics toward filaments and
  dense cores requires a heterogeneous model composed of at least an inner
  dense layer surrounded by two outer translucent layers.
\item The proposed model succeeds in reproducing integrated intensities
  within $20\,\%$ for all lines. The associated bias is small for the
  \thCO{} and \CeiO{} lines.
\item We quantitatively confirm that the different lines of the CO and
  \HCOp{} isotopologues are sensitive to different depths along the LoS:
  The emission from the \twCO{} \Jone{} line is dominated by the foreground
  layer (85\%). The integrated intensity of the \thCO{} \Jone{} and \Jtwo{}
  lines comes from all three layers at approximately the same share. The
  inner layer has the largest contribution to the total integrated line
  intensity of the \CeiO{} \Jone{} and \Jtwo{} lines, about 60\%. The
  emission from the \HCOp{} and \HthCOp{} \Jone{} lines is dominated by the
  outer layers (foreground and background), but the inner layer also
  contributes significantly. Moreover, the emission from the \HCOp{} and
  \HthCOp{} \Jone{} lines is sub-thermalized with an excitation temperature
  lower than 10\K{}.
\item Deduced chemical properties clearly show 1) the depletion of \CO{}
  isotopologues toward the dense core LoSs and 2) the increase of the
  \twC{}O and \HCOp{} column density as a function of \Av{} or $G_0/\Av$ in
  the warm translucent LoSs. The basic abundances of \twCO, \thCO, and
  \CeiO~defined in Eq.~\ref{eq:abundance} are reached at visual extinction
  between 3 and 7\magn{}, respectively. The typical mean abundance of
  \HCOp{} with respect to \HH{} is $\sim 1.3 \times 10^{-9}$, which is
  about two times lower than the usual value in diffuse
  clouds~\citep{Gerin19}.  The estimated column density ratio
  \mbox{\NthCO/\NCeiO} in the envelope increases with decreasing visual
  extinction and reaches $25$ in the pillar outskirts.
\item In addition to confirming the general rotation of the Horsehead
  pillar around its axis, we found the signature of accreting motions of
  the foreground layer onto the inner denser gas.
\item While the inferred \Tkin~of the envelope varies from $25$ to
  \mbox{$40\,\K$}, that of the inner layer drops to \mbox{$\sim 15\,\K$} in
  the western dense core. The estimated \nHH~in the inner layer is
  \mbox{$\sim 3\times10^4\,\pccm$} toward the filament, and it increases by
  a factor of ten toward dense cores.  The thermal pressure of the outer
  layer shows a decreasing east-west gradient.
\end{itemize}
Except for the abundance ratios in the inner layer that were fixed to ease
the estimation in the dense core LoSs, the parameters were estimated from
the data with confidence intervals. It will be useful to analyze other
datasets in order to check whether the used physical and chemical
hypotheses depend on the considered region. If they indeed depend on the
physical gas conditions or region, trying to process large maps would imply
automatically selecting the most suitable assumptions for a given LoS.
This would probably require some insight into the dominant chemical
processes as a function of the typical volume density and kinetic
temperature. Processing large maps will also require dealing with spectra
that present multiple velocity components.

%%%%%%%%%%%%%%%%%%%%%%%%%%%%%%%%%%%%%%%%%%%%%%%%%%%%%%%%%%%%%%%%%%%%%%%%%%%

\begin{acknowledgements}
  We thank the referee for useful comments that helped us improve the
  manuscript.
  % Observations
  This work is based on observations carried out under project numbers
  019-13, 022-14, 145-14, 122-15, 018-16, and finally the large program
  number 124-16 with the IRAM 30m telescope. IRAM is supported by INSU/CNRS
  (France), MPG (Germany) and IGN (Spain).
  % ANR DAOISM and PCMI.
  This work received support from the French Agence Nationale de la
  Recherche through the DAOISM grant ANR-21-CE31-0010, and from the
  Programme National ``Physique et Chimie du Milieu Interstellaire'' (PCMI)
  of CNRS/INSU with INC/INP, co-funded by CEA and CNES.
  % Javier & Miriam
  M.G.S.M. and J.R.G. thank the Spanish MICINN for funding support under
  grant PID2019-106110GB-I00.  M.G.S.M acknowledges support from the NSF
  under grant CAREER 2142300.
  % Darek
  Part of the research was carried out at the Jet Propulsion Laboratory,
  California Institute of Technology, under a contract with the National
  Aeronautics and Space Administration (80NM0018D0004).
  D.C.L. acknowledges financial support from the National Aeronautics and
  Space Administration (NASA) Astrophysics Data Analysis Program (ADAP).
\end{acknowledgements}

\bibliographystyle{aa} %
\bibliography{aa51567-24corr} %

\begin{appendix}

%%%%%%%%%%%%%%%%%%%%%%%%%%%%%%%%%%%%%%%%%%%%%%%%%%%%%%%%%%%%%%%%%%%%%%%%%%% 

\newcommand{\FigAvCorrection}{%
  \begin{figure}[h]
    \includegraphics[width=\linewidth]{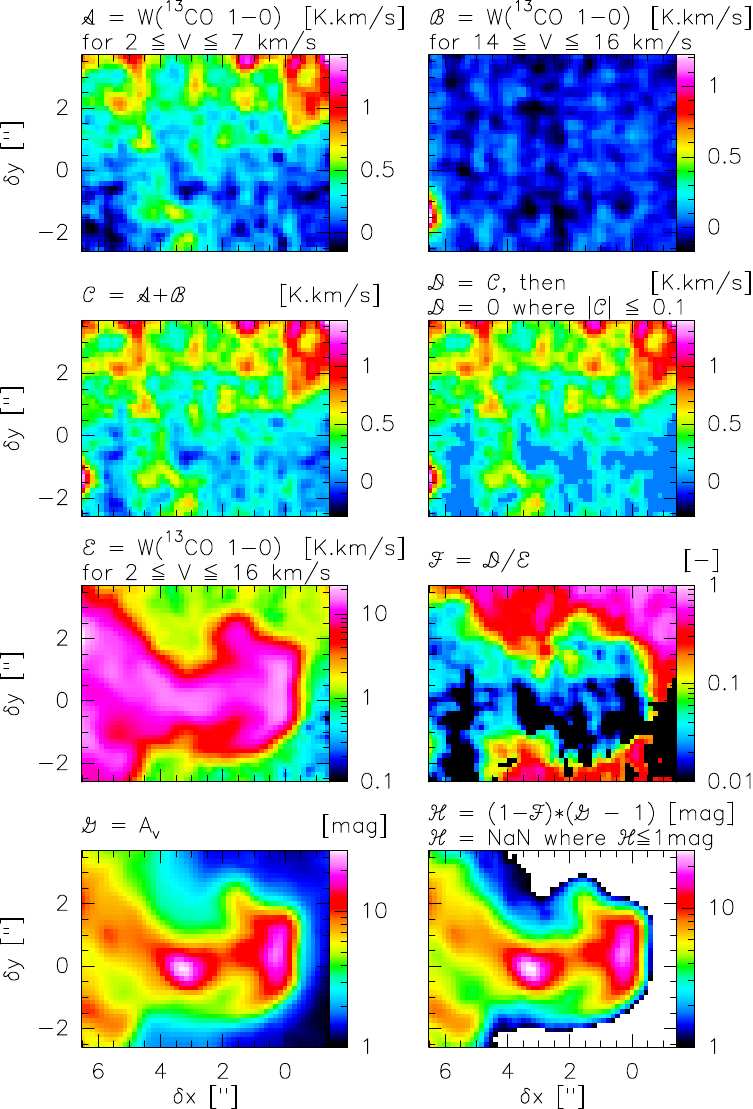}
    \caption{Steps to correct the visual extinction for 1) the line
      integrated emission that arises outside of the $[7,14\kms]$, and 2)
      the contribution from atomic hydrogen.}
    \label{fig:av:correction}
  \end{figure}
}

%%%%%%%%%%%%%%%%%%%%%%%%%%%%%%%%%%%%%%%%%%%%%%%%%%%%%%%%%%%%%%%%%%%%%%%%%%%

\section{Corrections to the measured visible extinction}
\label{app:av:correction}

\FigAvCorrection{}%

The visible extinction estimated from the dust spectral energy
distribution is a measure of all hydrogen atoms present along each line
of sight. In order to deduce the amount of molecular hydrogen associated
to the modeled lines, we thus needed to correct it for two effects that are
illustrated in Fig.~\ref{fig:av:correction}.

First,~\citet{Pety17} show that the southwestern edge of the Orion B GMC
has on average two velocity components: the brightest one centered around
$10.5\,\kms{}$ and a fainter one centered around $5\,\kms{}$. Each of them
has a typical FWHM of 4\kms. The model fitted in this paper still has
limitations to be able to fit multiple velocity components. We thus
restricted the studied velocity interval around the bright component,
namely, between 7 and $14\,\kms{}$. We used the method presented
in~\citet{einig23} to define a position-position-velocity mask of LoS
containing some signal on the \thCO\Jone~line. We choose this line as the
best compromise between optical depth and \SNR. We estimated that the
relevant velocity interval is \mbox{$[2,16]\,\kms$}. It is thus obvious
that part of the estimated visual extinction is associated with line
emission between \mbox{$[2,7]\,\kms$} and \mbox{$[14,16]\,\kms$}. We thus
computed the \thCO\Jone~integrated intensity inside these two
intervals. This integrated intensity map has negative values in regions
where the \SNR~is low. An estimation based on the negative intensity values
leads us to conclude that we can only trust the integrated intensity inside
the above velocity intervals when it is larger than $0.1\,\Kkms$.  We thus
set the integrated intensity to 0 when it is lower than $0.1\,\Kkms$. We
then computed the intensity integrated inside \mbox{$[2,16]\,\kms$}, and
the contribution of the \mbox{$[2,7]\,\kms$} and \mbox{$[14,16]\,\kms$}
velocity intervals to the \thCO\Jone~intensity integrated inside
\mbox{$[2,16]\,\kms$}. This allowed us to compute the fraction of the
\thCO\Jone~integrated intensity coming from the \mbox{$[7,14]\,\kms$}
velocity interval.

Second, part of the hydrogen may still be in atomic form. Looking at the
\HI~emission toward the Orion B GMC,~\citet{Pety17} estimate that about
\mbox{1\,\magn} of visual extinction is coming from atomic hydrogen.  We
thus subtracted \mbox{1\,\magn} to the total dust extinction, and we
applied the previously computed fraction to this dust extinction corrected
for the atomic contribution. This gives the amount of visual extinction
coming from molecular gas in the \mbox{$[7,14]\,\kms$} velocity
interval. As the low \Av~part is still uncertain, we additionally nullified
all the LoSs whose visual extinction associated with molecular gas is lower
than \mbox{1\,\magn}.  In this study, we use this ``molecular'' visual
extinction.

%%%%%%%%%%%%%%%%%%%%%%%%%%%%%%%%%%%%%%%%%%%%%%%%%%%%%%%%%%%%%%%%%%%%%%%%%%%

\section{Radiative transfer equation for a $M$-layer cloud}
\label{app:m-layers}

\begin{figure}[!htb]
  \includegraphics[width=\linewidth]{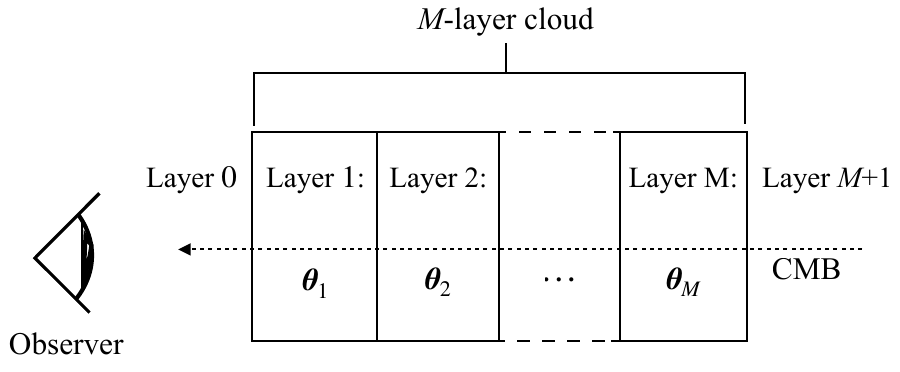}
  \caption{Sketch of a $M$-layer model of a line of sight.  Each slab has
    homogeneous physical and chemical conditions (characterized by the
    parameter vectors $\bthetal{i}$, for $1 \leq i \leq M$).  The emission
    from the CMB (layer $M+1$) happens at infinity, and it is assumed
    homogeneous and isotropic.}
  \label{fig:Mlayer}
\end{figure}

Let a $M$-layer cloud model of the LoS as depicted in \ref{fig:Mlayer}. A
simple generalization of the computation for a sandwich model (including
the ON-OFF calibration procedure) yields
\begin{eqnarray*}
  \funci{s}{M} &=& \sum_{i = 1}^{M}  J(\Texl{i}) \bracket{ 1 - \exp \rbracket{-\funci{\Psi}{i}}} \exp \rbracket{- \sum_{j = 1}^{i-1} \funci{\Psi}{j}} \\
               &-& J(\TCMB) \bracket{1 - \exp \rbracket{- \sum_{k = 1}^{M} \funci{\Psi}{k}}}.
                   \label{eq:s:M:layers}
\end{eqnarray*}
Assuming that there exist a layer 0 of opacity $\funci{\Psi}{0} = 0$, we
can rewrite this equation as
\begin{equation}
  \funci{s}{M} = \sum_{m = 0}^{M}  \bracket{J(\Texl{m+1}) - J(\Texl{m})} \exp \rbracket{-\sum_{k = 0}^{m} \funci{\Psi}{k}},
  \label{eq:s:M:layers:proposition}
\end{equation}
where we have written \mbox{$J(\TCMB) = J(\Texl{0}) = J(\Texl{M+1})$} only
for mathematical convenience (i.e., it is not physically
interpretable). This alternate expression simplifies the computation of
the Fisher matrix and Cramér-Rao bound.

Indeed, developing Eq.~\ref{eq:s:M:layers}, we have
\begin{equation*}
  \begin{array}{lll}
    \funci{s}{M} = & J(\Texl{1}) \bracket{ 1 - \exp \rbracket{-\Psi_1}} \exp \rbracket{-\Psi_0}
    \\ 
                   & +
                     \, J(\Texl{2}) \bracket{ 1 - \exp\rbracket{-\Psi_2}} \exp \rbracket{-\Psi_0-\Psi_1}
    \\
                   & + \cdots
    \\
                   & +
                     \, J(\Texl{M}) \bracket{ 1 - \exp\rbracket{-\Psi_M}} \exp \rbracket{-\sum_{j = 0}^{M-1} \funci{\Psi}{j}}
    \\ 
                   & -
                     \,  J(\TCMB) \bracket{1 - \exp \rbracket{-\sum_{k = 0}^{M} \funci{\Psi}{k}}}.
  \end{array}
\end{equation*}
Reorganizing the terms yields
\begin{equation}
  \begin{array}{lll}
    \funci{s}{M} = & J(\Texl{1}) \exp \rbracket{-\Psi_0} - J(\Texl{1}) \exp \rbracket{-\Psi_1 -\Psi_0}
    \\ 
                   & +
                     \, J(\Texl{2}) \exp \rbracket{-\Psi_0-\Psi_1} \\
                   & - J(\Texl{2})\exp\rbracket{-\Psi_2-\Psi_0-\Psi_1}
    \\
                   & + \cdots
    \\
                   & +
                     \, J(\Texl{M}) \rbracket{-\sum_{j = 0}^{M-1} \funci{\Psi}{j}} - J(\Texl{M}) \exp \rbracket{-\sum_{j = 0}^{M} \funci{\Psi}{j}}
    \\ 
                   & -
                     \,  J(\TCMB) + J(\TCMB) \exp \rbracket{-\sum_{k = 0}^{M} \funci{\Psi}{k}},
  \end{array}
  \label{eq:Mlayers:1}
\end{equation}
and
\begin{equation}
  \begin{array}{lll}
    \funci{s}{M} = & J(\Texl{1}) \exp \rbracket{-\Psi_0} - J(\Texl{1}) \exp \rbracket{-\Psi_1 -\Psi_0}
    \\ 
                   & +
                     \, J(\Texl{2}) \exp \rbracket{-\Psi_0-\Psi_1} \\
                        
                   & - J(\Texl{2}) \exp\rbracket{-\Psi_2-\Psi_0-\Psi_1}
    \\
                   & + \cdots
    \\
                   & +
                     \, J(\Texl{M}) \rbracket{-\sum_{j = 0}^{M-1} \funci{\Psi}{j}} - J(\Texl{M}) \exp \rbracket{-\sum_{j = 0}^{M} \funci{\Psi}{j}}
    \\ 
                   & -
                     \,  J(\Texl{0}) + J(\Texl{M+1}) \exp \rbracket{-\sum_{k = 0}^{M} \funci{\Psi}{k}}.
  \end{array}
  \label{eq:Mlayers:2}
\end{equation}
By factorizing each term
$\exp \rbracket{-\sum_{k = 0}^{m} \funci{\Psi}{k}}$ for $m \leqslant M$ in
Eq.~\ref{eq:Mlayers:2}, we get
\begin{equation*}
  \begin{array}{lll}
    \funci{s}{M} = & \bracket{J(\Texl{1}) - J(\Texl{0})} \exp \rbracket{-\Psi_0}
    \\ 
                   & +
                     \, \bracket{J(\Texl{2}) - J(\Texl{1})} \exp \rbracket{-\Psi_0-\Psi_1}
    \\
                   & + \cdots
    \\
                   & +
                     \, \bracket{J(\Texl{M+1}) - J(\Texl{M})} \exp \rbracket{-\sum_{k = 0}^{m} \funci{\Psi}{k}}.
  \end{array}
  \label{eq:Mlayers:3}
\end{equation*}
This form can be rewritten as Eq.~\ref{eq:s:M:layers:proposition}.

%%%%%%%%%%%%%%%%%%%%%%%%%%%%%%%%%%%%%%%%%%%%%%%%%%%%%%%%%%%%%%%%%%%%%%%%%%%

\section{Negative log-likelihood minimization}
\label{appendix:likelihood:optimization}

% The changes proposed by the editor modifies the scientific
% signification. We thus get back to our initial formulation.
One technical challenge of the proposed sandwich model is to define an
algorithm that converges within a reasonable amount of time to the
parameter vector $\btheta$ that minimizes the NLL:
\begin{equation}
  -\llog{\mathcal{L}(\btheta)}=
  \frac{1}{2}\sum_{l}
  \frac{\|\bvectori{x}{l}-\bvectori{s}{l}\|^2}{\sigma^2_{b,l}} +
  \emr{constant},
  \label{eq_NLL:app}
\end{equation}
where the sum is performed on all the observed lines $\bvectori{x}{l}$ and
the sandwich modeled lines $\bvectori{s}{l}(\btheta)$, parametrized by the
unknown vector $\btheta$ (see Eq.~\ref{eq:s:total}).  The challenge here is
that the number of unknowns is much larger than in~\citet{Roueff24} for the
same amount of observed information.

Lines of sight are processed independently of each other.  The main steps
of the minimization of Eq.~\ref{eq_NLL:app} are similar to these described
in~\citet{Roueff24}: An initial estimation of the centroid velocity is
followed by a grid search, and finally by a gradient descent.

\subsection{Initial estimation of the centroid velocity}
\label{sec:initial:estimate:cv}

We sum the lines of \CeiO~and \thCO~to get a bright and approximately thin
velocity spectrum. The velocity at which the intensity of the summed
spectrum is maximum provides an initial estimation of the centroid velocity
\CV~for all the lines and for both the inner and outer layers.

\subsection{Initial optimization with a grid search}
\label{app:grid:search}

We start by searching inside a grid in order to find a good first
approximation of the vector $\btheta$ that minimizes the NLL. We detail
here the case of the ``improved'' model (see
Sect.~\ref{section:chemical:model}). We need to explore a six-dimentional
space \mbox{\cbracket{\Tkin, \nHH, \NthCO, \NHCOp, \FWHM, \CV}} for the
inner layer, and a nine-dimentional space \mbox{\cbracket{\Tkin, \nHH,
    \NtwCO, \NthCO, \NCeiO, \NHCOp, \FWHM, \CV}} for the outer layer.
Performing an exhaustive grid search is not an option because it would be
too time-consuming: Testing ten values per dimension requires $10^{15}$
calculations of Eq.~\ref{eq_NLL:app}. We thus used an iterative stochastic
search in a gridded space, that is, a given number of ``walkers'' randomly
explore the $15-$dimensional space and converge to their closest local
minima.  Increasing the number of walkers allows the probability of getting
trapped in a local minimum to be decreased, while increasing the number of
walker steps is useful to ensure convergence.

To save CPU time, the opacity $(\tau_{l})$ and excitation temperature
$(\Texl{l})$ of each molecular species $l$ are pre-calculated with RADEX on
fixed grids.  As the sampling of parameter space is identical for the inner
and outer layers, only one four-dimentional grid per species needs to be
pre-computed. This avoids an excessive load on the RAM memory.  
More precisely, for each molecular line $l$, we pre-computed $\tau_{l}$ and
$\Texl{l}$ on the following regularly sampled grid:
\begin{itemize}
\item The \HH~volume density $\nHH$ is sampled between $10^{2}$ and
  $10^{6.5}\,\pccm$ in logarithmic steps of $0.1$ (46 values).
\item The kinetic temperature $\Tkin$ is sampled between $4$ and $113\,\K$
  in logarithmic steps of $0.05$ (30 values).
\item The column density of \thCO~is sampled between $10^{13}$ and
  $10^{18}\,\pscm$ in logarithmic steps of $0.05$ (102 values).
\item The other column densities \NtwCO, \NCeiO, \NHCOp, \NHthCOp~are
  sampled on the same grid except that they are shifted by the value fixed
  by the abundance ratios mentioned in Eq.~\ref{eq:abundance}.
\item The \FWHM~is sampled linearly between $0.25$ and $2\,\kms$ in steps
  of $0.05\,\kms$ (36 values).
\end{itemize}
During the random walk, the column densities are probed under the following
constraints. For the inner layer,
\begin{equation}
  \begin{array}{ll}
    \NtwCO/\NthCO = 10^{1.8} \sim 63,
    \\
    \NthCO/\NCeiO = 10^{0.9} \sim 7.9,
    \\
    10^{2.88} \sim 759 \leq \NthCO/\NHCOp \leq 10^{3.78} \sim 6000,
    \\
    \text{and} \quad %
    \NHCOp/\NtwCO=\NHthCOp/\NthCO.
    \label{eq_inner}
  \end{array}
\end{equation}
For the outer layer, the constraints are less severe:
\begin{equation}
  \begin{array}{ll}
    10^{1.3} \sim 20 \leq \NtwCO/\NthCO \leq 10^{1.9} \sim 80,
    \\
    10^{0.75} \sim 5.6 \leq \NthCO/\NCeiO \leq 10^{1.4} \sim 25.1,
    \\
    10^{2.88} \sim 759 \leq \NthCO/\NHCOp \leq 10^{3.78} \sim 6000,
    \\
    \text{and} \quad %
    \NHCOp/\NtwCO=\NHthCOp/\NthCO.
  \end{array}
  \label{eq_outer}
\end{equation}

Empirically, we found that the following four-stage algorithm provides a
satisfactory stochastic optimization in a reasonable amount of time (less
than one minute per pixel on a 2024 laptop). At each stage, we select
$10\,\%$ of the walkers to save CPU time.
\begin{enumerate}
\item We initialize the centroid velocities of the inner and outer layers
  to the value estimated in App.~\ref{sec:initial:estimate:cv}, and we fix
  them to these values during stage 1 and 2.  We then randomly draw the
  $\btheta$ position of $10^6$ walkers in the remaining $13-$dimensional
  space, and we select the $10^5$ walkers located at positions with the
  lowest NLL values. Indeed, these walkers are the ones who have the
  highest chance to be located close to the minimum.
\item Each selected walker moves in a randomly chosen direction with a
  randomly chosen step size to explore its vicinity. Its $\btheta$ position
  is then updated, if and only if the new position decreases the NLL
  value. Ten such steps are made during this exploratory stage, and at the
  end we select the $10^4$ walkers whose position corresponds to the lowest
  NLL values.
\item Each selected walker now makes $10^3$ steps. In this stage, the two
  dimensions of centroid velocities are added to the search space. The size
  of the step along the centroid velocity axes is set to
  $\pm 10~\unit{m/s}$.  At the end of the this stage, we select the $10^3$
  walkers located at positions with the lowest NLL values.
\item Each of the $10^3$ remaining walkers makes $5000$ new steps to
  finalize the convergence.
\end{enumerate}
Obviously, these settings can be adjusted differently. And this algorithm
does not provide any guarantee of convergence. To boost confidence in the
estimated $\btheta$, one can run the procedure several times to check that
the result remains unchanged. However, checking the number of walkers that
have converged to the minimum NLL value found already gives a useful
indication.

\subsection{Final optimization with a gradient descent}

The solution provided by the grid search can be refined with a gradient
descent algorithm for which the Hessian matrix is replaced by the Fisher
matrix.  At each iteration $i$, the vector $\btheta$ is updated such as
\begin{equation}
  \label{eq:gradient}
  \widehat{\btheta}^{i+1} = \widehat{\btheta}^i - \alpha\,
  \left(\bI_F(\widehat{\btheta}^i)\right)^{-1}
  \nabla_{\btheta}{-\llog{\mathcal{L}}(\widehat{\btheta}^i)},
\end{equation}
where $\bI_F(\btheta)$ is the Fisher Information Matrix computed at
$\btheta$ without considering calibration noise (see
Eq.~\ref{eq:IF:no:calibration}), $\nabla_{\btheta}{-\llog{\mathcal{L}}}$ is
the gradient
\begin{equation}
  \nabla_{\btheta}{-\llog{\mathcal{L}(\btheta)}}
  = \sum_{l} \frac{1}{{\sigma_{b, l}^2}} \frac{\partial \bvector{s}_{l}}{\partial \btheta}^\top \rbracket{\bvectori{x}{l} - \bvectori{s}{l}},
\end{equation}
and $\alpha\in[0,\,1]$ is a parameter to control the convergence speed. To
estimate $\alpha$, we compute $\mathcal{L}(\btheta)$ for
$\alpha= \cbracket{0.1,\,0.5,\,1}$ and we make a quadratic fit to get an
estimation of $\alpha$ that minimizes the NLL in the interval $[0.1, 1]$.
The gradient descent algorithm is stopped when the difference of NLL
between two consecutive iterations is smaller than $10^{-6}$ (our criterion
convergence), or when the number of iterations reaches $30$ (which is
enough for the considered problem). During the gradient descent, the
abundance ratios remain unchanged. If one decides to apply the gradient on
all the searched column density, the constraint from Eq.~\ref{eq_inner} and
\ref{eq_outer} would be difficult to satisfy, but more important, the
Fisher matrix could become singular.

%%%%%%%%%%%%%%%%%%%%%%%%%%%%%%%%%%%%%%%%%%%%%%%%%%%%%%%%%%%%%%%%%%%%%%%%%%%

\section{Fisher matrix calculation}
\label{appendix:Fisher:matrix}

We compute the Fisher matrix for the $M$-layer model described in the
previous section rather than to the limiting case of the sandwich model.

\subsection{Fisher matrix as a function of gradients
  $\frac{\partial \bs_M}{\partial \btheta_i}$}

To compute the Fisher matrix for several observed lines
\mbox{$\bvectori{x}{1}$, ..., $\bvectori{x}{L}$}, we just need to sum the
Fisher matrices of the different molecular lines
\begin{equation}
  \bI_F(\btheta;
  \sum_l \bvectori{x}{l})
  =\sum_l \bI_F(\btheta,\bvectori{x}{l}).
\end{equation}
This section explains the computations required to yield the Fisher matrix
for a single line $\bx_l$.

As discussed in Sect.~\ref{sec:constrained:optimization}, we consider the
case where abundance ratios with respect to \thCO~are fixed. The vector of
unknowns can be written as
\begin{equation}
  \btheta = \bracket{\bthetal{1}, \ldots, \bthetal{M}}, 
\end{equation}
where each layer of index $m$ is characterized by the vector of physical
parameters
\begin{equation}
  \bthetal{m}=[\llog{\nHHl{m}},\llog{\Tkinl{m}},\llog{\NthCOl{m}},\text{FWHM}_{m},\CVl{m}].  
\end{equation}
The Fisher matrix is therefore of size \mbox{$5M\times 5M$}.

In their Appendix~A.2,~\citet{Roueff24} showed that, when the calibration
noise is taken into account (see Eq.~\ref{eq:x}), the term $i,j$ of the
Fisher matrix can be written as
\begin{equation}
  \begin{array}{lll}
    \left[\bI_F(\btheta;\bvector{x}_l)\right]_{ij}&=&
                                                      \left(\frac{\sigma_l^2+\sigma_{c,l}^4(\bs_M^T\bs_M)}{\sigma_{b,l}^2\sigma_l^2}\right)
                                                      \left(\frac{\partial \bs_M^T}{\partial\theta_i}\frac{\partial
                                                      \bs_M}{\partial\theta_j}\right)
    \\
                                                  &+&
                                                      \left(
                                                      \frac{\sigma_{b,l}^2\sigma_{c,l}^4-\sigma_{c,l}^6(\bs^T\bs)}{\sigma_{b,l}^2\sigma_l^4}
                                                      -\frac{\sigma_{c,l}^2}{\sigma_{b,l}^2\sigma_l^2}
                                                      \right)
                                                      \left(\bs_M^T\frac{\partial\bs_M}{\partial\theta_i}\right)
                                                      \left(\bs_M^T\frac{\partial\bs_M}{\partial\theta_j}\right),
  \end{array}
  \label{eq:fisher:ij:1}
\end{equation}
where $\bs_M$ is defined in Eq.~\ref{eq:s:M:layers}, and
\mbox{$\sigma_l^2=\sigma_{b,l}^2+\sigma_{c,l}^2(\bs^T\bs)$}. When the
calibration noise is neglected, Eq.~\ref{eq:fisher:ij:1} simplifies into
\begin{equation}
  \left[\bI_F(\btheta;\bvector{x}_l)\right]_{ij}=
  \frac{1}{\sigma_{b,l}^2}
  \left(\frac{\partial \bs_M^T}{\partial\theta_i}\frac{\partial
      \bs_M}{\partial\theta_j}\right).
  \label{eq:IF:no:calibration}
\end{equation}
In both cases, only the gradients
\mbox{$\frac{\partial \bs_M^T}{\partial\theta_i}$} for all unknown
parameters $\theta_i$ are needed to complete the computations of the Fisher
matrix.

\subsection{Computation of $\frac{\partial \bs_M}{\partial \btheta_i}$}
  
We differentiate Eq.~\ref{eq:s:M:layers:proposition} to obtain for each
component $\thetal{i} \in \bthetal{i}$, and for $1 \leq i \leq M$
\begin{equation*}
  \begin{array}{lll}
    \frac{\partial \bs_M}{\partial \btheta_i} =
    &
      \frac{\partial J(\Texl{i})}{\partial \theta_i} \exp \rbracket{-\sum_{j = 0}^{i-1} \funci{\Psi}{j}}
    \\ 
    & -
      \, \frac{\partial J(\Texl{i})}{\partial \theta_i} \exp \rbracket{-\sum_{j = 0}^{i} \funci{\Psi}{j}}
    \\ 
    & -
      \, \sum_{m = i}^{M} \bracket{ J(\Texl{m+1}) - J(\Texl{m})} \exp \rbracket{-\sum_{k = 0}^{m} \funci{\Psi}{k}} \frac{\partial \funci{\Psi}{i}}{\partial \theta_i}.
  \end{array}
  \label{eq:Mlayers:3}
\end{equation*}
Finally,
\begin{equation}
  \begin{array}{ll}
    \frac{\partial \bs_M}{\partial \btheta_i} = & \frac{\partial
                                                  J(\Texl{i})}{\partial
                                                  \theta_i} \bracket{1
                                                  - \exp
                                                  \rbracket{-\funci{\Psi}{i}}}
                                                  \exp
                                                  \rbracket{-\sum_{j =
                                                  0}^{i-1} \funci{\Psi}{j}}
    \\ 
                                                & +
                                                  \, \frac{\partial \funci{\Psi}{i}}{\partial \theta_i}
                                                  \sum_{m = i}^{M} \bracket{ J(\Texl{m}) - J(\Texl{m+1})} \exp \rbracket{-\sum_{k = 0}^{m} \funci{\Psi}{k}},
  \end{array}
  \label{eq:Mlayers:4}
\end{equation}
where the terms $\frac{ \partial J(\Texl{i}) }{\partial \thetal{i}}$ and
$\frac{\partial \funci{\Psi}{i}}{\partial \thetal{i}}$ are expressed in
\ref{app:dJ:dtheta} and \ref{app:dpsi:dtheta} below.

\subsubsection{Computation of
  $\frac{ \partial J(\Texl{i}) }{\partial \thetal{i}}$}
\label{app:dJ:dtheta}
One has
\begin{align*}
  \frac{\partial J(\Texl{i})}{\partial \theta_i} = \frac{\partial J(\Texl{i})}{\partial \Texl{i}} \frac{\partial \Texl{i}}{\partial \theta_i}.
\end{align*}
Then,
\begin{align}
  \label{eq: general dJ/dtheta}
  \frac{\partial J(\Texl{i})}{\partial \theta_i} =
  \begin{cases}
    0 & \quad \text{if} \quad \theta_i = \CVl{i}, \\
    \rbracket{ \frac{h \nu}{k \Texl{i}}}^2 \frac{\exp \rbracket{\frac{h
          \nu}{k \Texl{i}}}}{ \bracket{ \exp \rbracket{\frac{h \nu}{k
            \Texl{i}}} -1 }^2} \frac{\partial \Texl{i}}{\partial \theta_i}
    & \quad \text{else}.
  \end{cases}
\end{align}

\subsubsection{Computation of
  $\frac{\partial \funci{\Psi}{i}}{\partial \thetal{i}}$}
\label{app:dpsi:dtheta}

One has
$\frac{\partial \funci{\Psi}{i}}{\partial \theta_i} = \frac{\partial
  \funci{\Psi}{i}}{\partial \taul{i}} \frac{\partial \taul{i}}{\partial
  \theta_i}$.  Then,
\begin{align}
  \label{eq: general dPsi/dtheta}
  \frac{\partial \funci{\Psi}{i}}{\partial \theta_i} =
  \begin{cases} 
    \bracket{ \frac{\partial \taul{i}}{\partial \theta_i} \,+\, \taul{i}
      \frac{\rbracket{V - \CVl{i}}^2}{\sigmaVl{i}^3} } \exp \rbracket{
      \frac{- \rbracket{V - \CVl{i}}^2}{2
        \sigmaVl{i}^2}} & \quad \text{if} \quad \theta_i = \sigmaVl{i}, \\
    \newline \\
    \taul{i} \frac{\rbracket{V - \CVl{i}}}{\sigmaVl{i}^2} \exp \rbracket{
      \frac{- \rbracket{V - \CVl{i}}^2}{2
        \sigmaVl{i}^2}} & \quad \text{else if} \quad \theta_i = \CVl{i}, \\
    \newline \\
    \frac{\partial \taul{i}}{\partial \theta_i} \exp \rbracket{ \frac{-
        \rbracket{V - \CVl{i}}^2}{2 \sigmaVl{i}^2}} & \quad \text{else}.
  \end{cases}
\end{align}

We now detail how the terms $\frac{\partial \Texl{i}}{\partial \theta_i}$
and $\frac{\partial \taul{i}}{\partial \theta_i}$ are computed. For
$\theta_i \in \cbracket{\llog{\nHHl{i}},\, \llog{\NthCOl{i}}}$,
$\frac{\partial \Texl{i}}{\partial \theta_i}$ and
$\frac{\partial \taul{i}}{\partial \theta_i}$ are numerically estimated
with RADEX by a finite difference technique with a step of 0.001 as it done
in~\citet{Roueff24}.

For $\theta_i = \log \Tkinl{i}$, one has
\begin{equation*}
  \llog {\Tkinl{i}} = \frac{\ln \rbracket{\Tkinl{i}}}{\ln \rbracket{10}}.
\end{equation*}
Then,
\begin{equation}
  \frac{\partial \Texl{i}}{\partial \llog{\Tkinl{i}}} =  \ln \rbracket{10} \, \Tkinl{i} \, \frac{\partial \Texl{i}}{\partial \Tkinl{i}}
\end{equation}
and
\begin{equation}
  \frac{\partial \taul{i}}{\partial \llog{\Tkinl{i}}} =  \ln \rbracket{10} \, \Tkinl{i} \, \frac{\partial \taul{i}}{\partial \Tkinl{i}},
\end{equation}
where $\frac{\partial \Texl{i}}{\partial \Tkinl{i}}$ and
$\frac{\partial \taul{i}}{\partial \Tkinl{i}}$ are numerically estimated
with RADEX by a finite difference technique with a step of
$\Tkinl{i}/1024$.
  
For $\theta_i = \sigmaVl{i}$, one has
\begin{equation*}
  \sigmaVl{i} = \frac{\FWHM_i}{\sqrt{8 \ln 2}}. 
\end{equation*}
Thus,
\begin{equation}
  \frac{\partial \Texl{i}}{\partial \sigmaVl{i}} = \sqrt{8 \ln 2}\, \frac{\partial \Texl{i}}{\partial \FWHM_i} %
  \quad \text{and} \quad %
  \frac{\partial \taul{i}}{\partial \sigmaVl{i}} = \sqrt{8 \ln 2}\, \frac{\partial \taul{i}}{\partial \FWHM_i},
\end{equation}
where the terms $\frac{\partial \Texl{i}}{\partial \FWHM_i}$ and
$\frac{\partial \taul{i}}{\partial \FWHM_i}$ are numerically estimated with
RADEX by a finite difference technique with a step of $\FWHM_i/128$.

For $\theta_i = \CVl{i}$, one has
\begin{equation*}
  \frac{\partial \Texl{i}}{\partial \CVl{i}} = 0
\end{equation*}
and
\begin{equation*}
  \frac{\partial \taul{i}}{\partial \CVl{i}} = 0.
\end{equation*}

\subsection{Application to the sandwich model}

In the case of the sandwich model, $M=3$ and $\btheta_1=\btheta_3$. One
thus has
\begin{equation*}
  \frac{\partial \bs_M}{\partial \btheta_\text{ou}}=\frac{\partial \bs_M}{\partial \btheta_1}+\frac{\partial \bs_M}{\partial \btheta_3} %
  \quad \text{and} \quad
  \frac{\partial \bs_M}{\partial \btheta_\text{in}}=\frac{\partial \bs_M}{\partial \btheta_2},
\end{equation*}
where $\btheta_\text{ou}$ and $\btheta_\text{in}$ are the vector of
unknowns of the outer and inner layers respectively.

\section{Cramér-Rao lower bounds for derived quantities}
\label{appendix:crb:derived:quantities}

\subsection{Thermal pressure}

Following~\citet{Roueff24}, we compute the \CRBm{} on the logarithm of the
thermal pressure \Pthl{m} for each layer $m$ of the sandwich model as
\begin{equation}
  \label{eq:log:pth}
  \begin{array}{ll}
    \text{\CRB}({\llog{\Pthl{m}}}) &= \text{\CRB}(\llog{\Tkinl{m}}) \\ 
                                   &+ \text{ \CRB}(\llog{\nHHl{m}}) \\
                                   &+ 2\,\text{\CRB}(\llog{\Tkinl{m}}, \llog{\nHHl{m}}),
  \end{array}
\end{equation}
where $\text{\CRB}(\llog{\Tkinl{m}}, \llog{\nHHl{m}})$ is the off-diagonal
term of the Fisher matrix.

\subsection{Centroid velocity difference
  $\Delta{\CV} = \CVl{\text{in}} - \CVl{\text{ou}}$}

Let $\CVl{\text{in}}$ and $\CVl{\text{ou}}$, the centroid velocities of the
inner and outer layers, respectively.  The \CRB{} on
$\Delta{\CV} = \CVl{\text{in}} - \CVl{\text{ou}}$ is
\begin{equation}
  \label{eq:crb:delta:cv}
  \text{\CRB}(\Delta{\CV}) = \text{\CRB}(\CVl{\text{in}}) + \text{\CRB}(\CVl{\text{ou}}) - 2\,\text{\CRB}(\CVl{\text{in}}, \CVl{\text{ou}}),
\end{equation}
where $\text{\CRB}(\CVl{\text{in}}, \CVl{\text{ou}})$ is the off-diagonal
term of the Fisher matrix.

\subsection{Total column density}

We derive the logarithm of the total column density of a species $X$, noted
$\llog{N_\emr{tot}(X)}$, from the estimation $\llog{N_{m}\emr{(X)}}$ with
\begin{equation}
  \label{eq:total:N:app}
  \llog{N_\emr{tot}(X)} = \llog{\sum_m 10^{\llog{N_{m}\emr{(X)}}}}.
\end{equation}
This section details how we compute the Cramér-Rao lower bound for
$\llog{N_\emr{tot}(X)}$.  The formula is generalized for any multilayer
model in \ref{app:crb:N:tot:M:layers} and we deal with the sandwich model
case in \ref{app:crb:N:tot:sandwich}.

\subsubsection{Case of an M-layer cloud}
\label{app:crb:N:tot:M:layers}

Each layer $m$ of the multilayer model is characterized by a five-parameter
vector
\begin{equation}
  {\bthetal{m}=\{\llog{\nHHl{m}},\, \llog{\Tkinl{m}},\, \llog{\NthCOl{m}},\,\text{FWHM}_{m},\CVl{m}\}}.
\end{equation}
The Fisher matrix dimension is $5M\times 5M$.  As explains by
\citet[][Sect.~{3.8}, Eq.~{3.30}]{kay:97}, one can then compute the lower
bound on the variance of $\alpha = g(\btheta)$ with
\begin{equation}
  \label{eq:CRLB:variable:change}
  \textrm{var} \left(\widehat \alpha \right) \geq \frac{\partial g(\btheta)}{\partial \btheta} \CRBm(\btheta) \frac{\partial g(\btheta)^T}{\partial \btheta},
\end{equation}
where $\frac{\partial g(\btheta)}{\partial \btheta}$ is the Jacobian
matrix.  Thus, for the total column density with Eq.~\ref{eq:total:N:app},
one has $g(\btheta) = \llog{\sum_{m = 0}^M 10^{\llog{N_{m}\emr{(X)}}}}$,
and
\begin{equation}
  \label{eq:CRLB:log:N:tot:Jacobian}
  \frac{\partial g(\btheta)}{\partial \btheta} = \left[ \frac{\partial g(\btheta)}{\partial \btheta_1}\, \cdots  \frac{\partial g(\btheta)}{\partial \btheta_{5M}}\right],
\end{equation}
where
\begin{align}
  \label{eq:CRLB:variable:change:Jacobian}
  \frac{\partial g(\btheta)}{\partial \theta_i} = 
  \begin{cases} 
    \frac{ 10^{\llog{N_{m}\emr{(X)}}}}{\sum_{m = 0}^M
      10^{\llog{N_{m}\emr{(X)}}}}
    & \quad \text{if} \quad \theta_i = \llog{N_{m}\emr{(X)}}, \forall m \in \llbracket 1, M\rrbracket, \\
    \newline \\
    0 & \quad \text{else}.
  \end{cases}
\end{align}
 
\subsubsection{Application to the sandwich model}
\label{app:crb:N:tot:sandwich}

Applying the result of the previous section leads to
\begin{equation}
  \label{eq:log:N:tot:sandwich}
  \llog{N_\emr{tot}(X)} = \llog{\left(2 \times 10^{\llog{N_{1}\emr{(X)}}} + 10^{\llog{N_{2}\emr{(X)}}}\right)}, %
\end{equation}
and
\begin{align}
  \label{eq:CRLB:variable:change:Jacobian:sandwich}
  \frac{\partial g(\btheta)}{\partial \btheta_i} = 
  \begin{cases} 
    2 \times \frac{ 10^{\llog{N_{1}\emr{(X)}}}}{\sum_{m = 0}^M 10^{\llog{N_{m}\emr{(X)}}}}  & \quad \text{if} \quad \theta_i = \llog{N_{1}\emr{(X)}}, \\
    \newline \\
    \frac{ 10^{\llog{N_{2}\emr{(X)}}}}{\sum_{m = 0}^M 10^{\llog{N_{m}\emr{(X)}}}}  & \quad \text{if} \quad \theta_i = \llog{N_{2}\emr{(X)}}, \\
    \newline \\
    0 & \quad \text{else}.
  \end{cases}
\end{align}

\subsection{$\FWHMl{\text{in}}/\FWHMl{\text{ou}}$}

We let $\FWHMl{\text{in}}$ and $\FWHMl{\text{ou}}$ be the FWHM of the inner
and the outer layers, respectively.  We use
Eq.~\ref{eq:CRLB:variable:change} to compute the \CRB{} on the FWHM ratio.
In this case, ${g(\btheta) = \FWHMl{\text{in}}/\FWHMl{\text{ou}}}$ and
\begin{align}
  \label{eq:CRLB:variable:change:Jacobian:sandwich:FWHM}
  \frac{\partial g(\btheta)}{\partial \btheta_i} = 
  \begin{cases}  1/\FWHMl{\text{ou}} & \quad \text{if} \quad \theta_i = \FWHMl{\text{in}}, \\
    \newline \\
    - \FWHMl{\text{in}}/\FWHMl{\text{ou}}^2 & \quad \text{if} \quad \theta_i = \FWHMl{\text{ou}}, \\
    \newline \\
    0 & \quad \text{else}.
  \end{cases}
\end{align}

%%%%%%%%%%%%%%%%%%%%%%%%%%%%%%%%%%%%%%%%%%%%%%%%%%%%%%%%%%%%%%%%%%%%%%%%%%% 

\section{Piecewise power-law fit of $N_{\mathrm{tot}}$ as a function of
  $A_v$}
\label{appendix:wmse}

We aim at fitting power-laws through the total column density as a function
\Av{} for different ranges of \Av{}. This means to make a linear regression
in the log-log space, that is, fitting
\begin{equation}
  y_n=\alpha\,(x_n-x_0)+\beta+b_n,
  \label{eq:wmse}
\end{equation}
where $x=\log(A_v)$, $y=\log N_{\mathrm{tot}}$, and $x_0$ is the center of
the $\log A_v$ bin and $b_n$ is an additive noise drawn from a white
Gaussian distribution. Equation~\ref{eq:wmse} can be written:
\begin{equation}
  \boldsymbol{y}=\boldsymbol{M} \,\boldsymbol{z} + \boldsymbol{b},
\end{equation}
where $\boldsymbol{y}=[y_1,...,y_N]^T$,
$\boldsymbol{z}=[\alpha,\,\beta]^T$, $\boldsymbol{b}=[b_1,...,b_N]^T$ and
\begin{equation}
  \boldsymbol{M}=
  \left(
    \begin{array}{c}
      \boldsymbol{m}_1\\
      ...
      \\
      \boldsymbol{m}_N
    \end{array}
  \right)
  \quad
  \text{with}
  \quad
  \boldsymbol{m}_n=(x_n-x_0,\,1).
\end{equation}
Moreover, to take into account the uncertainty on $\log N_{\mathrm{tot}}$,
we assume that the variance of $b_n$ is proportional to the CRB of
$\log\,N_{\mathrm{tot}}$.
Thus, $\boldsymbol{B} \sim {\cal{N}}(0,\lambda \,\boldsymbol{R})$ where
$\lambda$ is unknown and $\boldsymbol{R}$ is a diagonal matrix with the CRB
of $\log\,N_{\mathrm{tot}}$.  The weighted least square estimator of
$\boldsymbol{z}$ is then
\begin{equation}
  \widehat{\boldsymbol{z}}=(\boldsymbol{M}^T \boldsymbol{R}^{-1} \boldsymbol{M})^{-1} \boldsymbol{M}^T \boldsymbol{R}^{-1} \boldsymbol{y},
\end{equation}
and $\lambda$ can be estimated with
\begin{equation}
  \widehat{\lambda}= \frac{1}{N} \sum_n \frac{\left(
      y_n-\boldsymbol{m}_n^T
      \,\widehat{\boldsymbol{z}} \right)^2}{R_n},
\end{equation}
and the estimation of the covariance of $\boldsymbol{z}$ is
\begin{equation}
  \widehat{\boldsymbol{\Gamma}}=\widehat{\lambda} \left(\boldsymbol{M}^T \boldsymbol{R}^{-1}\boldsymbol{M}\right)^{-1}.
\end{equation}
% which thus provides the error bars on $a$ and $b$.
For a given $x$, the estimation of $y$ is
\begin{equation}
  \widehat{y}=\widehat\alpha\,(x-x_0)+\widehat\beta
\end{equation}
and its variance is
\begin{equation}
  \mathrm{var}(\widehat{y})=
  \mathrm{var}(\widehat\alpha)\,(x-x_0)^2
  +\mathrm{var} (\widehat\beta)
  +2 \mathrm{cov} (\widehat\alpha, \widehat\beta)\,(x-x_0)
\end{equation}
where
$\mathrm{var}(\widehat\alpha)=\left[\widehat{\boldsymbol{\Gamma}}\right]_{1,1}$
$\mathrm{var}(\widehat\beta)=\left[\widehat{\boldsymbol{\Gamma}}\right]_{2,2}$
$\mathrm{cov}(\widehat\alpha,
\widehat\beta)=\left[\widehat{\boldsymbol{\Gamma}}\right]_{1,2}$.

%%%%%%%%%%%%%%%%%%%%%%%%%%%%%%%%%%%%%%%%%%%%%%%%%%%%%%%%%%%%%%%%%%%%%%%%%%% 

\newcommand{\TabTafallaParameters}{%
  \begin{table}[h!]
    \caption{%
      Values of the parameters in the TUH model of the abundance profile as a
      function of the \HH{} column density.}
    \centering
    \begin{tabular}{c|c|c|c}
      \hline
      \hline
      Species 
      & $X_0$
      & $A_0/\magn{}$
      & $n_{\mathrm{fr}}/\pccm$
      \\
      \hline
      $\twCO$ 
      & $9.5\e{-5}$
      & $2$
      & $10^{5}$
      \\
      $\HCOp$ 
      & $1.5\e{-9}$
      & $2$
      & $2\e{5}$
      \\
      \hline
    \end{tabular}
    % }
    \label{tab:tafalla21:params}
  \end{table}
}

%%%%%%%%%%%%%%%%%%%%%%%%%%%%%%%%%%%%%%%%%%%%%%%%%%%%%%%%%%%%%%%%%%%%%%%%%%% 
\section{TUH model in a nutshell}
\label{app:tafalla}

\TabTafallaParameters{}%

For CO and \HCOp{}, \citep{Tafalla21} assume that the abundance averaged
along the LoS, \abmean{}, is defined as
\begin{equation}
  \abmean\bracket{\NHH} = X_0 \times f_\emr{out}\bracket{\NHH} \times f_\emr{in}\bracket{\NHH},
\end{equation}
where $X_0$ is a scaling constant, $f_\emr{out}$ mostly models the
photodissociation of the species $X$ by the far UV field in translucent 
gas, and $f_\emr{in}$ mostly models the molecular freeze-out onto dust
grains in cold dense cores. They use the following simple analytic
expressions
\begin{align}
  f_\emr{out}\bracket{\NHH} =
  \begin{cases}
    \exp \bracket{6\,(\Av-A_0)} & \quad \text{if } \Av \le A_0, \\
    1 & \quad \text{if } \Av > A_0,
  \end{cases}
\end{align}
and
\begin{equation}
  \label{eq:fin}
  f_\emr{in}\bracket{\NHH} = \exp \paren{-\frac{\nHH}{{n_\emr{fr}}}},
\end{equation}
where $A_0$ and $n_\emr{fr}$ are the visual extinction and volume
densities, which characterize the transition zone for the photodissociation
and the molecular freeze-out, respectively. They estimate the volume
density with
\begin{equation}
  \label{eq:nHH:Tafalla}
  \nHH = 2\e4\,\pccm \, \paren{\frac{\NHH}{10^{22}\,\pscm}}^{0.75}.  
\end{equation}

%%%%%%%%%%%%%%%%%%%%%%%%%%%%%%%%%%%%%%%%%%%%%%%%%%%%%%%%%%%%%%%%%%%%%%%%%%%

\newcommand{\FigSpectraDecompositionTwoChemical}{%
  \begin{figure*}[h]
    \centering %
    \includegraphics[width=0.685\linewidth]{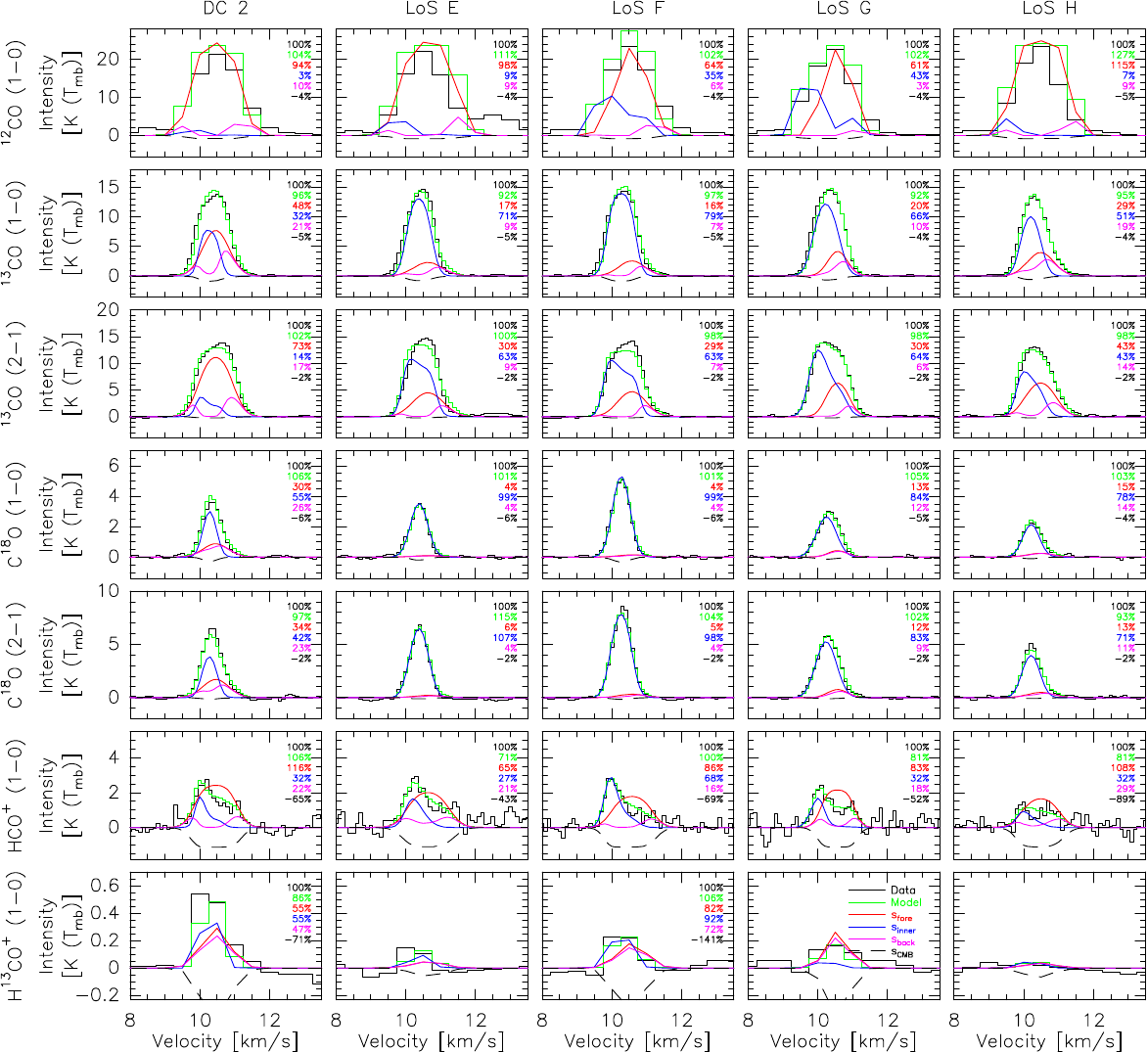}
    \caption{Same as Fig.~\ref{fig:spectra:decomposition:1:chemical} but
      for LoSs around the dense core 2.}
    \label{fig:spectra:decomposition:2:chemical}
  \end{figure*}
}
  
\newcommand{\FigSpectraDecompositionThreeChemical}{%
  \begin{figure*}[h]
    \centering %
    \includegraphics[width=0.685\linewidth]{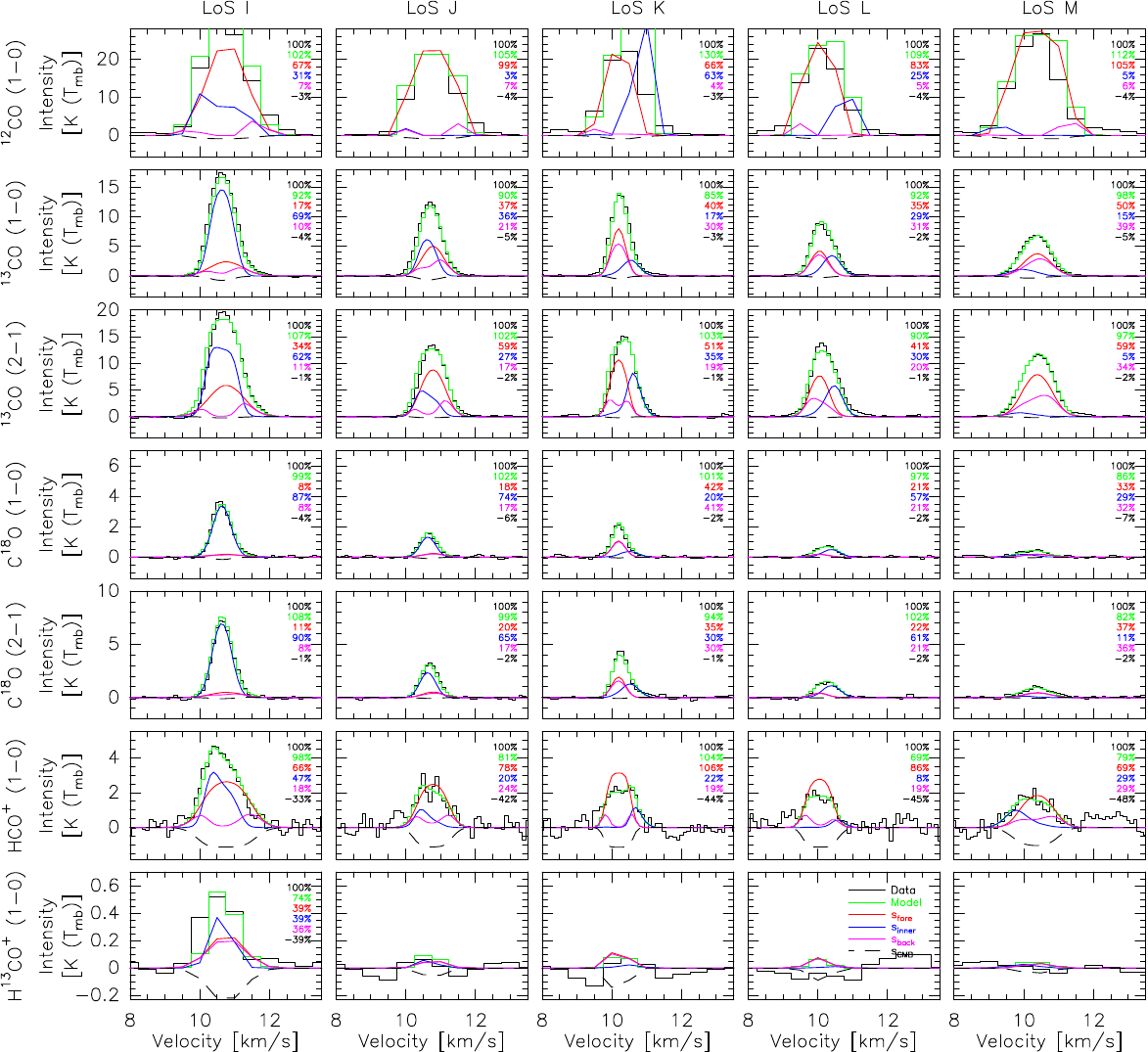}
    \caption{Same as Fig.~\ref{fig:spectra:decomposition:1:chemical} but
      for LoSs at the outskirt of the Horsehead nebula.}
    \label{fig:spectra:decomposition:3:chemical}
  \end{figure*}
}
  
%%%%%%%%%%%%%%%%%%%%%%%%%%%%%%%%%%%%%%%%%%%%%%%%%%%%%%%%%%%%%%%%%%%%%%%%%%%

\section{Supplementary figures}
\label{appendix:additional:figures}

\FigSpectraDecompositionTwoChemical{}%
\FigSpectraDecompositionThreeChemical{}%
\FigAbundancePDFsTwoChemical{}%

%%%%%%%%%%%%%%%%%%%%%%%%%%%%%%%%%%%%%%%%%%%%%%%%%%%%%%%%%%%%%%%%%%%%%%%%%%%

\end{appendix}

\end{document}